\documentclass[]{ss_areviews}

\usepackage[breaklinks]{hyperref}

\usepackage{natbib}

\usepackage{graphics}
\usepackage{times}
\usepackage{mathptm}

\newcommand{\bea}{\begin{eqnarray}}
\newcommand{\eea}{\end{eqnarray}}
\newcommand{\bc}{\begin{center}}
\newcommand{\ec}{\end{center}}

\renewcommand{\vec}[1]{ {\bf #1} }

\newcommand{\dd}{{\rm d}}

\usepackage[small,bf]{caption}

\begin{document}
\bibliographystyle{Astronomy}

\title{Smoothed Particle Hydrodynamics in Astrophysics}

\markboth{V. Springel}{Smoothed Particle Hydrodynamics}

\author{
Volker Springel
\affiliation{
Max-Planck-Institut f\"ur Astrophysik,
D-85741 Garching, Germany;\\
e-mail: volker@mpa-garching.mpg.de
}}

\jname{Annual Reviews of Astronomy and Astrophysics}
\jyear{2010}
\jvol{48}
\ARinfo{319:430}


\begin{abstract}
  This review discusses Smoothed Particle Hydrodynamics (SPH) in the
  astrophysical context, with a focus on inviscid gas dynamics. The
  particle-based SPH technique allows an intuitive and simple
  formulation of hydrodynamics that has excellent conservation
  properties and can be coupled to self-gravity easily and highly
  accurately. The Lagrangian character of SPH allows it to
  automatically adjust its resolution to the clumping of matter, a
  property that makes the scheme ideal for many applications in
  astrophysics, where often a large dynamic range in density is
  encountered. We discuss the derivation of the basic SPH equations in
  their modern formulation, and give an overview about extensions of
  SPH developed to treat physics such as radiative transfer, thermal
  conduction, relativistic dynamics or magnetic fields. We also
  briefly describe some of the most important applications areas of
  SPH in astrophysical research. Finally, we provide a critical
  discussion of the accuracy of SPH for different hydrodynamical
  problems, including measurements of its convergence rate for
  important classes of problems.
\end{abstract}

\maketitle

\section{INTRODUCTION}

Smoothed Particle Hydrodynamics (SPH) is a technique for approximating
the continuum dynamics of fluids through the use of particles, which
may also be viewed as interpolation points. SPH was originally
developed in astrophysics, as introduced by \protect\citet{Lucy1977}
and \protect\citet{Gingold1977} a good 30 years ago. Since then it has
found widespread use also in other areas of science and
engineering. In this review, I discuss SPH in its modern form, based
on a formulation derived from variational principals, giving SPH very
good conservative properties and making its derivation largely free of
ad-hoc choices that needed to be made in older versions of SPH.

A few excellent reviews have discussed SPH previously
\citep[e.g.][]{Monaghan1992,Monaghan2005,Dolag2008Review,Rosswog2009},
hence we will concentrate primarily on recent developments and on a
critical discussion of SPH's advantages and disadvantages, rather than
on giving a full historical account of the most important literature
on SPH. Also, we will for the most part restrict the discussion of SPH
to inviscid ideal gases, which is the relevant case for the most
common applications in astrophysics, especially in cosmology. Only in
passing we comment on some other important uses of SPH in astronomy
where fluids quite different from an ideal gas are modelled, e.g.~in
planet formation. Fully outside the scope of this review are the many
successful applications of SPH-based techniques in fields such as
geophysics or engineering.  For example, free-surface flows tend to be
very difficult to model with Eulerian methods, while this is
comparatively easy with SPH. As a result, there are many applications
of SPH to problems such as dam-braking, avalanches, and the like,
which we however will not discuss here.

Numerical simulations have become an important tool in astrophysical
research. For example, cosmological simulations of structure formation
within the $\Lambda$CDM model model have been instrumental to
understand the non-linear outcome of the initial conditions predicted
by the theory of inflation. By now, techniques to simulate
collisionless dark matter through the particle-based N-body method
\citep{Hockney1981} have fully matured and are comparatively well
understood.  However, to represent the collisional baryons as well,
the hydrodynamical fluid equations need to be solved, which represents
a much harder problem than the dark matter dynamics. On top of the
more complicated gasdynamics, additional physics, like radiation
fields, magnetic fields, or non-thermal particle components need to be
numerically followed as well to produce realistic models of the
formation and evolution of galaxies, stars or planets. There is
therefore ample need for robust, accurate, and efficient
hydrodynamical discretization techniques in astrophysics.

The principal idea of SPH is to treat hydrodynamics in a completely
mesh-free fashion, in terms of a set of sampling
particles. Hydrodynamical equations of motion are then derived for
these particles, yielding a quite simple and intuitive formulation of
gas dynamics. Moreover, it turns out that the particle representation
of SPH has excellent conservation properties.  Energy, linear
momentum, angular momentum, mass, and entropy (if no artificial
viscosity operates) are all simultaneously conserved. In addition,
there are no advection errors in SPH, and the scheme is fully Galilean
invariant, unlike alternative mesh-based Eulerian techniques.  Due to
its Lagrangian character, the local resolution of SPH follows the mass
flow automatically, a property that is extremely convenient in
representing the large density contrasts often encountered in
astrophysical problems.  Together with the ease in which SPH can be
combined with accurate treatments of self-gravity (suitable and
efficient gravity solvers for particles can be conveniently taken from
a cosmological N-body code developed for the representation of dark
matter), this has made the method very popular for studying a wide
array of problems in astrophysics, ranging from cosmological structure
growth driven by gravitational instability to studies of the
collisions of protoplanets.  Furthermore, additional `sub-resolution'
treatments of unresolved physical processes (such as star formation in
galaxy-scale simulations) can be intuitively added at the particle
level in SPH.

In this review, we first give, in Section~\ref{SecBasic}, a derivation
of what has become the standard formulation of SPH for ideal gases,
including also a description of the role of artificial viscosity.  We
then summarize some extensions of SPH to include additional physics
such as magnetic fields or thermal conduction in
Section~\ref{SecExtensions}, followed by a brief description of the
different application areas of SPH in astronomy in
Section~\ref{SecApplications}.  Despite the popularity of SPH, there
have been few systematic studies of the accuracy of SPH when compared
with the traditional Eulerian approaches. We therefore include a
discussion of the convergence, consistency and stability of standard
SPH in Section~\ref{SecAccuracy}, based on some of our own
tests. Finally, we give a discussion of potential future directions of
SPH development in Section~\ref{SecFuture}, and our conclusions in
Section~\ref{SecConclusions}.

\section{BASIC FORMULATION OF SPH FOR IDEAL GASES} \label{SecBasic}

\subsection{Kernel Interpolants}

At the heart of smoothed particle hydrodynamics lie so-called kernel
interpolants, which are discussed in detail by \citet{Gingold1982}. In
particular, we use a kernel summation interpolant for estimating the
density, which then determines the rest of the basic SPH equations
through the variational formalism.

For any field $F(\vec{r})$, we may define a smoothed interpolated
version, $F_s(\vec{r})$, through a convolution with a kernel
$W(\vec{r},h)$:
\begin{equation}
F_s(\vec{r}) = \int F(\vec{r})\, W(\vec{r} - \vec{r}',h)\,{\rm
d}\vec{r}'.
\label{eqk1}
\end{equation}
Here $h$ describes the characteristic width of the kernel, which is
normalized to unity and approximates a Dirac $\delta$-function in the
limit $h\to 0$. We further require that the kernel is symmetric and
sufficiently smooth to make it at least differentiable twice. One
possibility for $W$ is a Gaussian, which was in fact used by
\citet{Gingold1977}. However, most current SPH implementations are
based on kernels with a finite support. Usually a cubic spline is
adopted with $W(r, h)= w(\frac{r}{2h})$, and
\begin{equation}
w_{\rm 3D}(q) =\frac{8}{\pi} \left\{
\begin{array}{ll}
1-6 q^2 + 6 q^3, &
0\le  q \le\frac{1}{2} ,\\
2\left(1-q\right)^3, & \frac{1}{2}< q \le 1 ,\\
0 , & q>1 ,
\end{array}
\right.
\end{equation}
in three-dimensional normalization.  This kernel belongs to a broader
class of interpolation and smoothing kernels
\citep{Schoenberg1969,Hockney1981,Monaghan1985}. Note that in the
above most commonly used definition of the smoothing length $h$, the
kernel drops to zero at a distance of $r=2h$.  Through Taylor
expansion, it is easy to see that the kernel interpolant is at least
second-order accurate due to the symmetry of the kernel.

Suppose now we know the field at a set of points $\vec{r}_i$, i.e.
$F_i=F(\vec{r}_i)$. The points have an associated mass $m_i$ and
density $\rho_i$, such that $\Delta \vec{r}_i \sim m_i/\rho_i$ is
their associated finite volume element. Provided the points
sufficiently densely sample the kernel volume, we can approximate the
integral in Eqn.~(\ref{eqk1}) with the sum
\begin{equation}
F_s(\vec{r}) \simeq \sum_j \frac{m_j}{\rho_j}\,F_j\,
W(\vec{r} - \vec{r}_j,h).
\label{Eqkernelsum}
\end{equation}
This is effectively a Monte-Carlo integration, except that thanks to
the comparatively regular distribution of points encountered in
practice, the accuracy is considerably better than for a random
distribution of the sampling points.  In particular, for points in one
dimension with equal spacing $d$, one can show that for $h=d$ the sum
of Eqn.~(\ref{Eqkernelsum}) provides a second order accurate
approximation to the real underlying function. Unfortunately, for the
irregular yet somewhat ordered particle configurations encountered in
real applications, a formal error analysis is not straightforward. It
is clear however, that at the very least one should have $h\ge d$,
which translates to a minimum of $\sim 33$ neighbours in 3D.

Importantly, we see that the estimate for $F_s(\vec{r})$ is defined
everywhere (not only at the underlying points), and is differentiable
thanks to the differentiability of the kernel, albeit with a
considerably higher interpolation error for the derivative.  Moreover,
if we set $F(\vec{r}) = \rho(\vec{r})$, we obtain
\begin{equation}
\rho_s(\vec{r})  \simeq \sum_j  m_j  W (\vec{r}-\vec{r}_j, h),  \label{eqnDensEst}
\end{equation}
yielding a density estimate just based on the particle coordinates and
their masses.  In general, the smoothing length can be made variable
in space, $h=h(\vec{r},t)$, to account for variations in the sampling
density.  This adaptivity is one of the key advantages of SPH and is
essentially always used in practice. There are two options to
introduce the variability of $h$ into Eqn.~(\ref{eqnDensEst}). One is
by adopting $W(\vec{r}-\vec{r}_j, h(\vec{r}))$ as kernel, which
corresponds to the `scatter' approach \citep{Hernquist1989}. It has
the advantage that the volume integral of the smoothed field recovers
the total mass, $\int \rho_s(\vec{r})\,{\rm d}\vec{r} = \sum_i
m_i$. On the other hand, the so-called `gather' approach, where we use
$W(\vec{r}-\vec{r}_j, h(\vec{r}_i))$ as kernel in
Eqn.~(\ref{eqnDensEst}), requires only knowledge of the smoothing
length $h_i$ for estimating the density of particle $i$, which leads
to computationally convenient expressions when the variation of the
smoothing length is consistently included in the SPH equations of
motion. Since the density is only needed at the coordinates of the
particles and the total mass is conserved anyway (since it is tied to
the particles), it is not important that the volume integral of the
gather form of $\rho_s(\vec{r})$ exactly equals the total mass.

In the following we drop the subscript $s$ for indicating the smoothed
field, and adopt as SPH estimate of the density of particle $i$ the
expression
\begin{equation}
\rho_i  = \sum_{j=1}^N  m_j \, W (\vec{r}_i-\vec{r}_j, h_i).
\end{equation}
It is clear now why kernels with a finite support are preferred. They
allow the summation to be restricted to the $N_{\rm ngb}$ neighbors
that lie within the spherical region of radius $2h$ around the target
point $\vec{r}_i$, corresponding to a computational cost of order
${\cal O}(N_{\rm ngb}\,N)$ for the full density estimate.  Normally
this number $N_{\rm ngb}$ of neighbors within the support of the
kernel is approximately (or exactly) kept constant by choosing the
$h_i$ appropriately. $N_{\rm ngb}$ hence represents an important
parameter of the SPH method and needs to be made large enough to
provide sufficient sampling of the kernel volumes.  Kernels like the
Gaussian on the other hand would require a summation over all
particles $N$ for every target particle, resulting in a ${\cal
  O}(N^2)$ scaling of the computational cost.

If SPH was really a Monte-Carlo method, the accuracy expected from the
interpolation errors of the density estimate would be rather
problematic. But the errors are much smaller because the particles do
not sample the fluid in a Poissonian fashion. Instead, their distances
tend to equilibrate due to the pressure forces, which makes the
interpolation errors much smaller. Yet, they remain a significant
source of error in SPH and are ultimately the primary origin of the
noise inherent in SPH results.

Even though we have based most of the above discussion on the density,
the general kernel interpolation technique can also be applied to
other fields, and to the construction of differential operators. For
example, we may write down a smoothed velocity field and take its
derivative to estimate the local velocity divergence, yielding:
\begin{equation}
(\nabla\cdot\vec{v})_i = \sum_j \frac{m_j}{\rho_j} \vec{v}_j \cdot \nabla_i
  W(\vec{r}_i-\vec{r}_j, h) .
\end{equation}
However, an alternative estimate can be obtained by considering the
identity $\rho \nabla\cdot \vec{v} = \nabla(\rho\vec{v}) -
\vec{v}\cdot \nabla\rho$, and computing kernel estimates for the two
terms on the right hand side independently. Their difference then
yields
\begin{equation}
(\nabla\cdot\vec{v})_i = \frac{1}{\rho_i} 
\sum_j m_j (\vec{v}_j  - \vec{v}_i) \cdot \nabla_i W(\vec{r}_i-\vec{r}_j, h) .
\label{eqnVelDiv}
\end{equation}
This pair-wise formulation turns out to be more accurate in
practice. In particular, it has the advantage of always providing a
vanishing velocity divergence if all particle velocities are equal.

\subsection{Variational Derivation}

The Euler equations for inviscid gas dynamics in Lagrangian (comoving)
form are given by
\begin{equation}
\frac{{\rm d}\rho}{{\rm d}t} + \rho \,\nabla\cdot \vec{v} = 0 ,
\end{equation}
\begin{equation}
\frac{{\rm d}\vec{v}}{{\rm d}t} + \frac{\nabla P}{\rho} = 0 ,
\end{equation}
\begin{equation}
\frac{{\rm d}u}{{\rm d}t} + \frac{P}{\rho} \nabla \cdot \vec{v} = 0 ,
\end{equation}
where ${\rm d}/{\rm d}t  = \partial / \partial t + \vec{v}\cdot \nabla$
is the convective derivative. This system of partial
differential equations  expresses conservation of mass, momentum and
energy. \citet{Eckart1960} has shown that the Euler equations 
for an inviscid ideal gas follow from the Lagrangian
\begin{equation}
L = \int \rho\left(\frac{\vec{v}^2}{2} - u\right)  \,{\rm d}V .
\end{equation}
This opens up an interesting route for obtaining discretized equations
of motion for gas dynamics. Instead of working with the continuum
equations directly and trying to heuristically work out a set of
accurate difference formulas, one can discretize the Lagrangian and
then derive SPH equations of motion by applying the variational
principals of classical mechanics, an approach first proposed by
\citet{Gingold1982}. Using a Lagrangian also immediately guarantees
certain conservation laws and retains the geometric structure imposed
by Hamiltonian dynamics on phase space.

We here follow this elegant idea, which was first worked out by
\citet{Springel2002} with a consistent accounting of variable
smoothing lengths.  We start by discretizing the Lagrangian in terms
of fluid particles of mass $m_i$, yielding
\begin{equation}
L_{\rm SPH} = \sum_i \left(\frac{1}{2}m_i {\vec{v}_i^2} - m_i
u_i \right) , \label{eqndisctLgr}
\end{equation}
where it has been assumed that the thermal energy per unit mass of a
particle can be expressed through an entropic function $A_i$ of the
particle, which simply labels its specific thermodynamic entropy. The
pressure of the particles is
\begin{equation}
P_i = A_i \rho_i^\gamma  = (\gamma-1) \rho_i u_i,
\end{equation}
where $\gamma$ is the adiabatic index.  Note that for isentropic flow
(i.e.~in the absence of shocks, and without mixing or thermal
conduction) we expect the $A_i$ to be constant. We hence define $u_i$,
the thermal energy per unit mass, in terms of  the density estimate as
\begin{equation}
u_i(\rho_i) =  A_i \frac{\rho_i^{\gamma-1}}{\gamma-1} . \label{eqU}
\end{equation}
This raises the question of how the smoothing lengths $h_i$ needed for
estimating $\rho_i$ should be determined.  As we discussed above, we
would like to ensure adaptive kernel sizes, meaning that the number of
points in the kernel should be approximately constant. In much of the
older SPH literature, the number of neighbours was allowed to vary
within some (small) range around a target number. Sometimes the
smoothing length itself was evolved with a differential equation in
time, exploiting the continuity relation and the expectation that
$\rho h^3$ should be approximately constant
\citep[e.g][]{Steinmetz1993}. In case the number of neighbours outside
the kernel happened to fall outside the allowed range, $h$ was
suitably readjusted.  However, \citet{Nelson1994} pointed out that for
smoothing lengths varied in this way, the energy is not conserved
correctly.  They showed that the errors could be made smaller by
keeping the number of neighbours exactly constant, and they also
derived leading order correction terms (which became known as $\nabla
h$ terms) for the classic SPH equations of motion that could reduce
them still further.  In the modern formulation discussed below, these
$\nabla h$-terms do not occur; they are implicitly included at all
orders.

The central trick making this possible is to require that the mass in
the kernel volume should be constant, viz.
\begin{equation}
\rho_i  h_i^3 = const  \label{eqnconst}
\end{equation}
for three dimensions.  Since $\rho_i = \rho_i(\vec{r}_1, \vec{r}_2,
\ldots \vec{r}_N, h_i)$ is only a function of the particle coordinates
and of $h_i$, this equation implicitly defines the function $h_i =
h_i(\vec{r}_1, \vec{r}_2, \ldots \vec{r}_N)$ in terms of the particle
coordinates.

We can then proceed to derive the equations of motion from
\begin{equation}
 \frac{\dd}{\dd
t}\frac{\partial L}{\partial \dot \vec{r}_i} -\frac{\partial
L}{\partial \vec{r}_i} = 0.
\end{equation} 
This  first gives
\begin{equation}
 m_i\frac{\dd \vec{v}_i}{\dd t} = -
\sum_{j=1}^N m_j \frac{P_j}{\rho_j^2} \,\frac{\partial \rho_j}{\partial
\vec{r}_i},
\end{equation}
where the derivative $\partial \rho_j/\partial \vec{r}_i$ stands for
the total variation of the density with respect to the coordinate
$\vec{r}_i$, including any variation of $h_j$ this may entail. We can
hence write
\begin{equation}
\frac{\partial \rho_j}{\partial \vec{r}_i}  =
\vec{\nabla}_i \rho_j +  
\frac{\partial \rho_j}{\partial h_j}
\frac{\partial h_j}{\partial \vec{r}_i} ,
\label{eqA1}
\end{equation}
where the smoothing length is kept constant in the first derivative on
the right hand side (in our notation, the Nabla operator $\nabla_i
= \partial / \partial \vec{r}_i$ means differentiation with respect to
$\vec{r}_i$ holding the smoothing lengths constant).  On the other
hand, differentiation of $\rho_j h_j^3 = const$ with respect to
$\vec{r}_i$ yields
\begin{equation}
 \frac{\partial \rho_j}{\partial h_j}
\frac{\partial h_j}{\partial \vec{r}_i} 
\left[ 1+ \frac{3\,\rho_j}{h_j} \left(\frac{\partial \rho_j}{\partial
h_j}\right)^{-1} \right]  = - \vec{\nabla}_i\rho_j .
\label{eqA2}
\end{equation}
Combining equations (\ref{eqA1}) and (\ref{eqA2}) we then find
\begin{equation}
 \frac{\partial \rho_j}{\partial \vec{r}_i}  =
\left( 1+\frac{h_j}{3\rho_j} \frac{\partial \rho_j}{\partial h_j}    
\right)^{-1} \vec{\nabla}_i \rho_j \label{eqA3} .
\end{equation}
 Using 
\begin{equation} 
\nabla_i{ \rho_j} = m_i \nabla_i W_{ij}(h_j)
+\delta_{ij}\sum_{k=1}^N m_k \nabla_i W_{ki}(h_i) \, ,  \label{eqA4}
\end{equation}
 we finally
obtain the equations of motion 
\begin{equation}
\frac{\dd \vec{v}_i}{\dd t} = -
\sum_{j=1}^N m_j \left[ f_i \frac{P_i}{\rho_i^2} \nabla_i W_{ij}(h_i)
+ f_j \frac{P_j}{\rho_j^2} \nabla_i W_{ij}(h_j) \right], \label{eqEOM}
\label{eqnmot} 
\end{equation}
 where the $f_i$ are defined by 
\begin{equation}
 f_i = \left[ 1 +
\frac{h_i}{3\rho_i}\frac{\partial \rho_i}{\partial h_i} \right]^{-1} ,
\label{eqA5}
\end{equation} 
and the abbreviation $W_{ij}(h)= W(|\vec{r}_{i}-\vec{r}_{j}|, h)$ has
been used. Note that the correction factors $f_i$ can be easily
calculated alongside the density estimate, all that is required is an
additional summation to get $\partial \rho_i/ \partial \vec{r}_i$ for
each particle. This quantity is in fact also useful to get the correct
smoothing radii by iteratively solving $\rho_i h_i^3=const$ with a
Newton-Raphson iteration.

The equations of motion~(\ref{eqnmot}) for inviscid hydrodynamics are
remarkably simple. In essence, we have transformed a complicated
system of partial differential equations into a much simpler set of
ordinary differential equations.  Furthermore, we only have to solve
the momentum equation explicitly.  The mass conservation equation as
well as the total energy equation (and hence the thermal energy
equation) are already taken care of, because the particle masses and
their specific entropies stay constant for reversible gas
dynamics. However, later we will introduce an artificial viscosity
that is needed to allow a treatment of shocks. This will introduce
additional terms in the equation of motion and requires the time
integration of one thermodynamic quantity per particle, which can
either be chosen as entropy or thermal energy.  Indeed,
\citet{Monaghan2002} pointed out that the above formulation can also
be equivalently expressed in terms of thermal energy instead of
entropy. This follows by taking the time derivative of
Eqn.~(\ref{eqU}), which first yields
\begin{equation}
\frac{{\rm d}u_i}{{\rm d} t}
= \frac{P_i}{\rho_i} \sum_j \vec{v}_j \cdot \frac{\partial
\rho_i}{\partial \vec{r}_j} .
\end{equation}
Using equations (\ref{eqA3}) and (\ref{eqA4}) then gives the evolution
of the thermal energy as
\begin{equation}
\frac{{\rm d}u_i}{{\rm d} t}
=  f_i \frac{P_i}{\rho_i} \sum_j m_j (\vec{v}_i - \vec{v}_j) \cdot
\vec{\nabla}W_{ij}(h_i), \label{eqDuDt}
\end{equation}
which needs to be integrated along the equation of motion if one wants
to use the thermal energy as independent thermodynamic variable. There
is no difference however to using the entropy; the two are completely
equivalent in the variational formulation. This also solves the old
problem pointed out by \citet{Hernquist1993} that the classic SPH
equations did not properly conserve energy when the entropy was
integrated, and vice versa.  Arguably, it is numerically advantageous
to integrate the entropy though, as this is computationally cheaper
and eliminates time integration errors in solving Eqn.~(\ref{eqDuDt}).

Note that the above formulation readily fulfills the conservation laws
of energy, momentum and angular momentum. This can be shown based on
the discretized form of the equations, but it is also manifest due to
the symmetries of the Lagrangian that was used as a starting point.
The absence of an explicit time dependence gives the energy
conservation, the translational invariance implies momentum
conservation, and the rotational invariance gives angular momentum
conservation.

Other derivations of the SPH equations based on constructing kernel
interpolated versions of differential operators and applying them
directly to the Euler equations are also possible \citep[see,
e.g.,][]{Monaghan1992}. However, these derivations of `classic SPH'
are not unique in the sense that one is left with several different
possibilities for the equations and certain ad-hoc symmetrizations
need to be introduced. The choice for a particular formulation then
needs to rely on experimentally comparing the performance of many
different variants \citep{Thacker2000}.

\subsection{Artificial Viscosity} \label {SecVisco}

Even when starting from perfectly smooth initial conditions, the gas
dynamics described by the Euler equations may readily produce true
discontinuities in the form of shock waves and contact discontinuities
\citep{Landau1959}. At such fronts the differential form of the Euler
equations breaks down, and their integral form (equivalent to the
conservation laws) needs to be used.  At a shock front, this yields
the Rankine-Hugoniot jump conditions that relate the upstream and
downstream states of the fluid.  These relations show that the
specific entropy of the gas always increases at a shock front,
implying that in the shock layer itself the gas dynamics can no longer
be described as inviscid. In turn, this also implies that the
discretized SPH equations we derived above can not correctly describe
a shock, simply because they keep the entropy strictly constant.

One thus must allow for a modification of the dynamics at shocks and
somehow introduce the necessary dissipation. This is usually
accomplished in SPH by an artificial viscosity. Its purpose is to
dissipate kinetic energy into heat and to produce entropy in the
process. The usual approach is to parameterize the artificial
viscosity in terms of a friction force that damps the relative motion
of particles.  Through the viscosity, the shock is broadened into a
resolvable layer, something that makes a description of the dynamics
everywhere in terms of the differential form possible.  It may seem a
daunting task though to somehow tune the strength of the artificial
viscosity such that just the right amount of entropy is generated in a
shock. Fortunately, this is however relatively unproblematic. Provided
the viscosity is introduced into the dynamics in a conservative
fashion, the conservation laws themselves ensure that the right amount
of dissipation occurs at a shock front.

What is more problematic is to devise the viscosity such that it is
only active when there is really a shock present. If it also operates
outside of shocks, even if only at a weak level, the dynamics may begin to
deviate from that of an ideal gas.

The viscous force is most often added to the equation of motion as 
\begin{equation}
 \left. \frac{\dd \vec{v}_i}{\dd t}\right|_{\rm visc} 
=
-\sum_{j=1}^N m_j \Pi_{ij} \nabla_i\overline{W}_{ij} \, ,
\label{eqnvisc}
\end{equation}
where
\begin{equation}
 \overline{W}_{ij}= \frac{1}{2}\left[ W_{ij}(h_i) +
W_{ij} (h_j)\right] 
\end{equation} denotes
a symmetrized kernel, which some researchers prefer to define as
$\overline{W}_{ij}= W_{ij}( [h_i+h_j]/2)$.  Provided the viscosity
factor $\Pi_{ij}$ is symmetric in $i$ and $j$, the viscous force between
any pair of interacting particles will be antisymmetric and along the
line joining the particles. Hence linear momentum and angular momentum
are still preserved. In order to conserve total energy, we need to
compensate the work done against the viscous force in the thermal
reservoir, described either in terms of entropy,
\begin{equation}
\left. \frac{\dd A_i}{\dd t}\right|_{\rm visc} =
\frac{1}{2}\frac{\gamma-1}{\rho_i^{\gamma-1}}\sum_{j=1}^N m_j \Pi_{ij}
\vec{v}_{ij}\cdot\nabla_i \overline{W}_{ij} \,,
\label{eqnentropy}
\end{equation}
or in terms of thermal energy per unit mass,
\begin{equation}
\left. \frac{\dd u_i}{\dd t}\right|_{\rm visc} =  \frac{1}{2}
\sum_{j=1}^N m_j
 \Pi_{ij} \vec{v}_{ij}
\cdot \nabla_i \overline{W}_{ij}\, ,
\label{eqnu-as}
\end{equation}
where $\vec{v}_{ij}= \vec{v}_i - \vec{v}_j$. There is substantial
freedom in the detailed parametrization of the viscosity $\Pi_{ij}$.
The most commonly used formulation is an improved version of
the viscosity introduced  
by \citet{Monaghan1983},
\begin{equation}
\label{eqvisc}
\Pi_{ij}=\left\{
\begin{tabular}{cl}
${\left[-\alpha c_{ij} \mu_{ij} +\beta \mu_{ij}^2\right]}/{\rho_{ij}}$ &
\mbox{if
$\vec{v}_{ij}\cdot\vec{r}_{ij}<0$} \\
0 & \mbox{otherwise}, \\
\end{tabular}
\right. \label{eqnMoBals} \end{equation}
 with 
\begin{equation}
\mu_{ij}=\frac{h_{ij}\,\vec{v}_{ij}\cdot\vec{r}_{ij} }
{\left|\vec{r}_{ij}\right|^2 + \epsilon h_{ij}^2}.\label{egnMu} 
\end{equation}
Here $h_{ij}$ and $\rho_{ij}$ denote arithmetic means of the
corresponding quantities for the two particles $i$ and $j$, with
$c_{ij}$ giving the mean sound speed, whereas $\vec{r}_{ij}\equiv
\vec{r}_i - \vec{r}_j$.  The strength of the viscosity is regulated by
the parameters $\alpha$ and $\beta$, with typical values in the range
$\alpha\simeq 0.5-1.0$ and the frequent choice of $\beta=2\,\alpha$.
The parameter $\epsilon\simeq 0.01$ is introduced to protect against
singularities if two particles happen to get very close.

In this form, the artificial viscosity is basically a combination of a
bulk and a von Neumann-Richtmyer viscosity.  Historically, the
quadratic term in $\mu_{ij}$ has been added to the original
Monaghan-Gingold form to prevent particle penetration in high Mach
number shocks.  Note that the viscosity only acts for particles that
rapidly approach each other, hence the entropy production is always
positive definite. Also, the viscosity vanishes for solid-body
rotation, but not for pure shear flows. To cure this problem in shear
flows, \citet{Balsara1995} suggested adding a correction factor to the
viscosity, reducing its strength when the shear is strong. This can be
achieved by multiplying $\Pi_{ij}$ with a prefactor $(f_i^{\rm
  AV}+f_j^{\rm AV})/2$, where the factors
\begin{equation}
f_i^{\rm AV} = \frac{ |\nabla\cdot \vec{v}|_i}{ |\nabla\cdot \vec{v}|_i + 
|\nabla\times \vec{v}|_i}
\end{equation}
are meant to measure the rate of local compression in relation to the
strength of the local shear (estimated with formulas such as
Eqn.~\ref{eqnVelDiv}).  This Balsara switch has often been
successfully used in multi-dimensional flows and is enabled as default
in many SPH codes. We note however that it may be problematic
sometimes in cases where shocks and shear occur together, e.g.~in
oblique shocks in differentially rotating disks.

In some studies, alternative forms of viscosity have been tested. For
example, \citet{Monaghan1997} proposed a modified form of the
viscosity based on an analogy to the Riemann problem, which can be
written as
\begin{equation}
 \Pi_{ij} =
-\frac{\alpha}{2} \frac{ v_{ij}^{\rm sig} w_{ij}}
{\rho_{ij}} , \label{eqnViscNew} 
\end{equation} 
where $v_{ij}^{\rm sig} = \left[ c_{i} + c_{j} - 3 w_{ij} \right]$ is
an estimate of the signal velocity between two particles $i$ and $j$,
and $w_{ij}={\vec{v}_{ij}\cdot\vec{r}_{ij}} /
{\left|\vec{r}_{ij}\right|}$ is the relative velocity projected onto
the separation vector.  This viscosity is identical to
(\ref{eqnMoBals}) if one sets $\beta=3/2\, \alpha$ and replaces
$w_{ij}$ with $\mu_{ij}$.  The main difference between the two
viscosities lies therefore in the additional factor of $h_{ij}/r_{ij}$
that $\mu_{ij}$ carries with respect to $w_{ij}$. In equations
(\ref{eqnMoBals}) and (\ref{egnMu}), this factor weights the viscous
force towards particle pairs with small separations. In fact, after
multiplying with the kernel derivative, this weighting is strong
enough to make the viscous force in equation (\ref{egnMu}) diverge as
$\propto 1/r_{ij}$ for small pair separations up to the limit set by
the $\epsilon h_{ij}^2$ term.

\citet{Lombardi1999} have systematically tested different
parametrization of viscosity, but in general, the standard form was
found to work best. Recently the use of a tensor artificial viscosity
was conjectured by \citet{Owen2004} as part of an attempt to optimize
the spatial resolution of SPH. However, a disadvantage of this scheme
is that it breaks the strict conservation of angular momentum.

In attempting to reduce the numerical viscosity of SPH in regions away
from shocks, several studies have instead advanced the idea of keeping
the functional form of the artificial viscosity, but making the
viscosity strength parameter $\alpha$ variable in time. Such a scheme
was first suggested by \citet{Morris1997}, and was successfully
applied for studying astrophysical turbulence more faithfully in SPH
\citep{Dolag2005_turbulence}, and to follow neutron star mergers
\citep{Rosswog2005}. Adopting $\beta = 2\alpha$, one may evolve the
parameter $\alpha$ individually for each particle with an equation
such as
\begin{equation}
\frac{{\rm d}\alpha_i}{{\rm d}t} = - \frac{\alpha_i - \alpha_{\rm
    max}}{\tau_i} + S_i ,
\end{equation}
where $S_i$ is some source function meant to ramp up the viscosity
rapidly if a shock is detected, while the first term lets the
viscosity exponentially decay again to a prescribed minimum value
$\alpha_{\rm min}$ on a timescale $\tau_i$. So far, simple source
functions like $S_i = \max[-(\nabla\cdot\vec{v})_i, 0]$ and timescales
$\tau_i \simeq h_i/c_i$ have been explored and the viscosity
$\alpha_i$ has often also been prevented from becoming higher than
some prescribed maximum value $\alpha_{\rm max}$. It is clear that the
success of such a variable $\alpha$ scheme depends critically on an
appropriate source function. The form above can still not distinguish
purely adiabatic compression from that in a shock, so is not
completely free of creating unwanted viscosity.

\subsection{Coupling to Self-gravity}

Self-gravity is extremely important in many astrophysical flows, quite
in contrast to other areas of computational fluid dynamics, where only
external gravitational fields play a role. It is noteworthy that
Eulerian mesh-based approaches do not manifestly conserve total energy
if self-gravity is included \citep{Mueller1995,Springel2009}, but it
can be easily and accurately incorporated in SPH.

Arguably the best approach to account for gravity is to include it
directly into the discretized SPH Lagrangian, which has the advantage
of also allowing a consistent treatment of variable gravitational
softening lengths. This was first conducted by \citet{Price2007_adapt}
in the context of adaptive resolution N-body methods for collisionless
dynamics \citep[see also][]{Bagla2009}.

Let $\Phi(\vec{r}) = G \sum_i m_i \phi(\vec{r}-\vec{r}_i, \epsilon_i)$
be the gravitational field described by the SPH point set, where
$\epsilon_i$ is the gravitational softening length of particle $i$. We
then define the total gravitational self-energy of the system of SPH
particles as
\begin{equation}
E_{\rm pot} = \frac{1}{2} \sum_{i} m_i\Phi(\vec{r}_i) =  \frac{G}{2}
\sum_{ij}  m_i \, m_j \, \phi(r_{ij}, \epsilon_j).
\label{EqnEpot}
\end{equation}
For SPH including self-gravity, the Lagrangian then becomes
\begin{equation}
L_{\rm SPH} = \sum_i \left(\frac{1}{2}m_i {\vec{v}_i^2} - m_i
u_i \right)   -    \frac{G}{2}
\sum_{ij}  m_i \, m_j \, \phi(r_{ij}, \epsilon_j).
\label{eqLagGrav}
\end{equation}
As a result, the equation of motion acquires an additional term due to
the gravitational forces, given by:
\begin{eqnarray}
m_i\,\vec{a}_i^{\rm grav} & =&  - \frac{\partial E_{\rm
    pot}}{\partial\vec{r}_i} \nonumber \\
& = & - \sum_j G
m_i m_j
\frac{\vec{r}_{ij}}{r_{ij}} \frac{\left[ \phi'(r_{ij},\epsilon_i) +
    \phi'(r_{ij},\epsilon_j)\right]}{2} \nonumber \\
& & - \frac{1}{2} \sum_{jk} G m_j m_k  \frac{\partial
\phi(r_{jk},\epsilon_j)}{\partial
\epsilon}\,\frac{\partial \epsilon_j}{\partial \vec{r}_i} ,
\label{EqnGrAcc}
\end{eqnarray}
where $\phi'(r,\epsilon) = \partial\phi / \partial r$.

The first sum on the right hand side describes the ordinary
gravitational force, where the interaction is symmetrized by averaging
the forces in case the softening lengths between an interacting pair
are different. The second sum gives an additional force component that
arises when the gravitational softening lengths are a function of the
particle coordinates themselves, i.e.~if one invokes adaptive
gravitational softening. This term has to be included to make the
system properly conservative when spatially adaptive gravitational
softening lengths are used \citep{Price2007_adapt}.

In most cosmological SPH codes of structure formation, this has
usually not been done thus far and a fixed gravitational softening was
used for collisionless dark matter particles and SPH particles
alike. However, especially in the context of gravitational
fragmentation problems, it appears natural and also indicated by
accuracy considerations \citep{Bate1997} to tie the gravitational
softening length to the SPH smoothing length, even though some studies
also caution against this strategy \citep{Williams2004}.  Adopting
$\epsilon_i = h_i$, and determining the smoothing lengths as described
above, we can calculate the $\partial \epsilon_j / \partial \vec{r}_i
= \partial h_j / \partial \vec{r}_i$ term by means of equations
(\ref{eqA2}) and (\ref{eqA4}). Defining the quantities
\begin{equation}
\eta_j = 
\frac{h_j}{3\rho_j} f_j \sum_k m_k \frac{\partial\phi(\vec{r}_{jk},h_j)}
{\partial h},
\end{equation}
where the factors $f_i$ are defined as in Eqn.~(\ref{eqA5}), we can then
write the gravitational acceleration compactly as
\begin{eqnarray}
\left.\frac{{\rm d}\vec{v}_i}{{\rm d}t}\right|_{\rm grav} & =&  -
G \sum_j 
m_j
\frac{\vec{r}_{ij}}{r_{ij}} \frac{\left[ \phi'(r_{ij},h_i) +
    \phi'(r_{ij},h_j)\right]}{2} \nonumber \\
& & + \frac{G}{2} \sum_{j}  m_j \left[
\eta_i \nabla_i W_{ij}(h_i)   + \eta_j \nabla_i W_{ij}(h_j)
\right] .
\label{eqnadapsoft}
\end{eqnarray}
This acceleration has to be added to the ordinary SPH equation of
motion (\ref{eqEOM}) due to the pressure forces, which arises from the
first part of the Lagrangian (\ref{eqLagGrav}). Note that the factors
$\eta_j$ can be conveniently calculated alongside the SPH density
estimate. The calculation of the gravitational correction force,
i.e.~the second sum in Eqn.~(\ref{eqnadapsoft}), can then be done
together with the usual calculation of the SPH pressure forces. Unlike
in the primary gravitational force, here only the nearest neighbours
contribute.

We note that it appears natural to relate the functional form of the
gravitational softening to the shape of the SPH smoothing kernel, even
though this is not strictly necessary. If this is done, $\phi(r,h)$ is
determined as solution of Poisson's equation,
\begin{equation}
\nabla^2 \phi(r, h) = 4\pi W(r,h).
\end{equation}
The explicit functional form for $\phi(r,h)$ resulting from this
identification for the cubic spline kernel is also often employed in
collisionless N-body codes and can be found, for example, in the
appendix of \citet{Springel2001gadget}.

\subsection{Implementation Aspects}

The actual use of the discretized SPH equations in simulation models
requires a time-integration scheme.  The Hamiltonian character of SPH
allows in principle the use of symplectic integration schemes
\citep{Hairer2002,Springel2005} such as the leapfrog, which respects
the geometric phase-space constraints imposed by the conservation laws
and can prevent the build-up of secular integration errors with
time. However, since most hydrodynamical problems of interest are not
reversible anyway, this aspect is less important than in
dissipation-free collisionless dynamics. Hence any second-order
accurate time integration scheme, like a simple Runge-Kutta or
predictor-corrector scheme, may equally well be used.  It is common
practice to use individual timesteps for the SPH particles, often in a
block-structured scheme with particles arranged in a power-of-two
hierarchy of timesteps \citep{Hernquist1989}. This greatly increases
the efficiency of calculations in systems with large dynamic range in
timescale, but can make the optimum choice of timesteps tricky. For
example, the standard Courant timestep criterion of the form
\begin{equation}
\Delta t_i = C_{\rm CFL} \frac{h_i}{c_i},
\end{equation}
usually used in SPH, where $C_{\rm CFL}\sim 0.1-0.3$ is a
dimensionless parameter, may not guarantee fine enough time-stepping
ahead of a blast wave propagating into very cold gas. This problem can
be avoided with an improved method for determining the sizes of
individual timesteps \citep{Saitoh2009}.

If self-gravity is included, one can draw from the algorithms employed
in collisionless N-body dynamics to calculate the gravitational forces
efficiently, such as tree-methods \citep{Barnes1986}, or mesh-based
gravity solvers. The tree approach, which provides for a hierarchical
grouping of the particles, can also be used to address the primary
algorithmic requirement to write an efficient SPH code, namely the
need to efficiently find the neighboring particles inside the
smoothing kernel of a particle. If this is done naively, by computing
the distance to all other particles, a prohibitive ${\cal O}(N^2)$
scaling of the computational cost results. Using a range-searching
technique together with the tree, this cost can be reduced to ${\cal
  O}(N_{\rm ngb} N \log N)$, independent of the particle
clustering. To this end, a special walk of the tree is carried out for
neighbor searching in which a tree node is only opened if there is a
spatial overlap between the tree node and the smoothing kernel of the
target particle, otherwise the corresponding branch of the tree can be
immediately discarded.

These convenient properties of the tree algorithm have been exploited
for the development of a number of efficient SPH codes in astronomy,
several of them are publicly available and parallelized for
distributed memory machines.  Among them are {\small TreeSPH}
\citep{Hernquist1989,Katz1996}, {\small HYDRA} \citep{Pearce1997},
{\small GADGET} \citep{Springel2001,Springel2005}, {\small GASOLINE}
\citep{Wadsley2004}, {\small MAGMA} \citep{Rosswog2007}, and {\small
  VINE} \citep{Wetzstein2009}.

\section{EXTENSIONS OF SPH} \label{SecExtensions}

For many astrophysical applications, additional physical processes in
the gas phase besides inviscid hydrodynamics need to be modelled. This
includes, for example, magnetic fields, transport processes such as
thermal conduction or physical viscosity, or radiative transfer. SPH
also needs to be modified for the treatment of fluids that move at
relativistic speeds or in relativistically deep potentials. Below we
give a brief overview of some of the extensions of SPH that have been
developed to study this physics.

\subsection{Magnetic Fields}

Magnetic fields are ubiquitous in astrophysical plasmas, where the
conductivity can often be approximated as being effectively
infinite. In this limit one aims to simulate ideal, non-resistive
magneto-hydrodynamics (MHD), which is thought to be potentially very
important in many situations, in particular in star formation,
cosmological structure formation, or accretion disks. The equations of
ideal MHD are comprised of the induction equation
\begin{equation}
\frac{{\rm d}\vec{B}}{{\rm d} t} = 
(\vec{B}\cdot \nabla)\vec{v} - \vec{B}(\nabla \cdot \vec{v}) 
\end{equation}
and the magnetic Lorentz force. The latter can be obtained from the
Maxwell stress tensor
\begin{equation}
{\cal M}_{ij} = \frac{1}{4 \pi}\left(\vec{B}_i \vec{B}_j -
\frac{1}{2}\vec{B}^2 \delta_{ij}\right)
\end{equation}
as
\begin{equation}
\vec{F}_i = \frac { \partial {\cal M}_{ij}}{\partial x_j} .
\end{equation}
Working with the stress tensor is advantageous for deriving equations
of motion that are discretized in a symmetric fashion.  The magnetic
force $\vec{F}$ then has to be added to the usual forces from gas
pressure and the gravitational field.

First implementations of magnetic forces in SPH date back to
\citet{Gingold1977}, soon followed by full implementations of MHD in
SPH \citep{Phillips1985}.  However, a significant problem with MHD in
SPH has become apparent early on. The constraint equation $\nabla\cdot
\vec{B}=0$, which is maintained by the continuum form of the ideal MHD
equations, is in general not preserved by discretized versions of the
equations. Those tend to build up some $\nabla\cdot \vec{B} \ne 0$
error over time, corresponding to unphysical magnetic monopole
sources. In contrast, in the most modern mesh-based approaches to MHD,
so-called constrained transport schemes \citep{Evans1988} are able to
accurately evolve the discretized magnetic field while keeping a
vanishing divergence of the field.

Much of the recent research on developing improved SPH realizations of
MHD has therefore concentrated on constructing formulations that
either eliminate the $\nabla\cdot \vec{B} \ne 0$ error, or at least
keep it reasonably small \citep{Dolag1999,Price2004_MHD1,
  Price2004_MHD2, Price2005_MHD3, Price2009_MHDvec,Dolag2009}. To this
extent, different approaches have been investigated in the literature,
involving periodic field cleaning techniques
\citep[e.g.][]{Dedner2002}, formulations of the equations that
`diffuse away' the $\nabla\cdot \vec{B} \ne 0$ terms if they should
arise, or the use of the Euler potentials or the vector potential.
The latter may seem like the most obvious solution, since deriving the
magnetic field as $\vec{B}=\nabla \times \vec{A}$ from the vector
potential $\vec{A}$ will automatically create a divergence free
field. However, \cite{Price2009_MHDvec} explored the use of the vector
potential in SPH and concluded that there are substantial
instabilities in this approach, rendering it essentially useless in
practice.

Another seemingly clean solution lies in the so called-Euler
potentials (also known as Clebsch variables). One may write the
magnetic field as the cross product $\vec{B} = \nabla\alpha \times
\nabla\beta$ of the gradients of two scalar fields $\alpha$ and
$\beta$. In ideal MHD, where the magnetic flux is locked into the
flow, one obtains the correct field evolution by simply advecting
these Euler potentials $\alpha$ and $\beta$ with the motion of the
gas. This suggests a deceptively simple approach to ideal MHD; just
construct the fields $\alpha$ and $\beta$ for a given magnetic field,
and then move these scalars along with the gas. Although this has been
shown to work reasonably well in a number of simple test problems and
also has been used in some real-world applications
\citep{Price2007,Kotarba2009}, it likely has significant problems in
general MHD dynamics, as a result of the noise in SPH and the
inability of this scheme to account for any magnetic reconnection.
Furthermore, \citet{Brandenburg2010} points out that the use of Euler
potentials does not converge to a proper solution in hydromagnetic
turbulence.

Present formulations of MHD in SPH are hence back to discretizing the
classic magnetic induction equation. For example, \citet{Dolag2009}
give an implementation of SPH in the {\small GADGET} code where the
magnetic induction equation is adopted as
\begin{equation}
\frac{{\rm d}\vec{B}_i}{{\rm d}t} = \frac{f_i}{\rho_i}
\sum_j m_j \left\{ 
\vec{B}_i [\vec{v}_{ij}\cdot \nabla_iW_{ij}(h_i)]  -
\vec{v}_{ij}[\vec{B}_i\cdot\nabla_i W_{ij}(h_i)]\right\}
\end{equation}
and the acceleration due to magnetic forces as
\begin{equation}
\frac{{\rm d}\vec{v}_i}{{\rm d}t} = 
\sum_j m_j \left[ 
f_i \frac{{\cal M}_i}{\rho_i^2}  \cdot \nabla_iW_{ij}(h_i)]  +
f_j \frac{{\cal M}_j}{\rho_j^2}  \cdot \nabla_iW_{ij}(h_j)]  
\right],
\end{equation}
where the $f_i$ are the correction factors of Eqn.~(\ref{eqA5}), and
${\cal M}_i$ is the stress tensor of particle $i$.  Combined with
field cleaning techniques \citep{Borve2001} and artificial magnetic
dissipation to keep $\nabla\cdot \vec{B} \ne 0$ errors under control,
this leads to quite accurate results for many standard tests of MHD,
like magnetic shock tubes or the Orszang-Tang vortex. A similar
implementation of MHD in SPH is given in the independent {\small
  MAGMA} code by \citet{Rosswog2007}.

\subsection{Thermal Conduction}

Diffusion processes governed by variants of the  equation
\begin{equation}
\frac{{\rm d} Q}{{\rm d} t} = D\, \nabla^2 Q,
\end{equation}
where $Q$ is some conserved scalar field and $D$ is a diffusion
constant, require a discretization of the Laplace operator in
SPH. This could be done in principle by differentiating a kernel
interpolated version of $Q$ twice. However, such a discretization
turns out to be quite sensitive to the local particle distribution, or
in other words, it is fairly noisy. It is much better to use an
approximation of the $\nabla^2$ operator proposed first by
\citet{Brookshaw1985}, in the form
\begin{equation}
\left. \nabla^2 Q \right|_i =
- 2 \sum_j \frac{m_j}{\rho_j}\frac{Q_j - Q_i}{r_{ij}^2}\,
\vec{r}_{ij}\cdot 
\nabla_i \overline{W}_{ij} .
\end{equation}
\citet{Cleary1999} and \citet{Jubelgas2004} have used this to
construct implementations of thermal conduction in SPH which allow for
spatially variable conductivities. The heat conduction equation,
\begin{equation}
\frac{{\rm d} u}{{\rm d} t} = \frac{1}{\rho} \nabla( \kappa \nabla T),
\end{equation}
where $\kappa$ is the heat conductivity, 
can then be discretized in SPH as follows,
\begin{equation}
\frac{{\rm d} u_i}{{\rm d} t} =  \sum_j
\frac{m_j}{\rho_i \rho_j} \frac{(\kappa_i + \kappa_j)(T_i - T_j)}{r_{ij}^2}\,
\vec{r}_{ij}\cdot \nabla_i \overline{W}_{ij},
\end{equation}
with $\vec{r}_{ij}\equiv \vec{r}_i - \vec{r}_j$.  In this form, the
energy exchange between two particles is balanced on a pairwise basis,
and heat always flows from higher to lower temperature. Also, it is
easy to see that the total entropy increases in the process.  This
formulation has been used, for example, to study the influence of
Spitzer conductivity due to electron transport on the thermal
structure of the plasma in massive galaxy clusters \citep{Dolag2004}.

\subsection{Physical Viscosity}

If thermal conduction is not strongly suppressed by magnetic fields in
hot plasmas, then one also expects a residual physical viscosity (a
pure shear viscosity in this case). In general, real gases may possess
both physical bulk and physical shear viscosity. They are then
correctly described by the Navier-Stokes equations, and not the Euler
equations of inviscid gas dynamics that we have discussed thus
far. The stress tensor of the gas can be written as
\begin{equation}
\sigma_{ij} = \eta \left(\frac{\partial v_i}{\partial r_j} + 
 \frac{\partial v_j}{\partial r_i}  -\frac{2}{3}\delta_{ij} \frac{\partial
   v_k}{\partial r_k}\right) +\zeta\,\, \delta_{ij} \frac{ \partial
   v_k}{\partial r_k} ,
\end{equation}
where $\eta$ and $\zeta$ are the coefficients of shear and bulk
viscosity, respectively.  The Navier-Stokes equation including gravity
is then given by
\begin{equation}
\frac{{\rm d}\vec{v}}{{\rm d}t} = - \frac{\nabla P}{\rho} -\nabla\Phi
+\frac{1}{\rho} 
\nabla {\vec{\sigma}} .
\end{equation}
\citet{Sijacki2006} have suggested an SPH discretization of this
equation which estimates the shear viscosity of particle $i$ based on
\begin{equation}
\left. \frac{{\rm d} v_{\alpha}}{{\rm d} t}\right|_{i,{\rm shear}} 
= \sum_j
m_j 
\left[
\frac{\eta_i  \sigma_{\alpha\beta}|_i}{\rho_i^2}
[ \nabla_i {W}_{ij}(h_i)]|_\beta 
+ \frac{\eta_j  \sigma_{\alpha\beta}|_j}{\rho_j^2}
[ \nabla_i {W}_{ij}(h_j)]|_\beta 
\right],
\end{equation}
and the bulk viscosity as
\begin{equation}
\left. \frac{{\rm d} \vec{v}}{{\rm d} t}\right|_{i,{\rm bulk}} 
= \sum_j
m_j 
\left[
\frac{\zeta_i \nabla\cdot \vec{v}_i}{\rho_i^2}
 \nabla_i {W}_{ij}(h_i)
+ \frac{\zeta_j \nabla \cdot \vec{v}_j}{\rho_j^2}
\nabla_i {W}_{ij}(h_j)
\right].
\end{equation}
These viscous forces are antisymmetric, and together with a positive
definite entropy evolution
\begin{equation}
\frac{{\rm d}A_i}{{\rm d}t} = 
\frac{\gamma-1}{\rho_i^{\gamma-1}} 
\left[ \frac{1}{2} \frac{\eta_i}{\rho_i} \vec{\sigma}_i^2
+\frac{\zeta_i}{\rho_i} (\nabla\cdot \vec{v})^2 \right],
\end{equation}
they conserve total energy. Here the derivatives of the velocity can be
estimated with pair-wise formulations as in Eqn.~(\ref{eqnVelDiv}).

In some sense, it is actually simpler for SPH to solve the
Navier-Stokes equation than the Euler equations. As we discuss in more
detail in Section~\ref{SecGresho}, this is because the inclusion of an
artificial viscosity in SPH makes the scheme problematic for purely
inviscid flow, since this typically introduces a certain level of
spurious numerical viscosity also outside of shocks. If the gas has
however a sizable amount of intrinsic physical viscosity anyway, it
becomes much easier to correctly represent the fluid, provided the
physical viscosity is larger than the unavoidable numerical viscosity.

\subsection{Radiative Transfer}

In many astrophysical problems, like the reionization of the Universe
or in first star formation, one would like to self-consistently follow
the coupled system of hydrodynamical and radiative transfer equations,
allowing for a proper treatment of the backreaction of the radiation
on the fluid, and vice versa. Unfortunately, the radiative transfer
equation is in itself a high-dimensional partial differential equation
that is extremely challenging to solve in its full generality. The
task is to follow the evolution of the specific intensity field,
\begin{equation}
\frac{1}{c} \frac{\partial I_{\nu}}{\partial t} 
+ \vec{n}\cdot \nabla I_\nu
= - \kappa_\nu I_\nu + j_{\nu} ,
\end{equation}
which is not only a function of spatial coordinates, but also of
direction $\vec{n}$ and frequency of $\nu$. Here $j_{\nu}$ is a source
function, and $\kappa_{\nu}$ an absorption coefficient, which provide
an implicit coupling to the hydrodynamics.  Despite the formidable
challenge to develop efficient numerical schemes for the approximate
solution of the radiation-hydrodynamical set of equations, the
development of such codes based on SPH-based techniques has really
flourished in recent years, where several new schemes have been
proposed.

Perhaps the simplest and also oldest approaches are based on
flux-limited diffusion approximations to radiative transfer in SPH
\citep{Whitehouse2004, Whitehouse2005, Whitehouse2006, Fryer2006,
  Viau2006, Forgan2009}.  A refinement of this strategy has been
proposed by \citet{Petkova2009}, based on the optically thin variable
tensor approximations of \citet{Gnedin2001}. In this approach, the
radiative transfer equation is simplified in moment-based form,
allowing the radiative transfer to be described in terms of the local
energy density $J_\nu$ of the radiation,
\begin{equation}
J_\nu = \frac{1}{4 \pi} \int I_\nu \,  {\rm d}\Omega .
\end{equation}
This radiation energy density is then transported with an anisotropic
diffusion equation,
\begin{equation}
\frac{1}{c}\frac{\partial J_\nu}{\partial t} = 
\frac{\partial}{\partial r_j}
\left(
\frac{1}{\kappa_\nu}\frac{\partial J_\nu h^{ij}}{\partial r_i}\right) -
\kappa_\nu J_\nu + j_\nu
\end{equation}
where the matrix $h^{ij}$ is the so-called Eddington tensor, which is
symmetric and normalized to unit trace. The Eddington tensor encodes
information about the angle dependence of the local radiation field,
and in which direction, if any, it prefers to diffuse in case the
local radiation field is highly anisotropic.  In the ansatz of
\citet{Gnedin2001}, the Eddington tensors are simply estimated in an
optically thin approximation as a $1/r^2$-weighted sum over all
sources, which can be calculated efficiently for an arbitrary number
of sources with techniques familiar from the calculation of
gravitational fields.

\citet{Petkova2009} derive an SPH discretization of the transport part of this
 radiative transfer
approximation (described by the first term on the right hand side) as 
\begin{equation}
\frac{\partial{N_i}}{\partial t} = \sum_j w_{ij} (N_j-N_i), \label{EqnTrans}
\end{equation}
where the factors $w_{ij}$ are given by
\begin{equation}
w_{ij} = \frac{2 c m_{ij}}{\kappa_{ij} \rho_{ij}}
\frac{ \vec{r}_{ij}^{\rm T} \tilde\vec{h}_{ij} \nabla_i\overline{W}_{ij}}
{\vec{r}_{ij}^2} , \label{EqnAniso}
\end{equation}
and $N_i = c^2 J_\nu / (h_{\rm Planck}^4 \nu^3)$ is the photon number
associated with particle $i$.
The tensor 
\begin{equation}
\tilde\vec{h} = \frac{5}{2} \vec{h} - \frac{1}{2}{\rm Tr}(\vec{h}) 
\end{equation}
is a modified Eddington tensor such that the SPH discretization in
Eqn.~(\ref{EqnAniso}) corresponds to the correct anisotropic diffusion
operator.  The exchange described by Eqn.~(\ref{EqnTrans}) is photon
conserving, and since the matrix $w_{ij}$ is symmetric and positive
definite, reasonably fast iterative conjugate gradient solvers can be
used to integrate the diffusion problem implicitly in time in a stable
fashion.

A completely different approach to combine radiative transfer with SPH
has been described by \citet{Pawlik2008}, who transport the radiation
in terms of emission and reception cones, from particle to
particle. This effectively corresponds to a coarse discretization of
the solid angle around each particle. Yet another scheme is given by
\citet{Nayakshin2009}, who implemented a Monte-Carlo implementation of
photon transport in SPH. Here the photons are directly implemented as
virtual particles, thereby providing a Monte-Carlo sampling of the
transport equation. Although this method suffers from the usual
Monte-Carlo noise limitations, only few simplifying assumptions in the
radiative transfer equation itself have to be made, and it can
therefore be made increasingly more accurate simply by using more
Monte-Carlo photons.

Finally, \citet{Altay2008} suggested a ray-tracing scheme within SPH,
which essentially implements ideas that are known from so-called long-
and short characteristics methods for radiative transfer around point
sources in mesh codes. These approaches have the advantage of being
highly accurate, but their efficiency rapidly declines with an
increasing number of sources.

\subsection{Relativistic Dynamics}

It is also possible to derive SPH equations for relativistic dynamics
from a variational principle \citep{Monaghan2001}, both for special
and general relativistic dynamics. This is more elegant than
alternative derivations and avoids some problems inherent in other
approaches to relativistic dynamics in SPH
\citep{Monaghan1992,Laguna1993,Siegler2000,Faber2000,Ayal2001}.

The variational method requires discretizing the Lagrangian
\begin{equation}
L = -\int {T_{\mu\nu}U^{\nu}U^{\nu}}\,{\rm d}V ,
\end{equation}
where $U^{\mu}$ is the four-velocity, and $T_{\mu\nu}=(P+e) U_{\nu}U_{\nu} +
  P\,\eta_{\mu\nu}$ is the energy momentum tensor of a perfect fluid with pressure
  $P$ and rest-frame energy density 
\begin{equation}
e = n\, m_0 c^2 \left( 1+ \frac{u}{c^2}\right).
\end{equation}
Here a $(-1,1,1,1)$ signature for the flat metric tensor
$\eta^{\mu\nu}$ is used.  $n$ is the rest frame number density of
particles of mean molecular weight $m_0$, and $u$ is the rest-frame
thermal energy per unit mass.  In the following, we set $c=m_0=1$.

\citet{Rosswog2009} gives a detailed derivation of the resulting
equations of motion when the discretization is done in terms of fluid
parcels with a constant number $\nu_i$ of baryons in SPH-particle $i$
and when variable smoothing lengths are consistently included. In the
special-relativistic case, he derives the equations of motion as
\begin{equation}
  \frac{{\rm d} \vec{s}_i}{{\rm d}t} = 
-\sum_j \nu_j \left[ f_i \frac{P_i}{N_i^2} \nabla_i W_{ij}(h_i) 
+  f_j \frac{P_j}{N_j^2} \nabla_i W_{ij}(h_j) \right],
\end{equation}
where the generalized momentum $\vec{s}_i$ of particle $i$ is given by 
\begin{equation}
\vec{s}_i = \gamma_i \vec{v}_i \left( 1+ u_i + \frac{P_i}{n_i}\right),
\end{equation}
and $\gamma_i$ is the particle's Lorentz factor.  The baryon number
densities $N_i$ in the computing frame are estimated just as in
Eqn.~(\ref{eqnDensEst}), except for the replacements $\rho_i \to N_i$
and $m_j \to \nu_j$. Similarly, the correction factors $f_i$ are
defined as in Eqn.~(\ref{eqA5}) with the same replacements.

One also needs to complement the equation of motion with an energy
equation of the form
\begin{equation}
\frac{{\rm d}\epsilon_i}{{\rm d}t}
= - \sum_j \nu_j \left[ \frac{f_iP_i}{N_i^2} \vec{v}_i \cdot \nabla W_{ij}(h_i) 
 + \frac{f_j P_j}{N_j^2} \vec{v}_j \cdot \nabla W_{ij}(h_j) \right],
\end{equation}
where
\begin{equation}
\epsilon_i = \vec{v}_i \cdot \vec{s}_i +\frac{1+u_i}{\gamma_i}
\end{equation}
is a suitably defined relativistic energy variable.  These equations
conserve the total canonical momentum as well as the angular
momentum. A slight technical complication arises for this choice of
variables due to the need to recover the primitive fluid variables
after each timestep from the updated integration variables $\vec{s}_i$
and $\epsilon_i$, which requires finding the root of a non-linear
equation.

\section{APPLICATIONS OF SPH}  \label{SecApplications}

The versatility and simplicity of SPH has led to a wide range of
applications in astronomy, essentially in every field where
theoretical research with hydrodynamical simulations is carried
out. We here provide a brief, necessarily incomplete overview of some
of the most prominent topics that have been studied with SPH.

\subsection{Cosmological Structure Formation}

Among the most important successes of SPH in cosmology are simulations
that have clarified the origin of the Lyman-$\alpha$ forest in the
absorption spectra to distant quasars
\citep[e.g.][]{Hernquist1996,Dave1999}, which have been instrumental
for testing and interpreting the cold dark matter cosmology. Modern
versions of such calculations use detailed chemical models to study
the enrichment of the intragalactic medium \citep{Oppenheimer2006},
and the nature of the so-called warm-hot IGM, which presumably
contains a large fraction of all cosmic baryons \citep{Dave2001}.

Cosmological simulations with hydrodynamics are also used to study
galaxy formation in its full cosmological context, directly from the
primordial fluctuation spectrum predicted by inflationary theory. This
requires the inclusion of radiative cooling, and sub-resolution models
for the treatment of star formation and associated energy feedback
processes from supernovae explosions or galactic winds. A large
variety of such models have been proposed
\citep[e.g.][]{Springel2003_multi}, some also involve rather
substantial changes of SPH, for example the `decoupled' version of SPH
by \citet{Marri2003} for the treatment of multiphase structure in the
ISM.  Despite the uncertainties such modelling involves and the huge
challenge to numerical resolution this problem entails, important
theoretical results on the clustering of galaxies, the evolution of
the cosmic star formation rate, or the efficiency of galaxy formation
as a function of halo mass have been reached.

One important goal of such simulations has been the formation of disk
galaxies in a cosmological context, which has however proven to
represent a significant challenge. In recent years, substantial
progress in this area has been achieved, however, where some simulated
galaxies have become quite close now to the morphology of real spiral
galaxies \citep{Governato2007,Scannapieco2008}. Still, the ratio of
the stellar bulge to disk components found in simulations is typically
much higher than in observations, and the physical or numerical origin
for this discrepancy is debated.

The situation is better for clusters of galaxies, where the most
recent SPH simulations have been quite successful in reproducing the
primary scaling relations that are observed
\citep{Borgani2004,Puchwein2008,McCarthy2009}. This comes after a long
struggle in the literature with the cooling-flow problem, namely the
fact that simulated clusters tend to efficiently cool out much of
their gas at the centre, an effect that is not present in observations
in the predicted strength. The modern solution is to attribute this to
a non-gravitational heat source in the centre of galaxies, which is
thought to arise from active galactic nuclei. This important feedback
channel has recently been incorporated in SPH simulations of galaxy
clusters \citep{Sijacki2008}.

SPH simulations of cosmic large-scale structure and galaxy clusters
are also regularly used to study the Sunyaev-Zeldovich effect
\citep{daSilva2000}, the build-up of magnetic fields
\citep{Dolag1999,Dolag2002,Dolag2005}, or even the production of
non-thermal particle populations in the form of cosmic rays
\citep{Pfrommer2007,Jubelgas2008}. The latter also required the
development of a technique to detect shock waves on the fly in SPH,
and to measure their Mach number \citep{Pfrommer2006}.

\subsection{Galaxy Mergers}

The hierarchical theory of galaxy formation predicts frequent mergers
of galaxies, leading to the build-up of ever larger systems. Such
galaxy interactions are also prominently observed in many systems,
such as the `Antennae Galaxies' NGC 4038/4039. The merger hypothesis
\citep{Toomre1972} suggests that the coalescence of two spiral
galaxies leads to an elliptical remnant galaxy, thereby playing a
central role in explaining Hubble's tuning fork diagram for the
morphology of galaxies. SPH simulations of merging galaxies have been
instrumental in understanding this process.

In pioneering work by \citet{Hernquist1989Natur}, \citet{Barnes1991}
and \citet{Mihos1994,Mihos1996}, the occurrence of central nuclear
starbursts during mergers was studied in detail. Recently, the growth
of supermassive black holes has been added to the simulations
\citep{Springel2005_BH}, making it possible to study the co-evolution
of black holes and stellar bulges in galaxies. \citet{DiMatteo2005}
demonstrated that the energy output associated with accretion
regulates the black hole growth, and establishes the tight observed
relation between black hole masses and bulge velocity
dispersions. Subsequently, these simulations have been used to develop
comprehensive models of spheroid formation and for interpreting the
evolution and properties of the cosmic quasar population
\citep{Hopkins2005,Hopkins2006}.

\subsection{Star Formation and Stellar Encounters}

On smaller scales, many SPH-based simulations have studied the
fragmentation of molecular clouds, star formation, and the initial
mass function \citep{Bate1998,Klessen1998,Bate2005,Smith2009}. This
includes also simulations of the formation of the first stars in the
Universe \citep{Bromm2002,Yoshida2006,Clark2008}, which are thought to
be very special objects.

In this area, important numerical issues related to SPH have been
examined as well. \citet{Bate1997} show that the Jeans mass needs to
be resolved to guarantee reliable results for gravitational
fragmentation, a result that was strengthened by
\citet{Whitworth1998}, who showed that provided $h\sim \epsilon$ and
the Jeans mass is resolved, only physical fragmentation should
occur. Special on-the-fly particle splitting methods have been
developed in order to guarantee that the Jeans mass stays resolved
over the coarse of a simulation
\citep{Kitsionas2002,Kitsionas2007}. Also, the ability of SPH to
represent driven isothermal turbulence has been studied
\citep{Klessen2000}.
 
Pioneering work on stellar collisions with SPH has been performed by
\citet{Benz1987} and \citet{Benz1990}. Some of the most recent work in
this area studied quite sophisticated problems, such as relativistic
neutron star mergers with an approximate treatment of general
relativity and sophisticated nuclear matter equations of state
\citep{Oechslin2007}, or the triggering of sub-luminous supernovae
type Ia through collisions of white dwarfs \citep{Pakmor2009}.

\subsection{Planet Formation and Accretion Disks}

SPH has been successfully employed to study giant planet formation
through the fragmentation of protoplanetary disks
\citep{Mayer2002,Mayer2004}, and to explore the interaction of high-
and low-mass planets embedded in protoplanetary discs
\citep{Bate2003,Lufkin2004}.

A number of works have also used SPH to model accretion disks
\citep{Simpson1995}, including studies of spiral shocks in 3D disks
\citep{Yukawa1997}, accretion of gas onto black holes in viscous
accretion disks \citep{Lanzafame1998}, or cataclysmic variable systems
\citep{Wood2005}.

Another interesting application of SPH lies in simulations of
materials with different equations of state, corresponding to rocky
and icy materials \citep{Benz1999}. This allows the prediction of the
outcome of collisions of bodies with sizes from the scale of
centimeters to hundreds of kilometers, which has important
implications for the fate of small fragments in the Solar system or in
protoplanetary disks. Finally, such simulation techniques allow
calculations of the collision of protoplanets, culminating in
numerical simulations that showed how the Moon might have formed by an
impact between protoearth and an object a tenth of its mass
\citep{Benz1986}.

\section{CONVERGENCE, CONSISTENCY AND STABILITY OF SPH} \label{SecAccuracy}

There have been a few code comparisons in the literature between SPH
and Eulerian hydrodynamics \citep{Frenk1999,Oshea2005,Tasker2008}, but
very few formal studies of the accuracy of SPH have been carried
out. Even most code papers on SPH report only circumstantial evidence
for SPH's accuracy.  What is especially missing are rigorous studies
of the convergence rate of SPH towards known analytic solutions, which
is ultimately one of the most sensitive tests of the accuracy of a
numerical method. For example, this may involve measuring the error in
the result of an SPH calculation in terms of an L1 error norm relative
to a known solution, as it is often done in tests of Eulerian
hydrodynamics codes. There is no a priori reason why SPH should not be
subjected to equally sensitive tests to establish whether the error
becomes smaller with increasing resolution (convergence), and whether
the convergence occurs towards the correct physical solution
(consistency).  Such tests can also provide a good basis to compare
the efficiency of different numerical approaches for a particular type
of problem with each other.

In this section, we discuss a number of tests of SPH in one and two
dimensions in order to provide a basic characterization of the
accuracy of `standard SPH' as outlined in Section~\ref{SecBasic}. Our
discussion is in particular meant to critically address some of the
weaknesses of standard SPH, both to clarify the origin of the
inaccuracies and to guide the ongoing search for improvements in the
approach.  We note that there is already a large body of literature
with suggestions for improvements of `standard SPH', ranging from
minor modifications, say in the parametrization of the artificial
viscosity, to more radical changes, such as outfitting SPH with a
Riemann solver or replacing the kernel-interpolation technique with a
density estimate based on a Voronoi tessellation. We shall summarize
some of these attempts in Section~\ref{SecFuture}, but refrain from
testing them in detail here.

We note that for all test results reported below, it is well possible
that small modifications in the numerical parameters of the code that
was used \citep[{\small GADGET2},][]{Springel2005} may lead to
slightly improved results. However, we expect that this may only
reduce the error by a constant factor as a function of resolution, but
is unlikely to significantly improve the order or convergence.  The
former is very helpful of course, but ultimately represents only
cosmetic improvements of the results. What is fundamentally much more
important is the latter, the order of convergence of a numerical
scheme.

\subsection{One-dimensional Sound Waves}

\begin{figure}
\begin{center}
\resizebox{12.0cm}{!}{\includegraphics{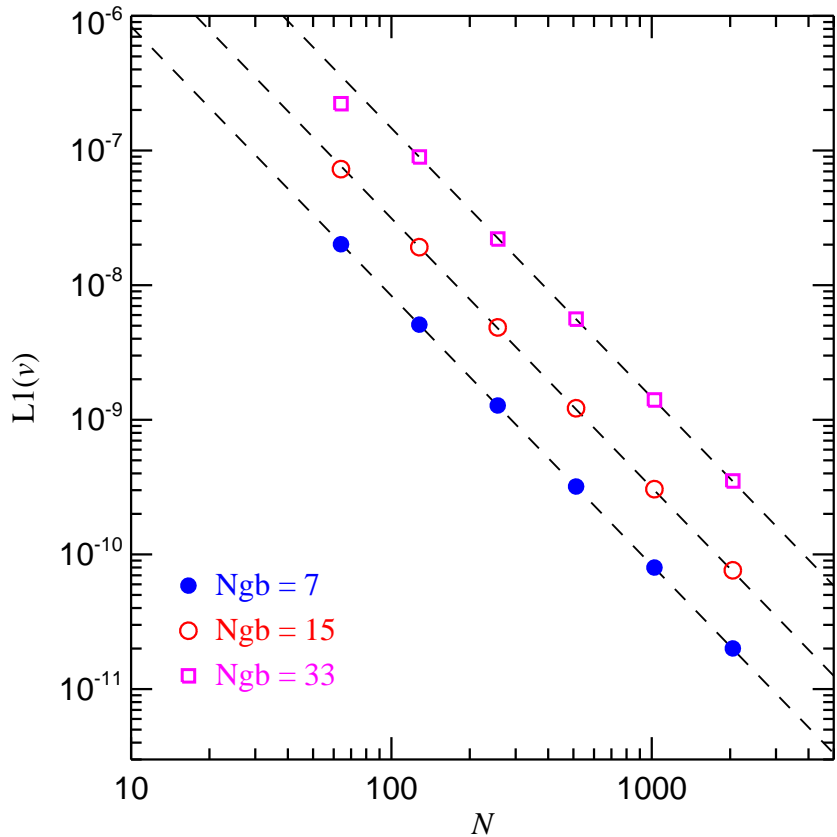}}%
\vspace*{-0.5cm}
\end{center}
\caption{Convergence rate for a traveling sound wave in 1D,
  calculated in SPH without artificial viscosity. The solution
  (symbols) converges with second order accuracy as ${\rm L1} \propto
  N^{-2}$ (this power law is shown with dashed lines), for different
  numbers of neighbours.
  \label{FigL1Accoustic}}
\end{figure}

Based on analytic reasoning, \citet{Rasio2000} argued that convergence
of SPH for one-dimensional sound waves requires increasing both the
number of smoothing neighbours and the number of particles, with the
latter increasing faster such that the smoothing length and hence the
smallest resolved scale decreases.  In fact, he showed that in this
limit the dispersion relation of sound waves in 1D is correctly
reproduced by SPH for all wavelengths.  He also pointed out that if
the number of neighbours is kept constant and only the number of
particles is increased, then one may converge to an incorrect physical
limit, implying that SPH is inconsistent in this case.

We note that in practice the requirement of consistency may not be
crucially important, provided the numerical result converges to a
solution that is `close enough' to the correct physical solution. For
example, a common error in SPH for a low number of neighbours is that
the density estimate carries a small bias of up to a percent or
so. Although this automatically means that the {\em exact} physical
value of the density is systematically missed by a small amount, in
the majority of astrophysical applications of SPH the resulting error
will be subdominant compared to other errors or approximations in the
physical modelling, and is therefore not really of concern. In our
convergence tests below we therefore stick with the common practice of
keeping the number of smoothing neighbours constant. Note that because
we compare to known analytic solutions, possible errors in the
dispersion relation will anyway be picked up by the error measures.

We begin with the elementary test of a simple acoustic wave that
travels through a periodic box. To avoid any wave steepening, we
consider a very small wave amplitude of $\Delta\rho/\rho=10^{-6}$.
The pressure of the gas is set to $P=3/5$ at unit density, such that
the adiabatic sound speed is $c_s=1$ for a gas with $\gamma=5/3$.  We
let the wave travel once through the box of unit length, and compare
the velocity profile of the final result at time $t=1.0$ with the
initial conditions in terms of an L1 error norm. We define the latter
as
\begin{equation}
{\rm L1} = \frac{1}{N} \sum_i |v_i - v(x_i)|,
\end{equation}
where $N$ is the number of SPH particles, $v_i$ is the numerical
solution for the velocity of particle $i$, and $v(x_i)$ is the
expected analytic solution for the problem, which is here identical to
the initial conditions. Note that we deliberately pick the velocity as
error measure, because the density estimate can be subject to a bias
that would then completely dominate the error norm. The velocity
information on the other hand will tell us more faithfully how the
wave has moved, and whether it properly returned to its original
state.

In Figure~\ref{FigL1Accoustic}, we show results for the L1 error norm
as a function of the number $N$ of equal-mass particles used to sample
the domain, for different numbers of smoothing
neighbours. Interestingly, the results show second-order convergence
of the code, with ${\rm L1}\propto N^{-2}$, as expected in a
second-order accurate scheme for smooth solutions without
discontinuities.  This convergence rate is the same as the one
obtained for this problem with standard state-of-the-art second-order
Eulerian methods \citep[e.g.][]{Springel2009,Stone2008}, which is
reassuring. Note that in this test a larger number of neighbours
simply reduces the effective spatial resolution but does not lead to
an advantage in the L1 velocity norm. Nevertheless, the density is
more accurately reproduced for a larger number of neighbors. Because
the sound speed depends only on temperature (which is set as part of
the initial conditions), the density error apparently does not
appreciably affect the travel speed of acoustic waves in this test.

\begin{table}
\begin{center}
\begin{tabular}{lcccccc}
\hline
& $\rho_1$ & $P_1$ & $v_1$ & $\rho_2$ & $P_2$ & $v_2$ \\
\hline
Problem 1 & 1.0  &  1.0 &   0.0     & 0.125  & 0.1  &   0.0  \\
Problem 2 &  1.0  &  0.4 &   -2.0    &  1.0    &  0.4 &  2.0    \\
Problem 3 & 1.0  &     1000.0 & 0.0 &  1.0    &    0.01  & 0.0   \\
\hline
\end{tabular}
\end{center}
\caption{Parameters of the one-dimensional Riemann problems (for $\gamma=1.4$)
  examined here.
 \label{TabRiemann}}
\end{table}

\subsection{One-dimensional Riemann Problems}

\begin{figure}
\begin{center}
\resizebox{4.5cm}{!}{\includegraphics{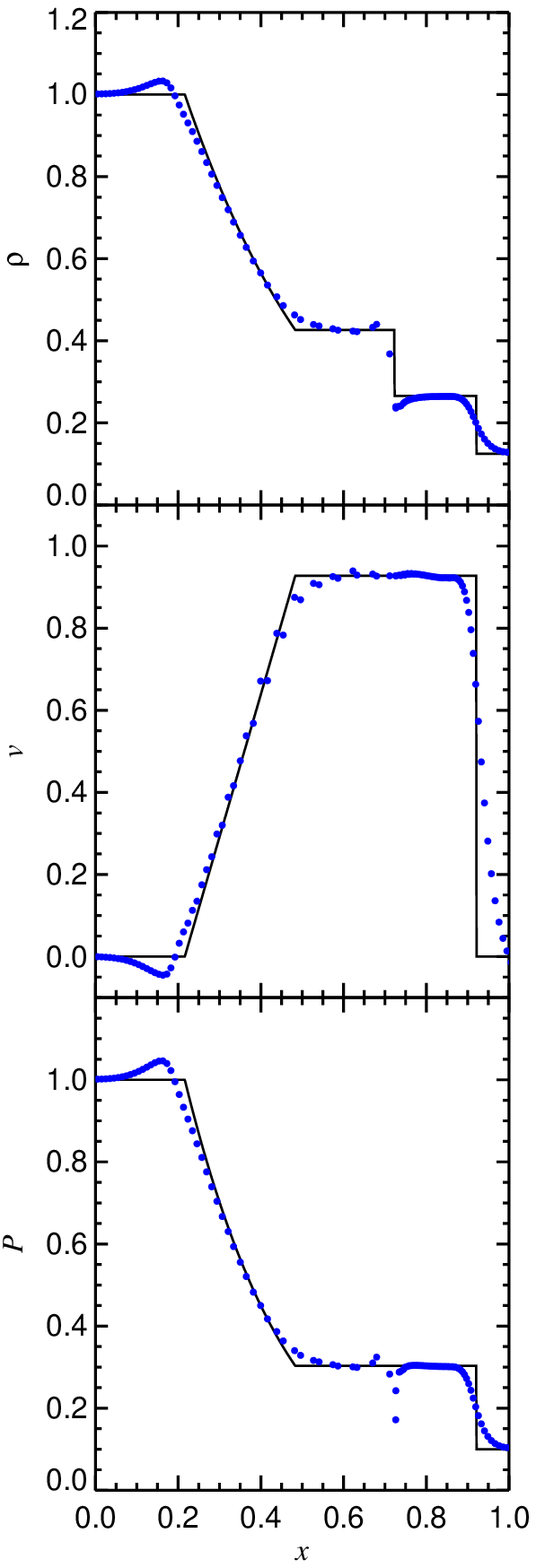}}%
\resizebox{4.5cm}{!}{\includegraphics{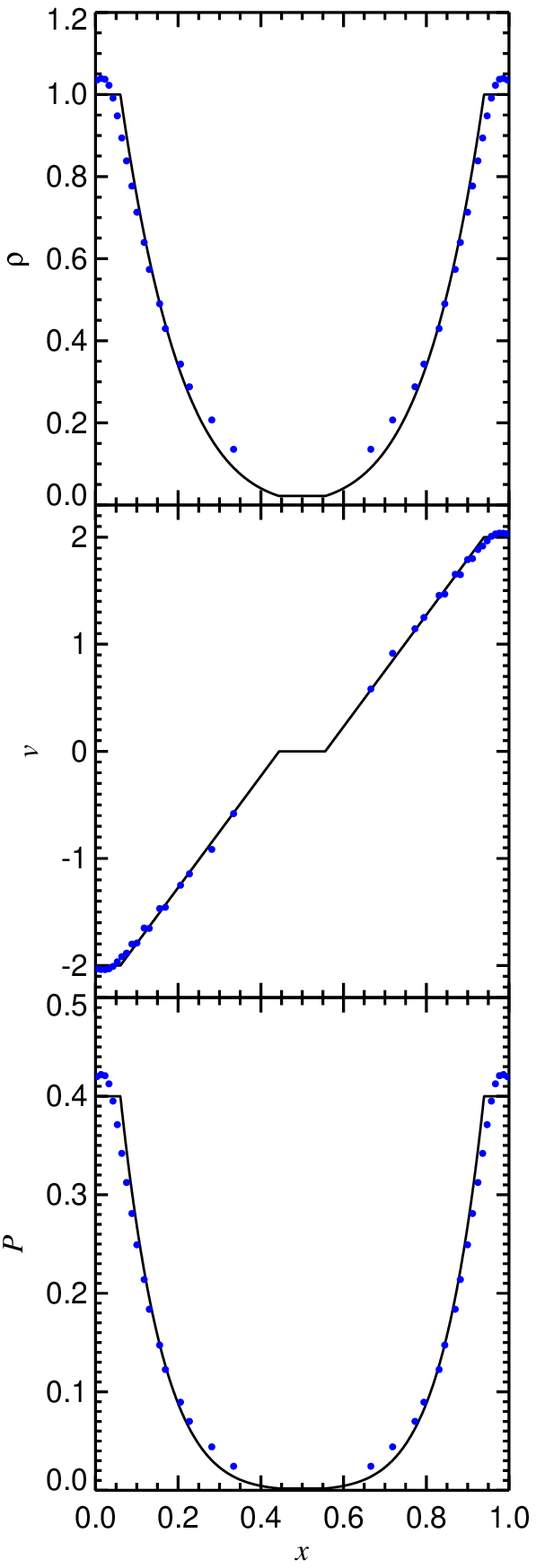}}%
\resizebox{4.5cm}{!}{\includegraphics{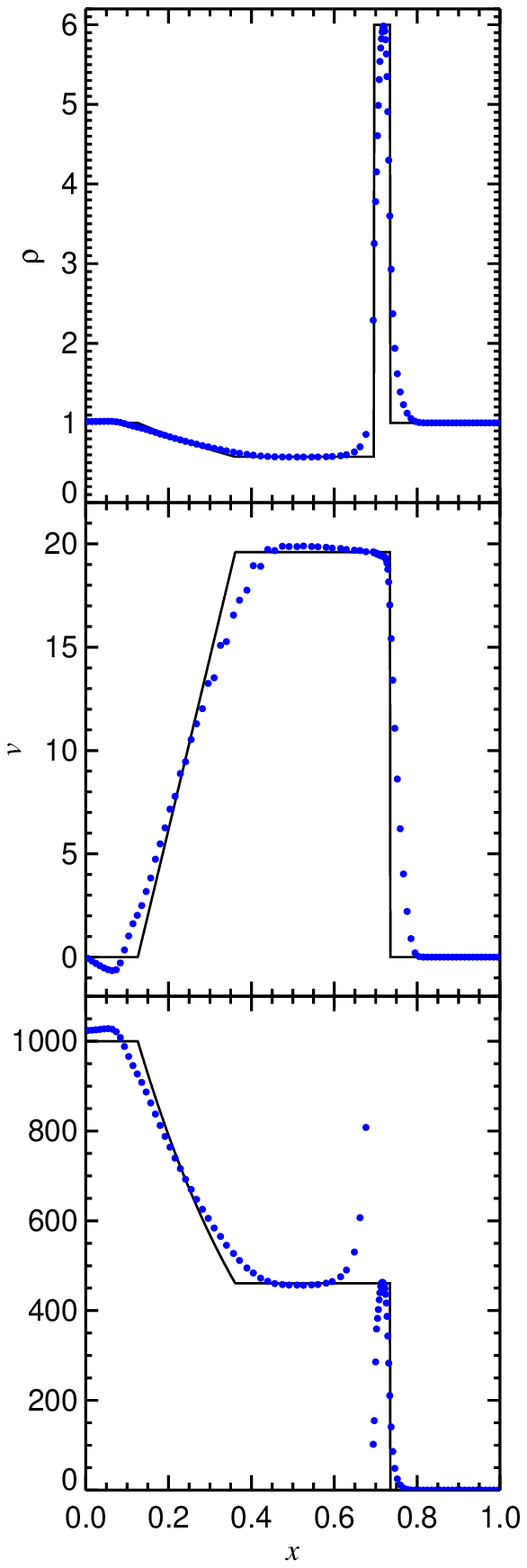}}
\vspace*{-0.5cm}
\end{center}
\caption{Different one-dimensional Riemann problems, calculated with a
  resolution of 100 points in the unit domain, and 7 smoothing
  neighbours. The three columns show results for the initial
  conditions of the problems 1, 2 and 3 as specified in
  Table~\ref{TabRiemann}. Symbols represent the SPH particles, solid
  lines the analytic solutions for density, velocity and pressure,
  from top to bottom.
  \label{Fig1DShocks}}
\end{figure}

\begin{figure}
\begin{center}
\resizebox{12.0cm}{!}{\includegraphics{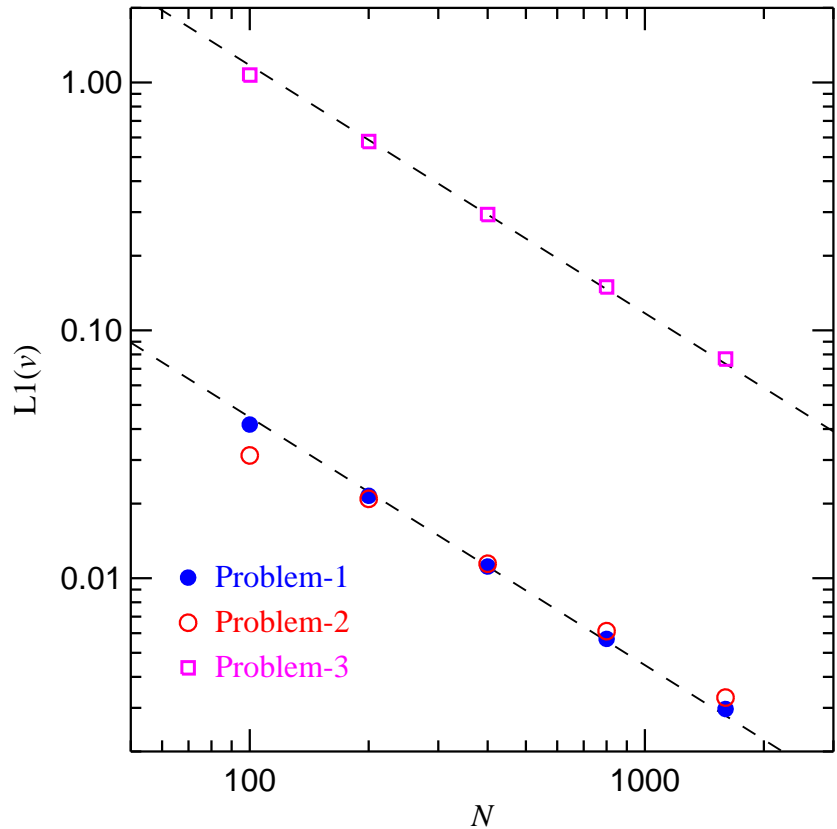}}%
\vspace*{-0.5cm}
\end{center}
\caption{Convergence rate of one-dimensional Riemann problems (as
  labeled), calculated with a resolution of 100 points in the unit
  domain, using 7 smoothing neighbours. The dashed lines indicate a
  ${\rm L1}\propto N^{-1}$ scaling of the error.
  \label{Fig1DShockConvergence}}
\end{figure}

Let us now consider a few Riemann problems in one dimension. Their
initial conditions are characterized by two piece-wise constant states
that meet discontinuously at $x=0.5$ at time $t=0$.  The subsequent
evolution then gives rise to a set of self-similar waves, containing
always one contact wave, which is sandwiched on the left and the right
by either a shock wave or a rarefaction wave. The ability to correctly
reproduce the non-linear outcome of arbitrary Riemann problems is the
backbone of any hydrodynamical method.

In Figure~\ref{Fig1DShocks} we show SPH results for three different
Riemann-type problems, with initial conditions characterized by
triples of density, pressure and velocity for either side, as listed
in Table~\ref{TabRiemann}.  For all these problems \citep[which are
taken from][]{Toro1997}, $\gamma=1.4$ has been assumed. Problem 1
gives rise to a comparatively weak shock, which has been studied in
very similar form (but with a different value of $P_2=0.1795$) in a
number of previous tests of SPH \citep[][among
others]{Hernquist1989,Rasio1991,Wadsley2004,Springel2005}.  In the
second problem, the gas is suddenly ripped apart with large supersonic
velocity, giving rise to a pair of strong rarefaction waves. Finally,
problem 3 is a again a Sod shock type problem but involves an
extremely strong shock.

The results shown in Figure~\ref{Fig1DShocks} are based on 100 points
with initially equal spacing in the unit domain, 7 smoothing
neighbours, and a standard artificial viscosity setting. All problems
are treated qualitatively correctly by SPH, with some inaccuracies at
the contact discontinuities.  Characteristically, shocks and contact
discontinuities are broadened over 2-3 smoothing lengths, and
rarefaction waves show a small over- and underestimate at their high-
and low-density sides, respectively. Also, there is a pressure
``blip'' seen at the contact discontinuity. However, the properties of
the postshock flow are correct, and the artificial viscosity has
successfully suppressed all postshock oscillations. Also, the errors
become progressively smaller as the resolution is increased. This is
seen explicitly in Figure~\ref{Fig1DShockConvergence}, where we show
results for the L1 error norm for the velocity as a function of
resolution. The error declines as ${\rm L1}\propto N^{-1}$, which is
expected due to the reduced order of the scheme at the
discontinuities. The same convergence rate for the shock-problem is
obtained with Eulerian approaches, as they too exhibit only first
order accuracy around discontinuities in the solutions.

Often the discussion of the numerical accuracy of shocks focuses on
the sharpness with which they are represented, and on that basis SPH
has frequently been portrayed as being inferior in comparison with
Eulerian methods. However, it should be noted that in both approaches
the numerical shock width is always many orders of magnitude larger
than the true width of the physical shock layer.  What is much more
relevant than the width is therefore that the properties of the
post-shock flow are correct, which is the case in SPH.  Also, the
numerical thickness of the shocks may be reduced arbitrarily by using
more particles, and is hence entirely a matter of resolution. We also
point out that contact discontinuities can often become quite broad in
Eulerian methods; in fact, their width increases with growing
advection speed, unlike in the Galilean-invariant SPH.  As far as
one-dimensional hydrodynamics is concerned, SPH can hence be
characterized as being quite accurate, and in particular, its
convergence rate appears competitive with second-order Eulerian
schemes.

\subsection{Two-dimensional Shock Waves}

\begin{figure}
\begin{center}
\resizebox{6.5cm}{!}{\includegraphics{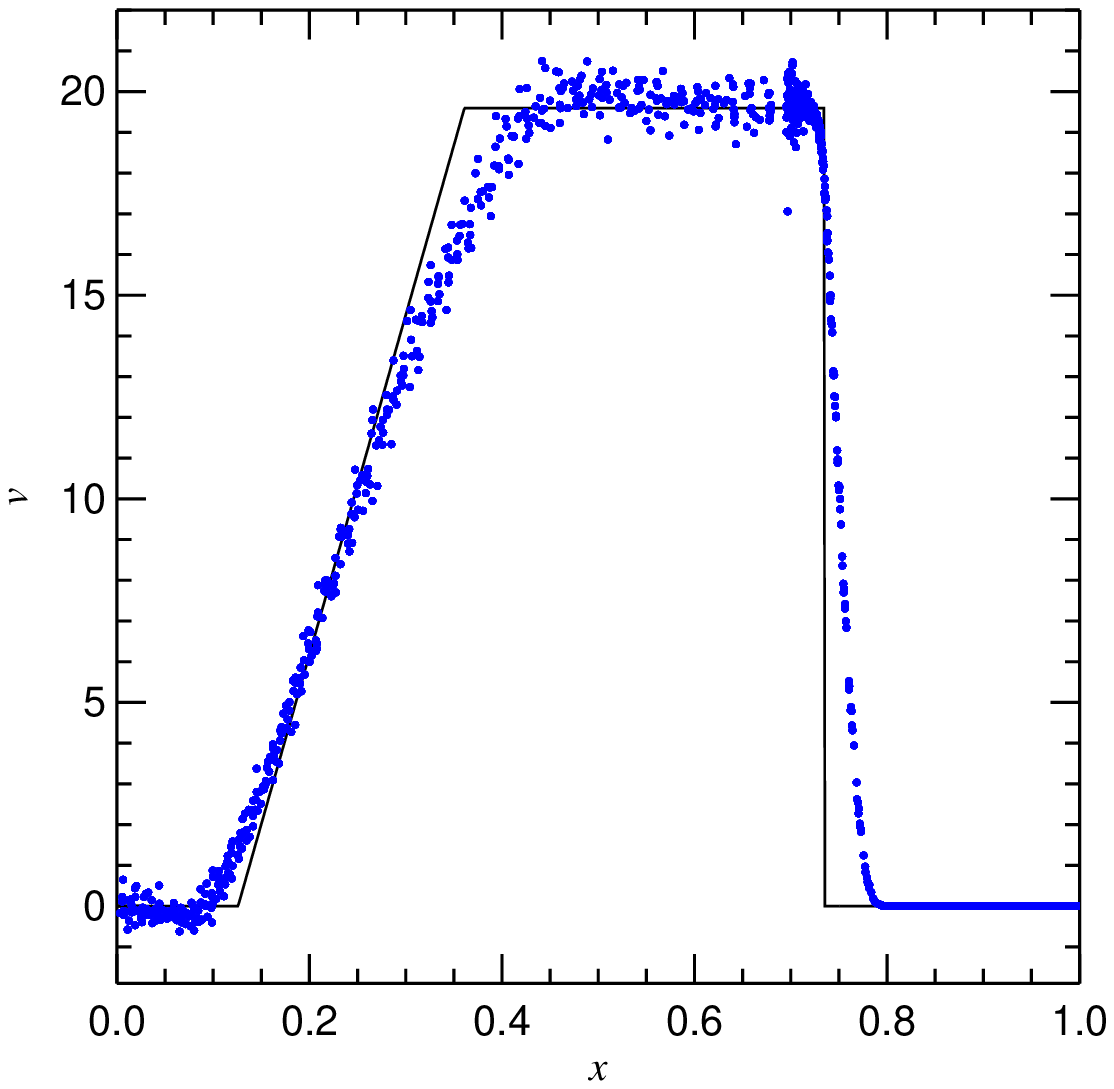}}%
\resizebox{6.5cm}{!}{\includegraphics{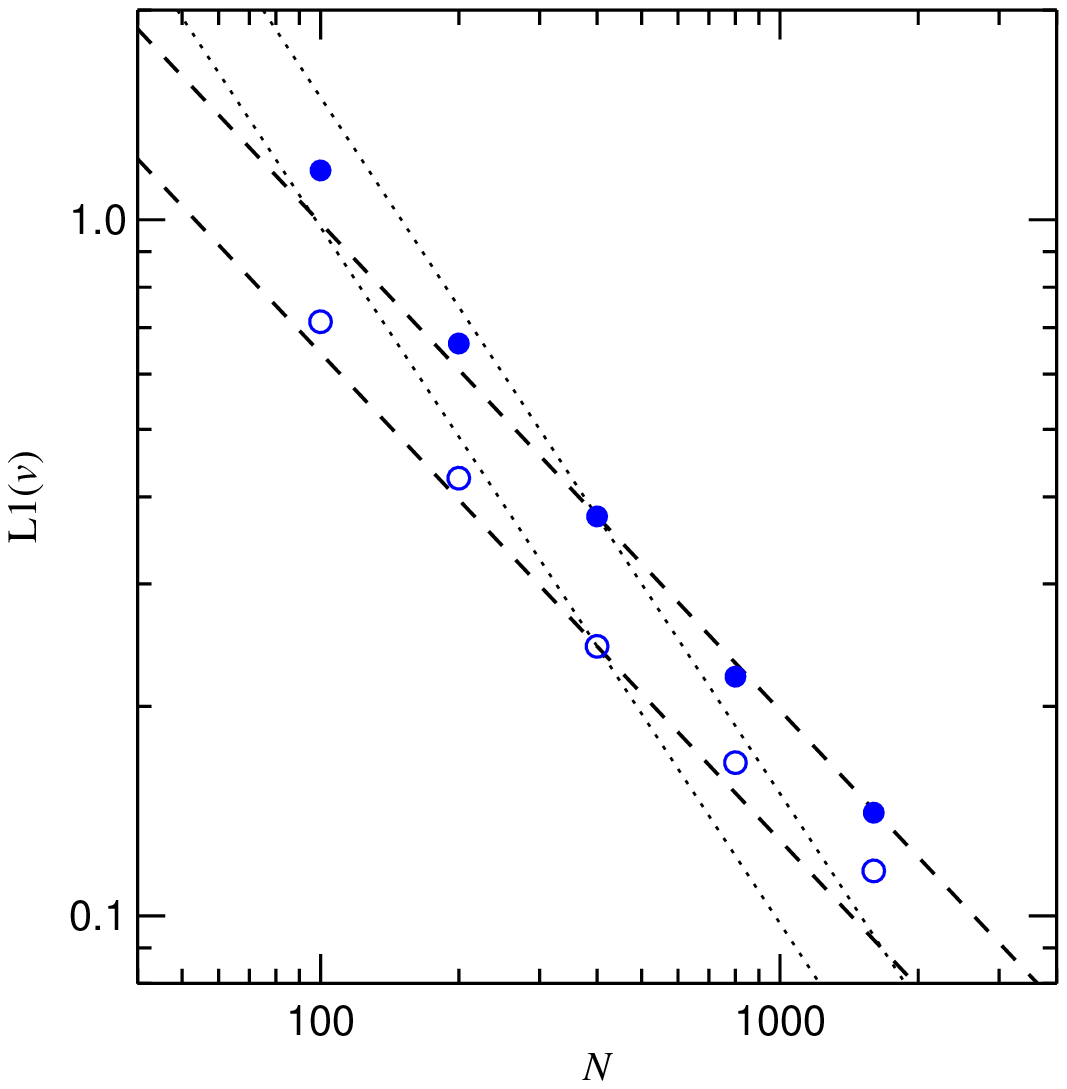}}%
\vspace*{-0.5cm}
\end{center}
\caption{Velocity profile and convergence rate of a two-dimensional
  strong shock problem. In the left panel, symbols show the
  velocities of all SPH particles, compared to the analytic result
  (solid line).  In the right panel, the open circles show the L1
  error for a binned result (using $N$ bins in the $x$-direction in
  each case), the filled circles give the error when individual
  particles are considered.  In this 2D problem, the ideal $N^{-1}$
  convergence rate (dotted lines) of 1D simulations is reduced to
  approximately $L1\propto N^{-0.7}$ (dashed lines).
  \label{Fig2DShockConvergence}}
\end{figure}

Multi-dimensional hydrodynamics adds much additional complexity, such
as shear flows, fluid instabilities, and turbulence. We here first
briefly examine whether a problem with one dimensional symmetry, again
a Riemann problem, can be equally well represented with SPH in
multiple dimensions. For definiteness, we study again the strong shock
of problem 3 from the previous section, but this time with the
$y$-dimension added to the initial set-up. This is taken as a regular
Cartesian $N \times N$ particle grid in the unit square, with periodic
boundaries in the $y$-direction, and reflective boundaries in the
$x$-direction.

In the left panel panel of Figure~\ref{Fig2DShockConvergence}, we show
the $x$ velocities of all particles of a run at $100\times 100$
resolution and compare them to the analytic solution. Relative to the
corresponding 1D-result shown in the middle-right panel of
Figure~\ref{Fig1DShocks}, the primary difference is a considerably
increased `noise' in the particle velocities.  This noise is a generic
feature of multi-dimensional flows simulated with SPH. It can be
reduced by an enlarged artificial viscosity or a larger number of
neighbours, but typically tends to be much larger than in 1D
calculations. Presumably, this is just a consequence of the higher
degree of freedom in the particle motion.

Unsurprisingly, the noise has a negative impact on the convergence
rate of planar shocks in multi-dimensional SPH. This is shown for the
same problem in Figure~\ref{Fig2DShockConvergence}, where we measure
${\rm L1}\propto N^{-0.7}$ instead of the ideal ${\rm L1}\propto
N^{-1.0}$. It also does not appear to help to first bin the particles
and then to compute the L1 error as a difference between the averaged
and the analytical result. Although this reduces the absolute size of
the error substantially and is certainly warranted to eliminate the
intrinsic noise, it does not affect the convergence rate itself (see
Fig.~\ref{Fig2DShockConvergence}).  Nevertheless, the noise does not
destroy the principal correctness of the solution obtained with SPH
for multi-dimensional planar Riemann problems.

\subsection{Shear Flows} \label{SecGresho}

We now a discuss a yet more demanding two-dimensional problem, the vortex test 
of Gresho \citep{Gresho1990,Liska2003}.
It consists of a triangular
azimuthal velocity profile
\begin{equation}
v_\phi(r) = \left\{
\begin{array}{ll}
5r & {\rm for}\;\;\; 0\le r <0.2 \\
2-5r & {\rm for}\;\;\; 0.2\le r <0.4 \\
0 & {\rm for}\;\;\; r \ge 0.4 \\
\end{array}
\right.
\end{equation}
in a gas of constant density equal to $\rho=1$ and adiabatic index
of $\gamma=5/3$.  By adopting the pressure profile
\begin{equation}
P(r) = \left\{
\begin{array}{ll}
5 + 25/2 r^2 & {\rm for}\;\;\; 0\le r <0.2 \\
9 + 25/2 r^2 - \\\;\;\;\;\;\;20 r + 4 \ln (r/0.2) & {\rm for}\;\;\; 0.2\le r <0.4 \\
3+4\ln 2 & {\rm for}\;\;\; r \ge 0.4 \\
\end{array}
\right.
\end{equation}
the centrifugal force is balanced by the pressure gradient, and the vortex becomes
independent of time.

\begin{figure}
\begin{center}
\resizebox{4.5cm}{!}{\includegraphics{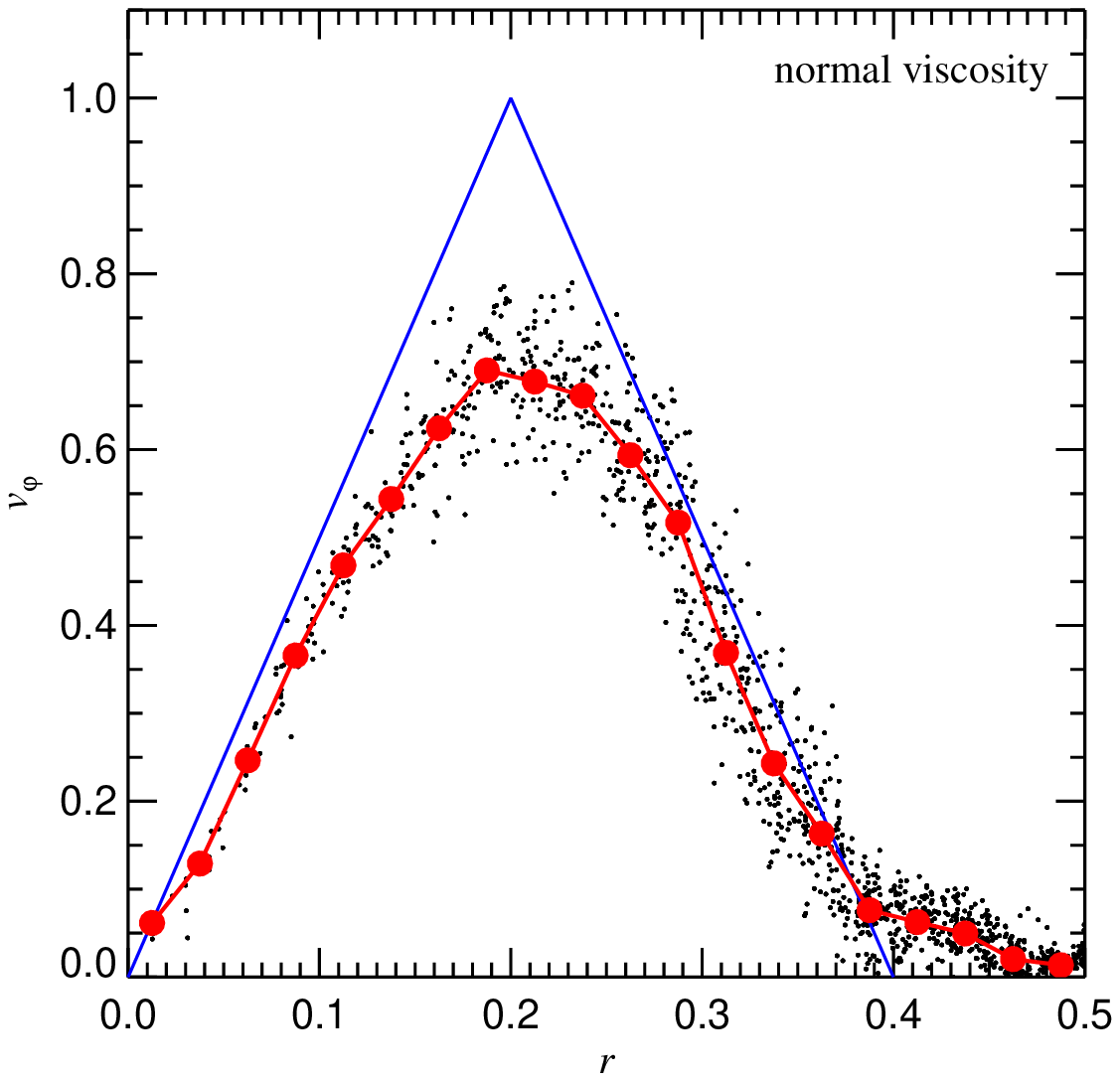}}%
\resizebox{4.5cm}{!}{\includegraphics{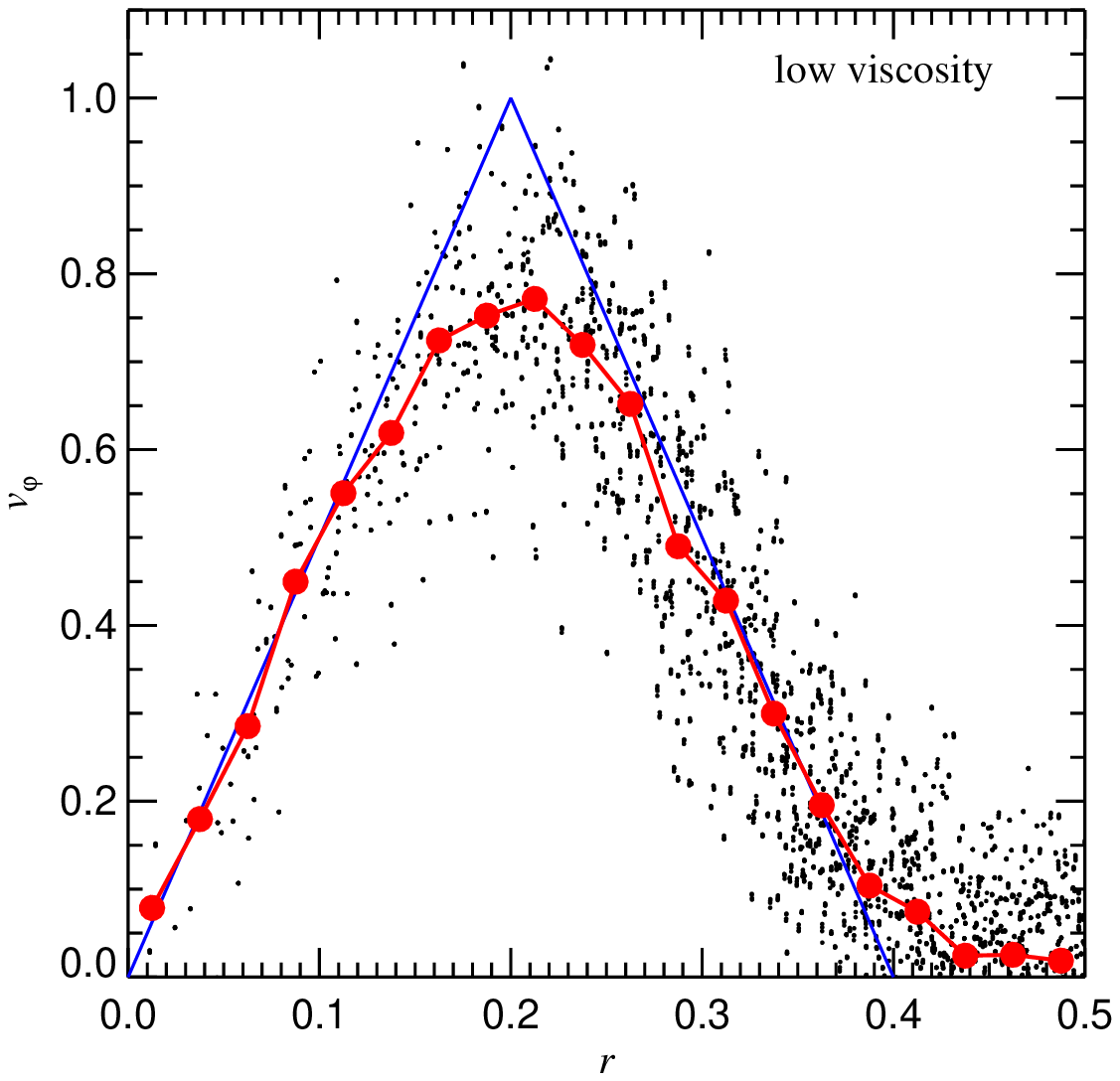}}%
\resizebox{4.5cm}{!}{\includegraphics{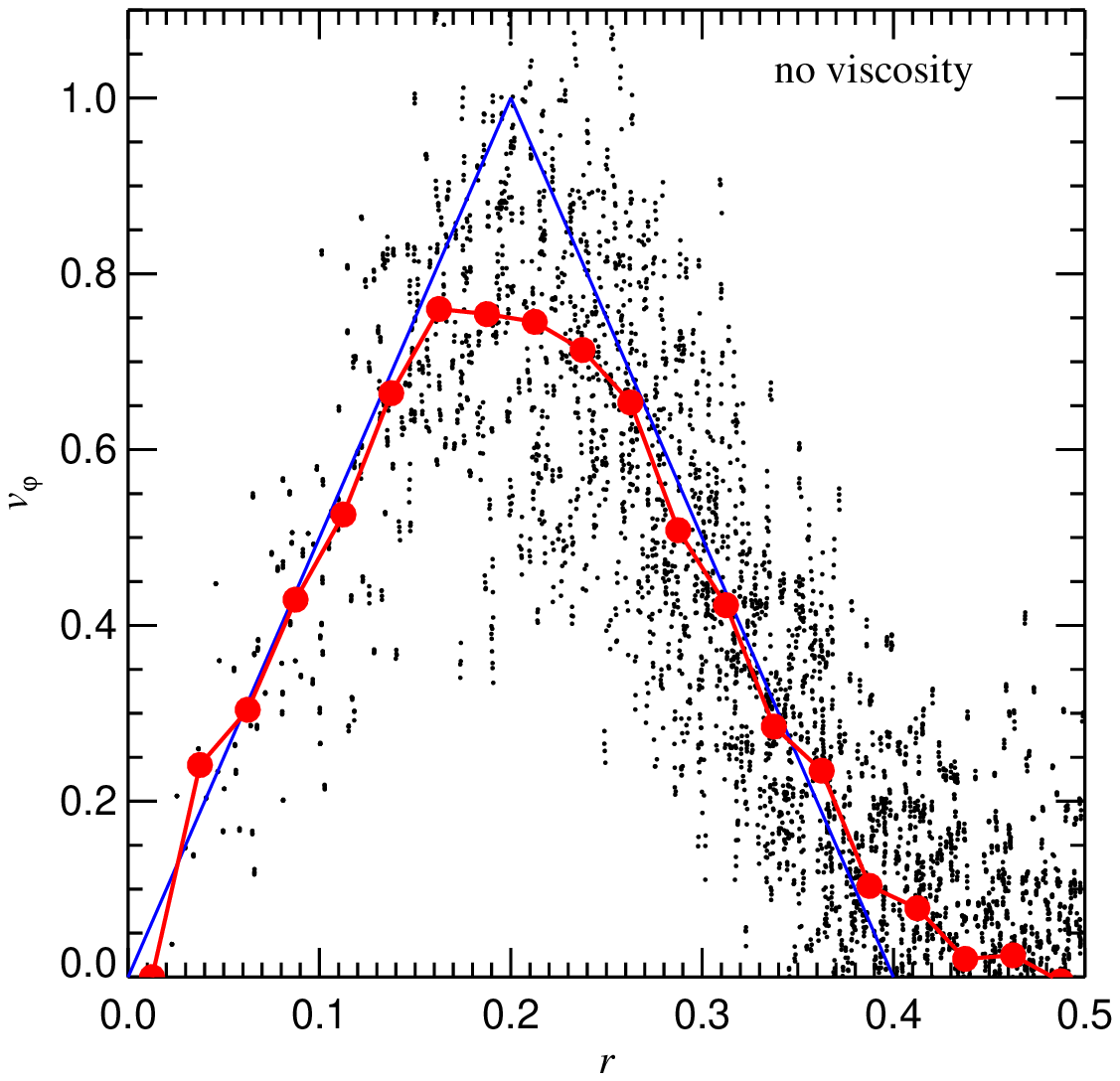}}
\vspace*{-0.5cm}
\end{center}
\caption{Velocity profile in the Gresho vortex test at time $t=1.0$,
  at a base resolution of $80\times 80$, for different viscosity
  settings. Here `low viscosity' corresponds to $\alpha=0.05$, and
  `normal viscosity' to $\alpha=1.0$, with the Balsara switch enabled
  in both cases. The small dots are the azimuthal velocities of
  individual SPH particles, while the red circles show binned
  results. The blue triangular shaped profile is the analytic
  (stationary) solution.
 \label{FigGreshoProfiles}}
\end{figure}

In Figure~\ref{FigGreshoProfiles}, we compare the results for the
azimuthal velocity profiles at time $t=1.0$ for three different runs,
carried out with $80\times 80$ particles in the unit domain for
different settings of the artificial viscosity. The left-most panel
shows the outcome for a standard viscosity of $\alpha=1.0$, the middle
panel is for $\alpha=0.05$ and the right panel did not use any
artificial viscosity at all. We see that in all three cases
substantial noise in the velocity profile develops, but it is clearly
largest in the simulation without viscosity. On the other hand, one
can see that the average velocity profile of the run without viscosity
is actually closest to the expected stationary solution (blue lines),
while the standard viscosity run already shows a reduced angular
frequency in the solid-body part of the rotation in the inner part of
the vortex. Despite the use of the Balsara switch in this problem, the
velocity noise, and as a consequence the noisy estimates of divergence
and curl, have produced enough residual viscosity to lead to
appreciable angular momentum transport.

\begin{figure}
\begin{center}
\resizebox{12.5cm}{!}{\includegraphics{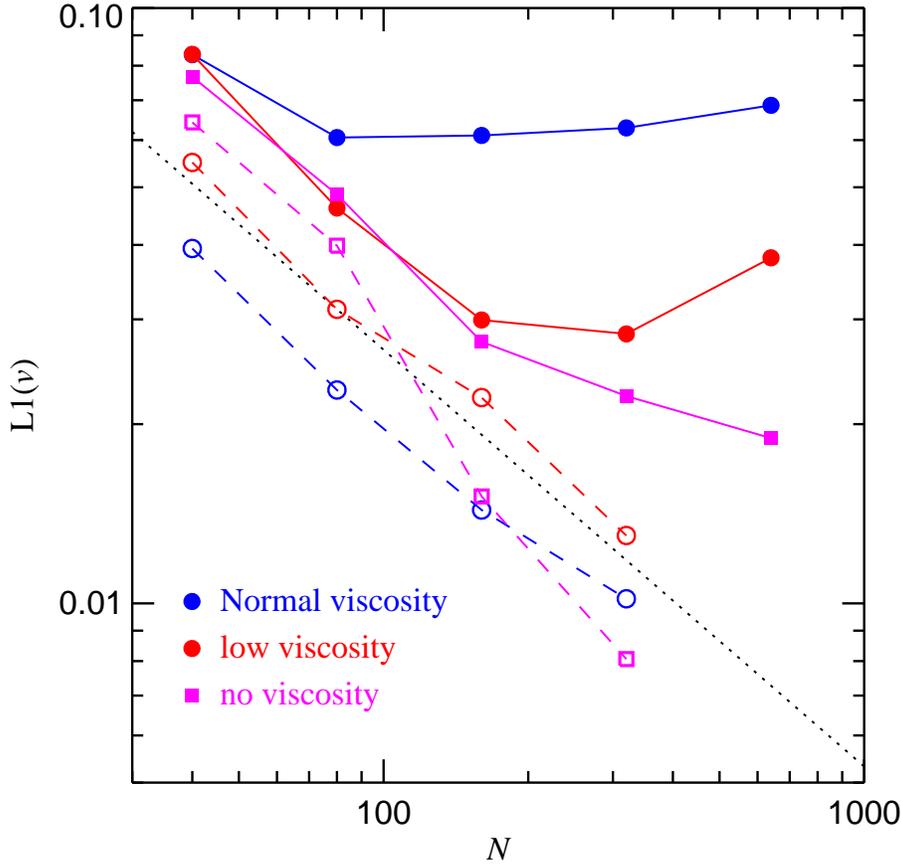}}%
\vspace*{-0.5cm}
\end{center}
\caption{Convergence rate of the Gresho problem, in terms of the
  binned azimuthal velocity, for different viscosity settings. Here
  the open symbols measure the L1 error towards the highest resolution
  result, while the filled symbols measure the difference with respect
  to the analytic solution. The dotted line indicates a ${\rm
    L1}\propto N^{-0.7}$ convergence.
  \label{FigGreshoConv}}
\end{figure}

This is corroborated by the results of convergence tests for this
problem. In Figure~\ref{FigGreshoConv}, we consider the L1 error norm
of the binned azimuthal velocity profile (based on $N/2$ bins,
linearly placed in the radial range 0 to 0.5). Interestingly, the two
runs with non-vanishing viscosity do not converge to the analytic
solution with higher resolution. Instead, they converge to a different
solution. If we approximately identify this solution as the highest
resolution result in each case, then we see that the lower resolution
calculations converge to it with ${\rm L1}\propto N^{-0.7}$. What is
important to note is that the effects of the artificial viscosity on
the shear flow do not diminish with higher resolution. Instead, the
solutions behave as if one had simulated a fluid with some residual
shear viscosity instead. Only the run without viscosity appears to
converge to the correct physical solution, albeit at a very low rate
for high $N$.  In any case, it appears that the convergence rate of
SPH for this problem is considerably worse than the ${\rm L1}\propto
N^{-1.4}$ measured for a moving-mesh code and Eulerian codes by
\citet{Springel2009}.

\subsection{Contact Discontinuities and Fluid Instabilities}

\begin{figure}
\begin{center}
\resizebox{9.0cm}{!}{\includegraphics{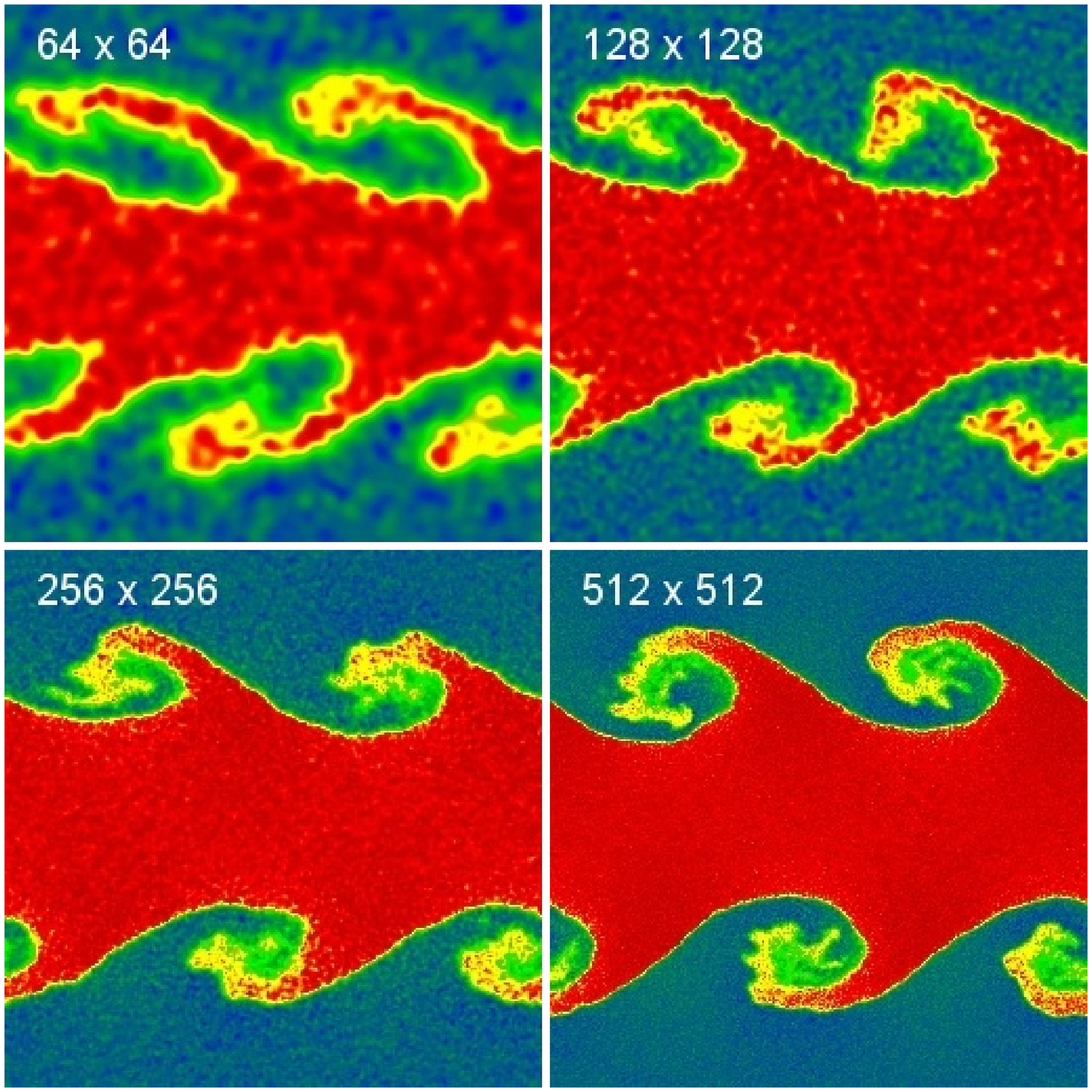}}\\
\vspace*{-0.5cm}
\end{center}
\caption{Density fields of the Kelvin-Helmholtz instability test at
  $t=2.0$, simulated with SPH for different resolutions as labeled.
  \label{FigKHImages}}
\end{figure}

\begin{figure}
\begin{center}
\resizebox{6.5cm}{!}{\includegraphics{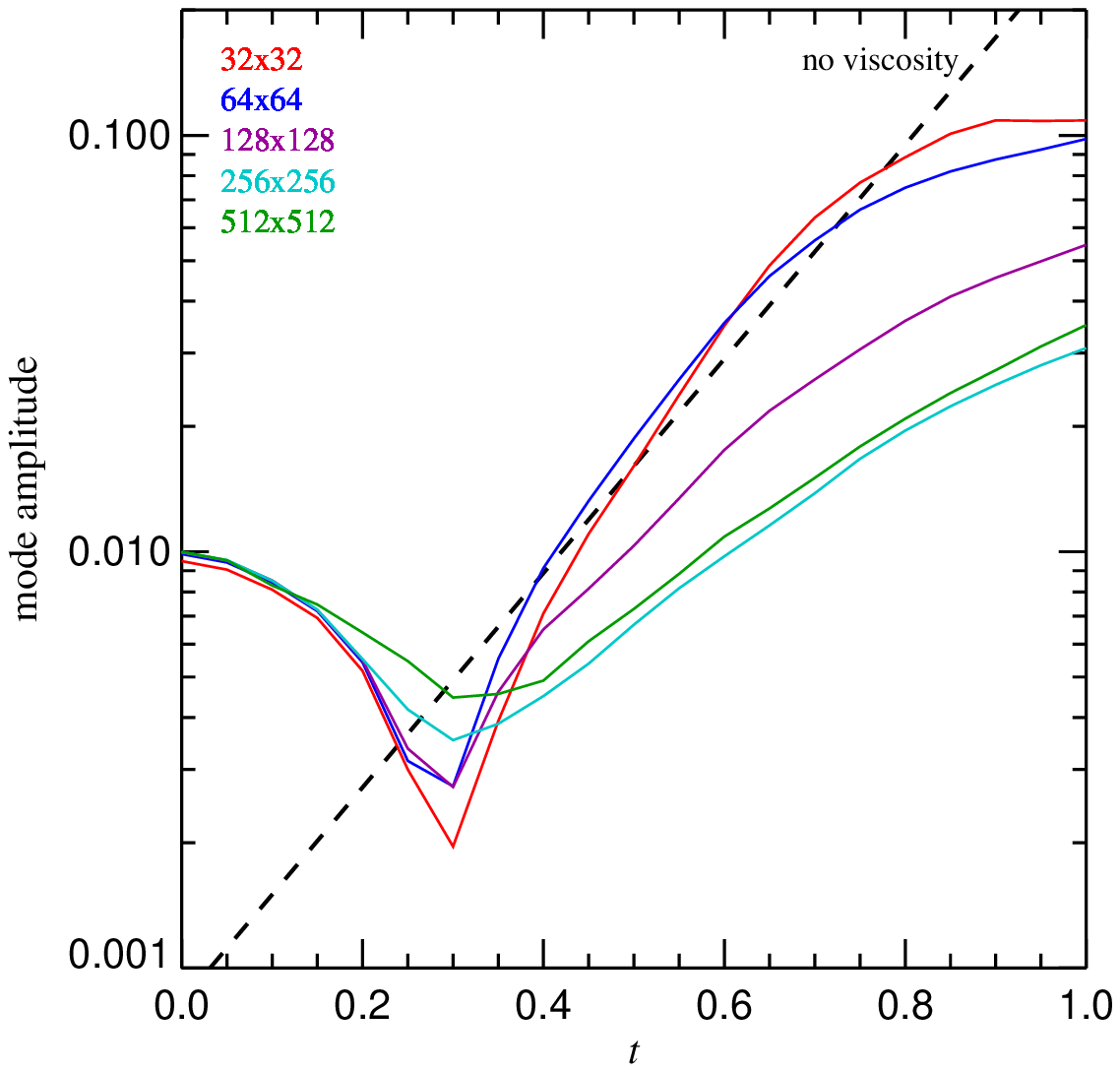}} %
\resizebox{6.5cm}{!}{\includegraphics{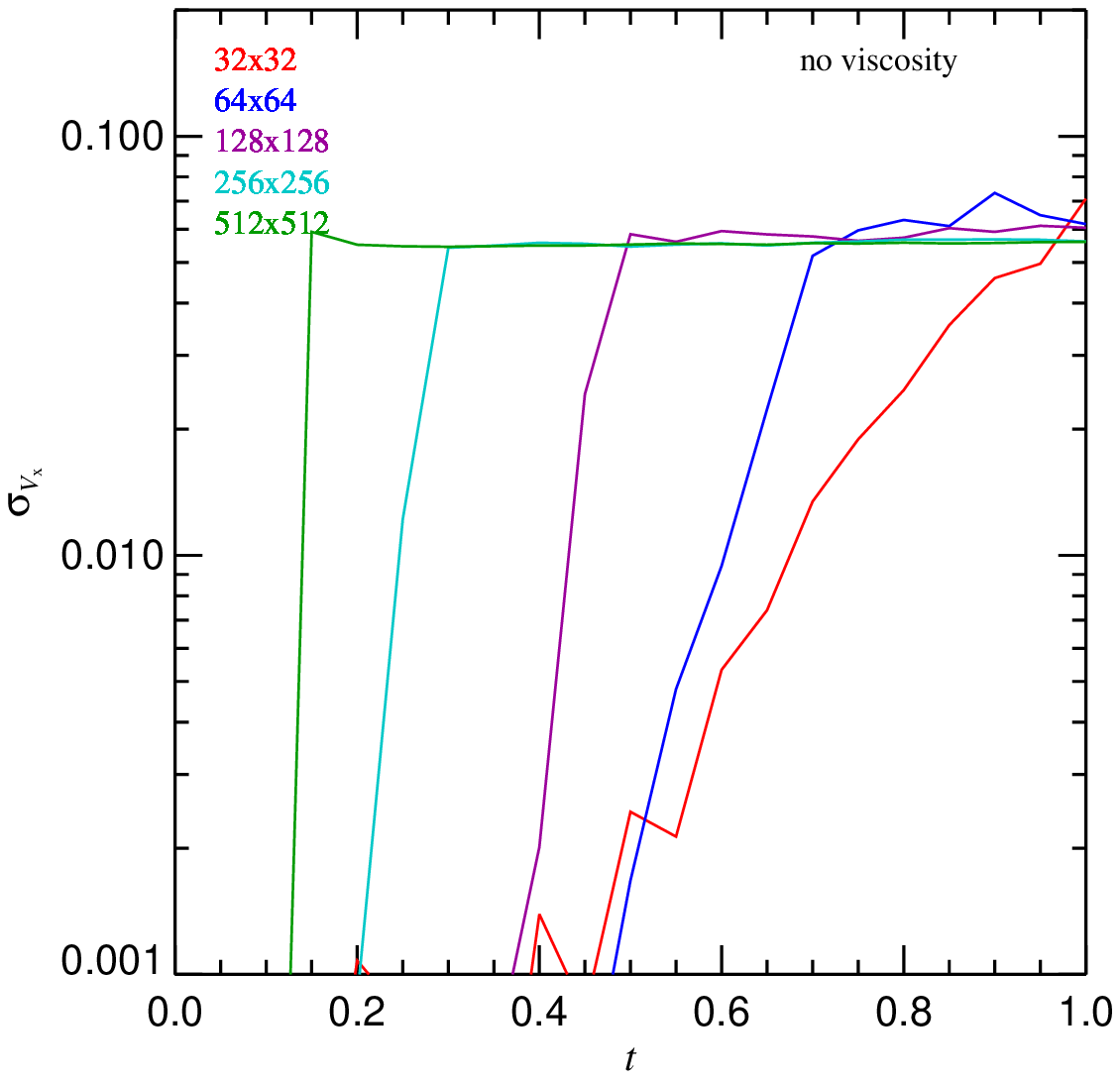}}
\vspace*{-0.5cm}
\end{center}
\caption{Growth rate of KH instabilities in SPH test simulations, for
  different resolutions as labeled. The left panel gives the
  amplitude of the initially excited velocity mode, compared to the
  expected linear growth rate (dashed line).  The right panel shows
  the velocity dispersion of the SPH particles in the $x$-direction as
  a function of time, for a thin layer close to the midplane of the
  box. This dispersion is an indication of the level of velocity noise
  in the calculation.
  \label{FigKH_growth_novisc}}
\end{figure}

\begin{figure}
\begin{center}
\resizebox{6.5cm}{!}{\includegraphics{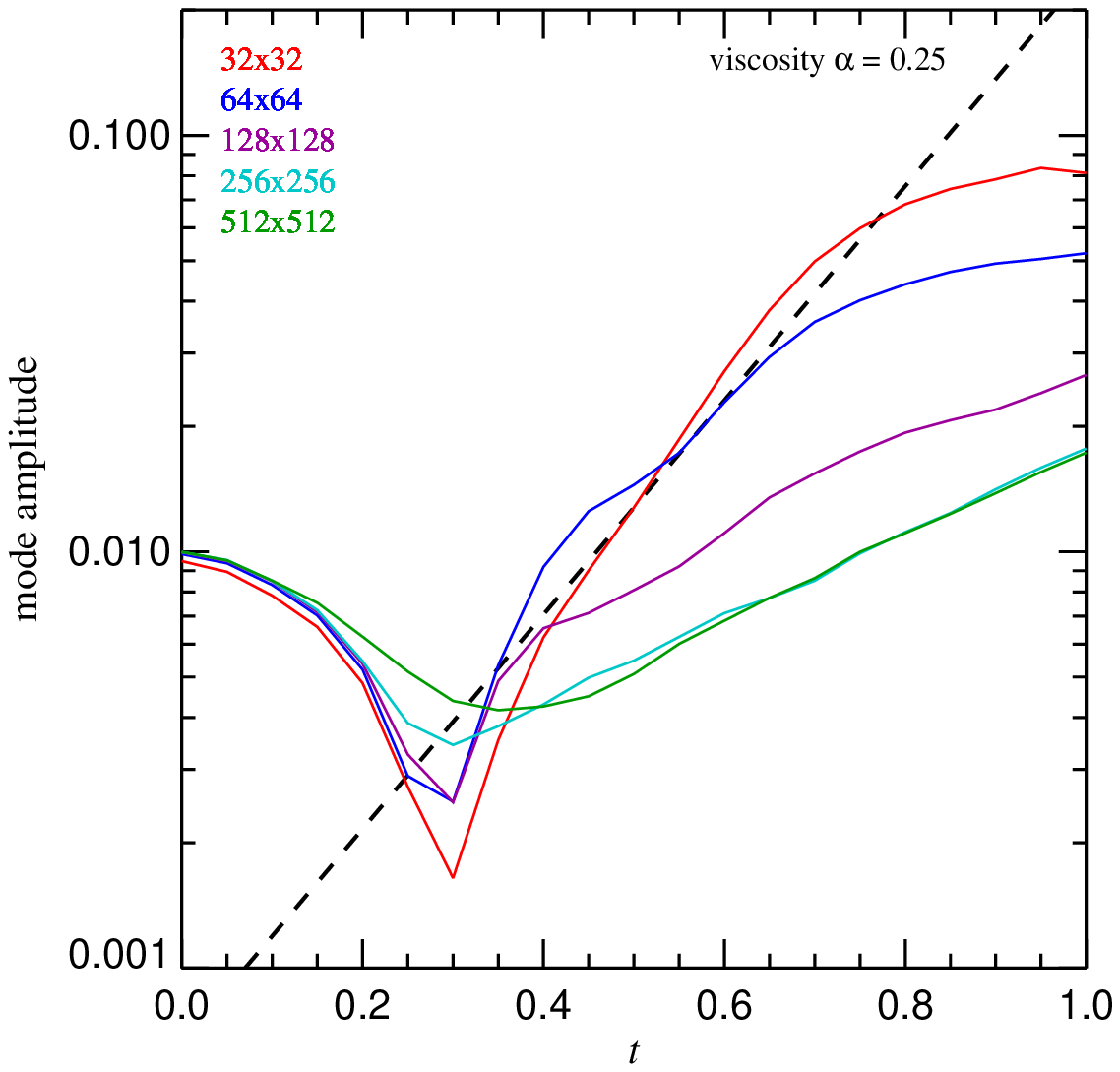}} %
\resizebox{6.5cm}{!}{\includegraphics{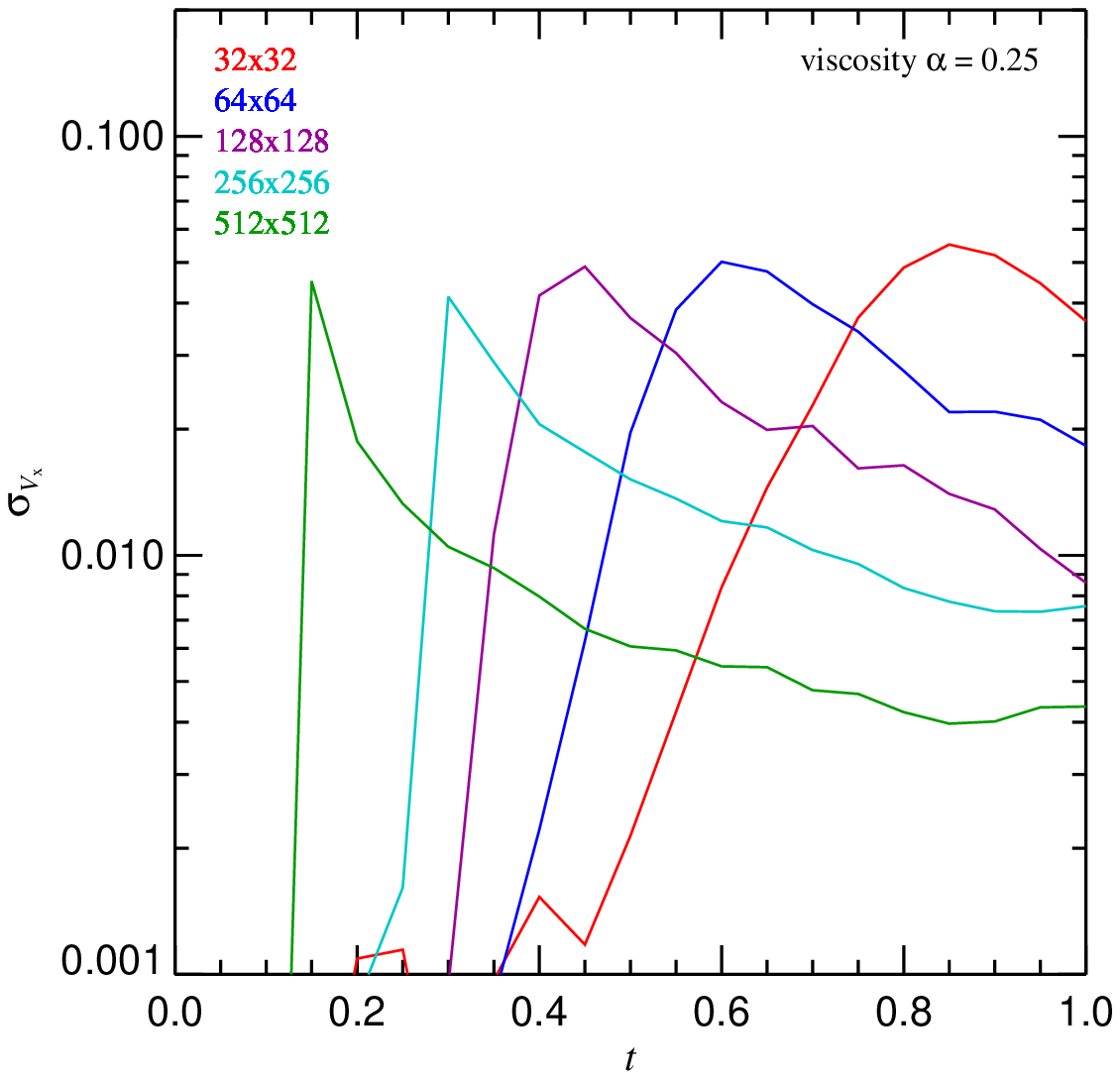}}
\vspace*{-0.5cm}
\end{center}
\caption{The same as in Figure~\ref{FigKH_growth_novisc}, but for
  simulations that include an artificial viscosity of moderate
  strength with $\alpha=0.25$.
  \label{FigKH_growth_visc}}
\end{figure}

An issue that has created considerable attention recently is the
question of whether SPH can properly resolve fluid instabilities, such
as the Kelvin-Helmholtz (KH) instability in shear flows.  For a set-up
with equal particle masses and a sharp initial density contrast of
$1:2$ at a contact discontinuity, \citet{Agertz2007} did not observe
any growth of the instability, whereas for a vanishing density jump
the fluids would start to mix. \citet{Agertz2007} attributed these
problems to substantial errors in the pressure gradient estimates at
the contact discontinuity. A number of recent studies have addressed
this problem as well, proposing modifications of SPH designed to
improve the KH results. In particular, \citet{Price2008_KH} has
suggested adding an artificial heat conduction to smooth out the phase
boundary, which indeed improved the results. A different approach was
followed by \citet{Read2009}, who employed a different kernel, a much
enlarged number of neighbors, and a modified density estimation
formula to obtain a better representation of mixing of different
phases in shear layers. We here do not examine these modifications but
want to shine more light onto the nature of the inaccuracies of
standard SPH for the KH problem.

We adopt the KH test parameters of \citet{Springel2009}, i.e.~a
central phase in the region $|y-0.5| < 0.25$ is given density
$\rho_2=2$ and velocity $v_x = 0.5$, while the rest of the gas has
density $\rho_1=1$ and velocity $v_x=-0.5$, all at the same pressure
of $P=2.5$ with $\gamma=5/3$.  In order to soften the transition at
the two interfaces, we follow \citet{Robertson2009} and impose a
smooth transition in the vertical profiles of density $\rho(y)$, shear
velocity $v_x(y)$, and specific entropy $A(y) =
P/\rho(y)^{\gamma}$. For example, the initial density structure is
adopted as
\begin{equation}
\rho(y) = \rho_1 + \frac{
  \rho_2-\rho_1}{\left[1+\exp(-2(y-0.25)/\sigma)\right]
\left[1+\exp(2(y-0.75)/\sigma)\right]},
\end{equation}
with $\sigma=0.025$, and similar adoptions are made for the other
profiles.  The transition removes the sharp discontinuities from the
initial conditions, such that they can in principle be fully resolved
by the numerical scheme. This regularization makes the problem
well-posed for convergence studies, which is particularly important
for the KH instability, where smallest scales grow fastest.

For simplicity, we consider a Cartesian particle grid in the unit
domain with periodic boundaries, and trigger the instability by
applying the velocity perturbation throughout the $y$-domain, as
\begin{equation}
v_y(x)= v_0 \, \sin( k\,x),
\end{equation}
with an initial amplitude of $v_0=0.01$ and a wave number of $k= 2
\times (2\pi)/L$.  We can easily study the growth of this velocity
mode by measuring its amplitude through a Fourier transform of the
two-dimensional $v_y$ velocity field.

In Figure~\ref{FigKHImages}, we show density maps of the evolved
fields at time $t=2.0$ for different SPH simulations, using 27
neighbours and no artificial viscosity. As can be seen, the KH
instability clearly develops and produces the characteristic wave-like
features. The density field is visibly noisy though, as a result of
the absence of artificial viscosity. If the latter is added, the noise
is substantially reduced, but the KH billows look much more anemic and
have a reduced amplitude.

An interesting question is whether the KH instability actually grows
with the right rate in these SPH calculations. This is examined in
Figure~\ref{FigKH_growth_novisc}, where the amplitude of the excited
$v_y$ mode of the velocity field is shown as a function of time, for
the case without viscosity. The growth is compared to the expected
exponential growth rate $v_y \propto \exp(t/\tau_{\rm KH})$ shown as a
solid line, where
\begin{equation}
\tau_{\rm KH} = \frac{\rho_1 + \rho_2}{ |v_2-v_1|\,k \sqrt{\rho_1 \rho_2} }
\end{equation}
is the KH growth timescale for an inviscid gas.  After an initial
transient phase (which is expected because the initial perturbation
was not set-up self-consistently), the two low-resolution calculations
with $32^2$ and $64^2$ particles do actually reproduce this growth
rate quite accurately for some time, but then the growth rate suddenly
slows down considerably. Curiously, the higher resolution simulations
never quite reach the expected growth rate. So what is going on here?

A hint is obtained by the results shown in the right panel of
Fig.~\ref{FigKH_growth_novisc}, where the $v_x$ velocity dispersion of
the particles in a narrow strip around $y\simeq 0.5$ is shown, as a
function of time. We in principle expect this to remain close to zero
until very late in the evolution. However, what we actually observe is
a sudden increase of this dispersion, when the regularity and
coherence of the initial Cartesian particle grid is finally lost. At
this moment, SPH develops its velocity noise that is characteristic
even of smooth flows. Interestingly, we observe that the development
of this velocity noise closely coincides in time with the termination
of the correct linear growth rate. The natural interpretation is that
as long as the particles are still quite regularly ordered, the
analytic linear KH growth rate is calculated accurately by SPH, but at
late times the developing noise introduces a substantial degradation
of the accuracy that reduces the proper growth rate.

This suggests that perhaps the use of the right amount of artificial
viscosity may yield an accurate growth rate even at late times if it
prevents the detrimental noise effects.  In
Figure~\ref{FigKH_growth_visc}, the equivalent results are shown with
artificial viscosity included. Here the behaviour is rather similar
initially. As long as the regular order of the particles is still
approximately maintained, the artificial viscosity is not changing
anything as it is reduced by the Balsara switch to a very low
level. Once the initial order decays, the velocity dispersion suddenly
increases, first to the same level that one would have without
viscosity. Only after some time, the viscosity damps out this noise,
but this comes at the price of also damping the growth rate even
further, since the fluid behaves now in a slightly viscous fashion.

We conclude that whereas KH fluid instabilities do occur in SPH,
calculating them with high accuracy is a challenge due to the
significant noise present the multi-dimensional velocity
field. Getting rid off this noise with artificial viscosity tends to
reduce the KH growth rate below the analytic expectation, making it
hard to achieve truly inviscid behavior. More work on this problem is
therefore clearly warranted in the future.

\subsection{Surface Tension}

In the standard energy and entropy conserving formulation of SPH, two
phases of particles with different specific entropies tend to avoid
mixing, simply because this is energetically disfavored. To understand
the origin of this effect, let us consider a simple virtual experiment
where a box of volume $V$ is filled in one half with gas of density
$\rho_1$, and in the other half with gas of density $\rho_2$. We
assume an equal pressure $P$, such that the SPH particles in the
different phases will have specific entropies $A_1=P/\rho_1^\gamma$
and $A_2=P/\rho_2^\gamma$. Then the total thermal energy in the box
will be $E_{\rm therm} = u_1 M_1 + u_2 M_2 = P V / (\gamma-1)$. The
numerical estimate for a particular SPH realization may be slightly
different from this due to intermediate density values in the
transition layer, but this effect can be made arbitrarily small if a
large number of particles is used.  Now imagine that we rearrange the
particles by homogeneously spreading them throughout the volume, in
some regular fashion (say as two interleaved grids), but keeping their
initial entropies. Provided the SPH smoothing lengths are large
enough, each particle will then estimate the mean density
$\overline{\rho}=(\rho_1+\rho_2)/2$ as its new density, throughout the
volume.  The new thermal energy per unit mass for particles of species
1 will then be $u_1' = A_1\overline{\rho}^{\gamma-1} / (\gamma -1 )$
and similarly for particles of species 2. As a result, the new thermal
energy becomes $E_{\rm therm}' = \left[ P V /(\gamma-1)\right]
\left[(\rho_1 + \rho_2)/2\right]^{\gamma-1} \left[\rho_1^{1-\gamma} +
  \rho_2^{1-\gamma}\right]/2$, which is larger than the original
energy if $\rho_1 \ne \rho_2$.  Clearly then, the mixing of the
particle set in this fashion is energetically forbidden, and an energy
conserving code will resist it. This resistance appears as an
artificial surface tension term in SPH.

We note that the mixing can be accomplished at constant thermal
energy, but this requires that the entropies of the particles and
hence their temperature estimates are made equal. The final entropy is
then $\overline{A} = P / \overline{\rho}^\gamma$, which corresponds to
a total thermodynamic entropy that is larger than in the unmixed
state. Mixing the phases in this irreversible fashion hence requires
creating the relevant amount of mixing entropy, but for this no source
is foreseen in the entropy-conserving formulation of standard
SPH. Including artificial thermal conduction \citep[as
in][]{Price2008_KH} is one interesting approach to address this
problem, as this process equilibrates the temperatures while
conserving energy and increasing the entropy.

\begin{figure}
\begin{center}
\resizebox{6.5cm}{!}{\includegraphics{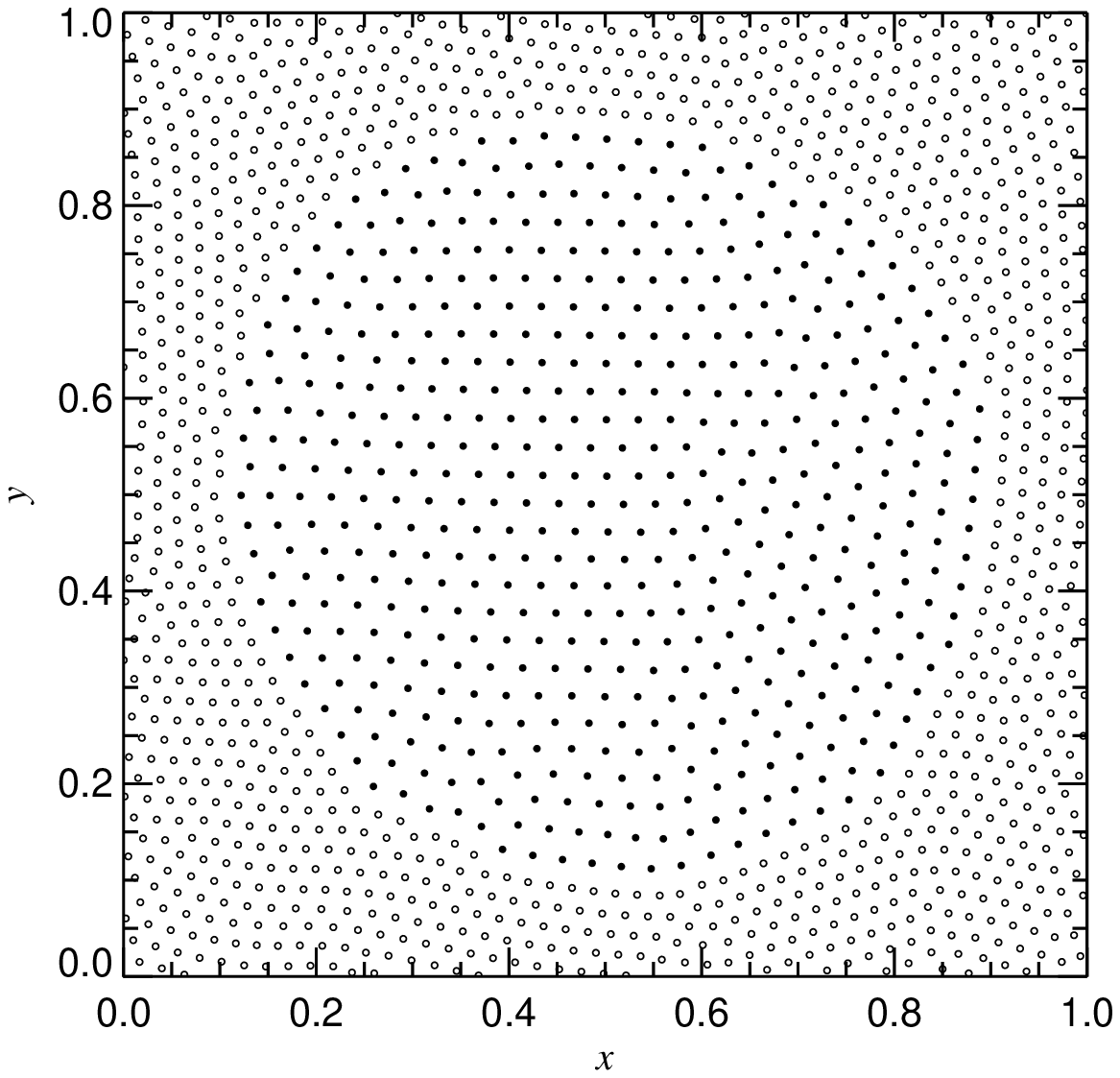}}%
\resizebox{6.5cm}{!}{\includegraphics{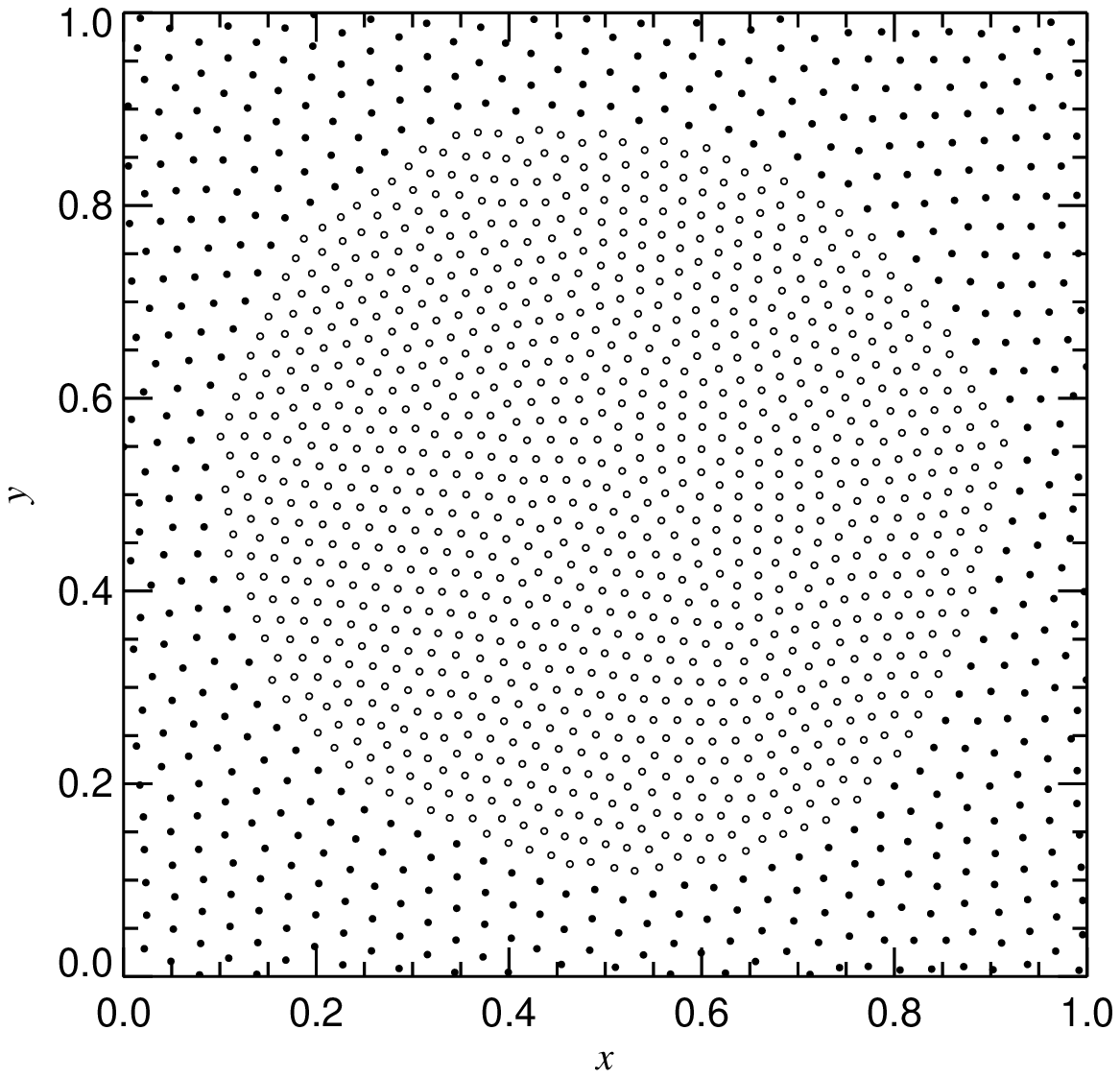}}
\vspace*{-0.5cm}
\end{center}
\caption{Equilibrium configurations of particles in 2D for two phases
  of different entropy, corresponding to a jump in density by a factor
  of 2. Both in the left and in the right case, the final pressure in
  the inside of the sphere is slightly higher than the one on the
  outside. This pressure difference offsets the spurious surface
  tension present at the interface. \label{FigTension}}
\end{figure}

The presence of some level of surface tension in SPH can be
demonstrated experimentally through simple settling tests
\citep{Hess2009}.  For example, one possibility is to set-up an
overdense spherical region and let it relax to an equilibrium
distribution (this can be done by keeping the particle entropies
fixed, and by adding an artificial decay of the velocities). Then the
pressure inside of the sphere in the final relaxed state is found to
be slightly higher than outside, as it needs to be to balance the
surface tension.  In fact, according to the Young-Laplace equation,
the expected pressure difference is
\begin{equation}
\Delta P = \sigma \left (\frac{1}{R_1}+\frac{1}{R_2}\right),
\end{equation}
where $\sigma$ is the surface tension, and $R_1$ and $R_2$ are the two
principal radii of curvature of the surface. In
Figure~\ref{FigTension}, we show the outcome of such an experiment in
2D, for densities $\rho_2=2$ and $\rho_1 = 1$ realized with equal mass
particles, a pressure of $P=2.5$, and 13 smoothing
neighbors. Independent of whether the high or low density phase is
arranged to be inside the sphere, the inner region shows a small
pressure difference relative to the outer region, which is $\Delta P
\sim 0.002$ at an effective resolution of $N^2 = 128^2$ for the lower
density phase. For a radius of $R_1=0.4$ of the sphere, this then
implies a surface tension of $\sigma = 0.0025$. We have confirmed that
this surface tension varies with one-dimensional resolution as $\sigma
\propto 1/N$, i.e.~it declines with higher spatial resolution. On the
other hand, it increases for higher density ratios, and for a larger
number of smoothing neighbours.

One consequence of this tension is that SPH in principle may support
capillary waves at interfaces, with dispersion relation $\omega^2 =
\sigma k^3 /(\rho_1 + \rho_2)$.  In the context of the
Kelvin-Helmholtz problem considered above, we note that surface
tension may also modify the growth rate of small wavelength
perturbations. In the presence of surface tension, the KH growth
timescale becomes
\begin{equation}
t_{\rm KH} = \left[ \frac{k^2\rho_1\rho_2(v_2-v_1)^2}{(\rho_1+\rho_2)^2} -
  \frac{\sigma \,k^3}{\rho_1+\rho_2}\right]^{-1/2} ,
\end{equation}
and waves with wavelength
\begin{equation}
\lambda < \lambda_{\rm crit} = 2\pi \,\frac{\rho_1+\rho_2}{\rho_1\rho_2}\frac{\sigma}{(v_2-v_1)^2} 
\end{equation}
will not grow at all. Using the value measured for $\sigma$ for the
numerical set-up we considered above, we obtain $\lambda_{\rm crit}/d
\simeq 3.0 / (v_2-v_1)^2$, where $d$ is the mean particle
separation. For a shear of $v_2-v_1 = 1$, we hence expect only waves
with wavelengths up to a couple of particle separations to be
suppressed. In particular, this effect should not have influenced the
KH tests carried out in the previous section. However, for small
shear, the effect can be much more of a problem, and here may
stabilize SPH against the KH instability.

\subsection{The Tensile Instability}

\begin{figure}
\begin{center}
\resizebox{4.5cm}{!}{\includegraphics{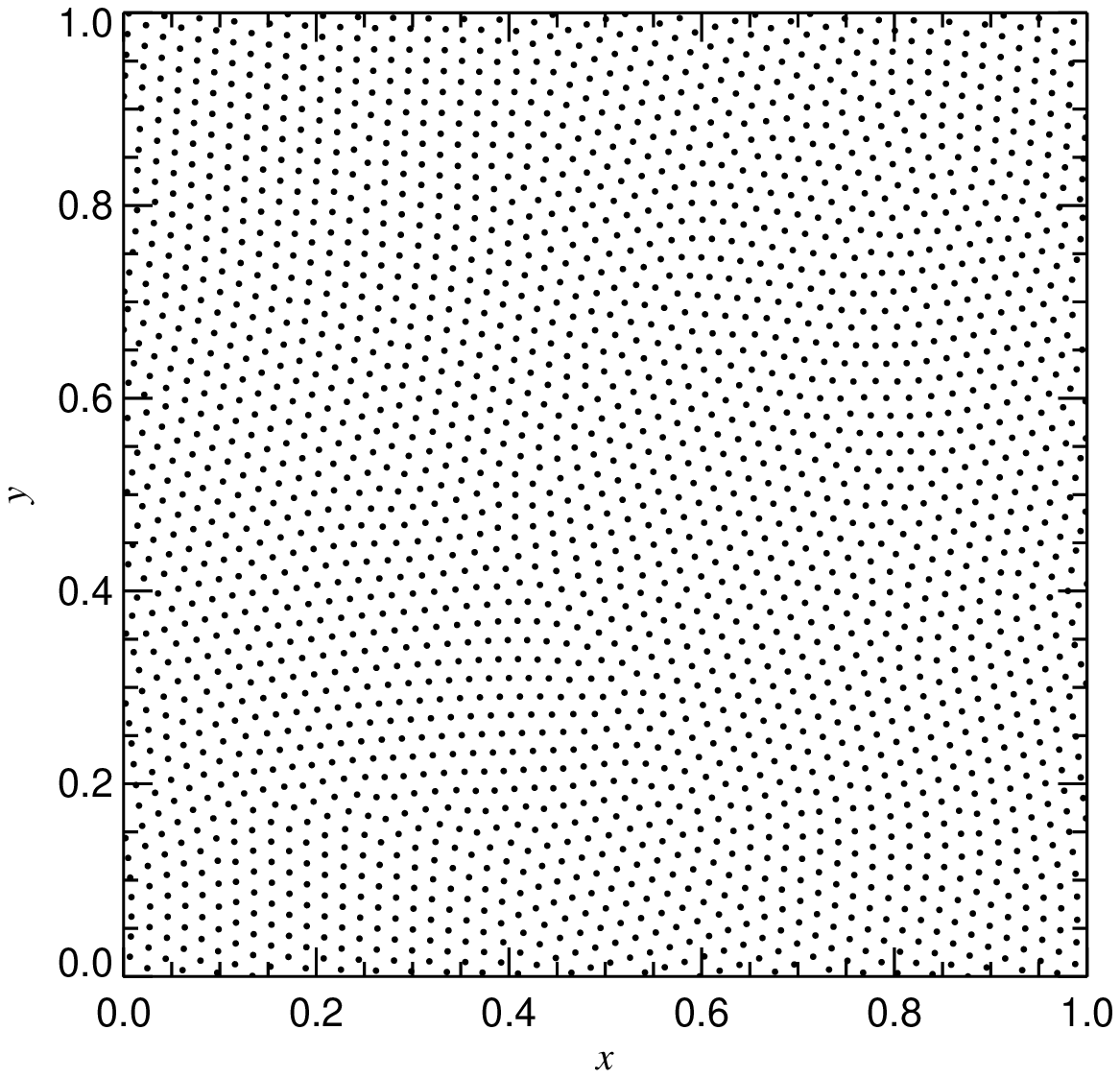}}%
\resizebox{4.5cm}{!}{\includegraphics{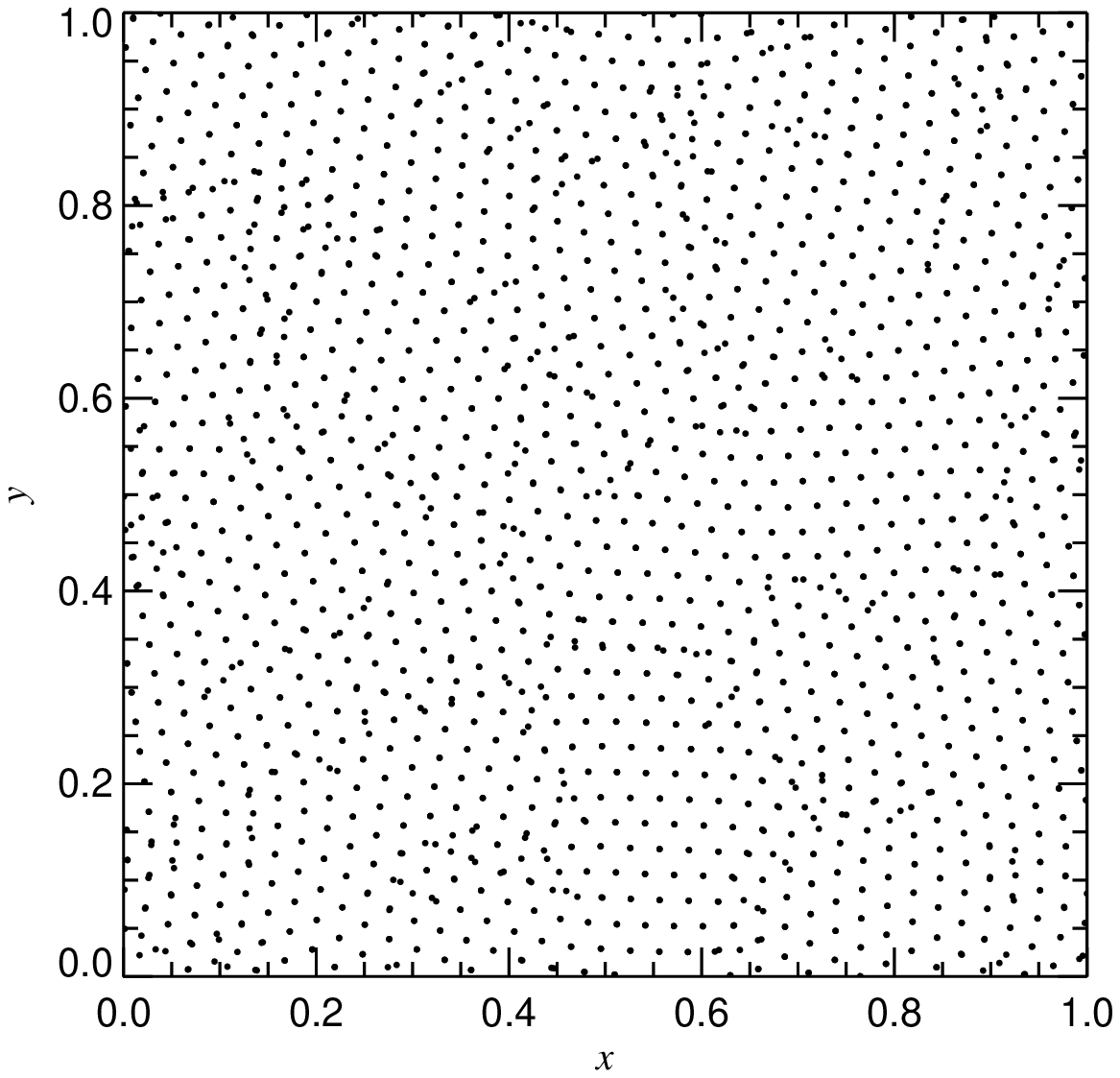}}%
\resizebox{4.5cm}{!}{\includegraphics{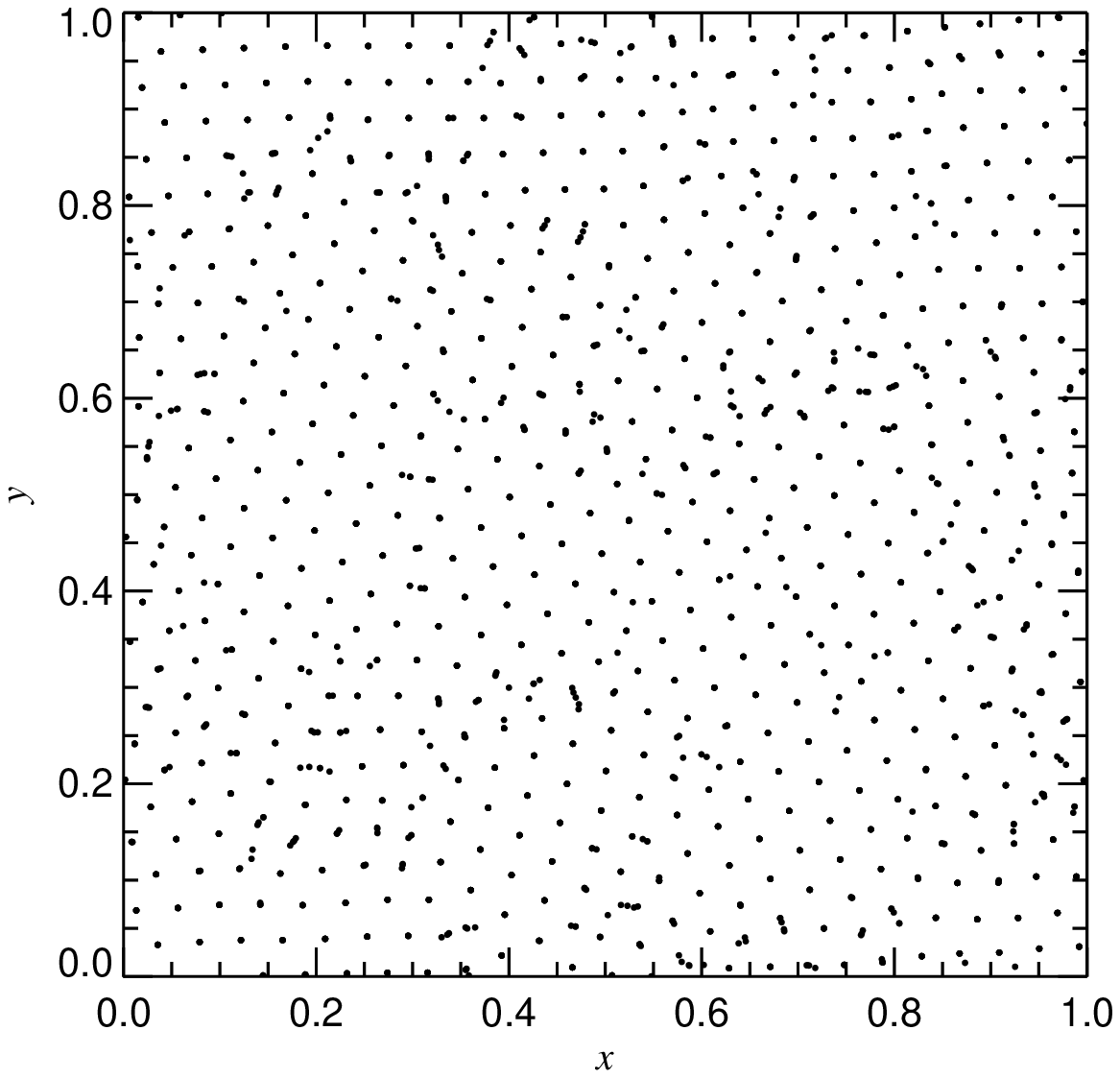}}
\vspace*{-0.8cm}
\end{center}
\caption{Clumping instability in a 2D settling test, using 13 (left
  panel), 27 (middle panel), or 55 (right panel) smoothing
  neighbours. In all three cases, the number of particles used
  ($50^2$) is the same.  It is seen that for a large number of
  smoothing neighbours, particles clump into groups, thereby reducing
  the effective spatial resolution of the scheme.
  \label{FigClumping}}
\end{figure}

First described by \citet{Swegle1995}, the clumping or tensile
instability is a well known nuisance in SPH that occurs if a large
number of smoothing neighbours is used. In this case, it can happen
that the net force between a close particle pair is not repulsive but
attractive, simply because the kernel gradient tends to become shallow
for close separations such that the pair may be further compressed by
other surrounding particles. As a result, particles may clump
together, thereby reducing the effective spatial resolution available
for the calculation.

Settling tests in which an initially random distribution of particles
with equal and fixed entropies is evolved towards an equilibrium state
under the influence of a friction force are a quite sensitive tool to
reveal the presence of this instability. We show an example of the
outcome of such a test in Figure~\ref{FigClumping}, where $50^2$
particles were randomly placed in a periodic 2D box of unit length,
and then evolved until a good equilibrium of the pressure forces was
reached, corresponding to a low energy state of the system. The
leftmost panel in Fig.~\ref{FigClumping} shows the result for 13
smoothing neighbours, the middle panel for 27 neighbours, and the
right panel for 55 neighbours. While for a low number of neighbours a
nicely regular, hexagonal particle distribution is obtained, for a
large number of smoothing neighbours many particle clumps are formed,
strongly reducing the number of independent sampling points for the
fluid.

Taken at face value, this seems to make attempts to use a large number
of neighbors to increase the accuracy of SPH calculations a futile
exercise. However, we note that the clumping instability is
essentially never observed in this extreme form in real applications
of SPH. This is because the dissipation added in the settling test is
not present in real application, and there the dynamics is usually
also not `cold' enough to create many artificial SPH particle clumps
or string-like configurations. Nevertheless, a number of suggestions
have been made to fix the tensile instability through modifications of
the SPH formalism. \citet{Steinmetz1996} simply invoke an artificially
steepened first derivative of the kernel to obtain a larger repulsive
pressure force at small separations.  \citet{Monaghan2000} proposed
instead an additional artificial pressure, which has the advantage of
maintaining the conservative properties of SPH, but it introduces some
small error in the dispersion relation. Finally, \citet{Read2009}
suggest the use of a peaked kernel, thereby ensuring large repulsive
pressure forces at small separations, at the price of a lower order of
the density estimate.

\section{ALTERNATIVE FORMULATIONS OF SPH AND FUTURE DIRECTIONS} \label{SecFuture}

There are many avenues for modifications of the standard SPH method
outlined in Section~\ref{SecBasic}, even though this usually implies
giving up the elegance of the Lagrangian derivation in favor of a more
heuristic construction of the discretization scheme. However, what
ultimately matters in the end is the achieved accuracy, and hence
there is certainly a lot of merit in investigating such alternative
schemes.  We here briefly mention some of the proposed modifications.

\subsection{Modifications of the Kernel Estimation Scheme} 

Changing the basic SPH kernel shape has been investigated in a number
of studies \citep{Monaghan1985,Fulk1996,Cabezon2008}, but generally no
kernel has been found that performed significantly better than the
cubic spline. In particular, higher order kernels may both be positive
and negative, which invokes numerous stability problems.

A few studies \citep{Shapiro1996,Owen1998} proposed anisotropic
kernels that adapt to local flow features. Together with the use of a
tensorial artificial viscosity, this promises an optimum use of the
available spatial resolution, and hence ultimately higher accuracy for
a given number of particles.

In the context of galaxy formation simulations and the so-called
overcooling problem, it has been pointed out that SPH's density
estimate in the vicinity of contact discontinuities may be problematic
\citep{Pearce1999}. This led \citet{Ritchie2001} to suggest a modified
density estimate specifically designed to work better across contact
discontinuities. To this extent, particles are weighted with their
thermal energies in the density estimate, which works well across
contacts. In particular, this removes the pressure blip that is often
found in contact waves that develop in Riemann problems. This idea has
also been recently reactivated by \citet{Read2009}, combined with the
suggestion of a much larger number of neighbors and a modified shape
of the smoothing kernel.

A more radical approach has been proposed by \citet{Borve2001}, who
introduced the so-called regularized SPH formalism. In it the
discretization errors of the ordinary kernel estimation are reduced by
introducing a new concept of intermediate interpolation cells. They
are also used for periodic redefinitions of the whole particle
distribution to obtain a better numerical description of the dynamics,
in particular of MHD related problems.

Another attempt to reduce the errors associated with kernel sums is
given by \citet{Imaeda2002} in terms of their `consistent velocity
method'. The central idea of this scheme is to achieve consistency
between the continuity equation and the density estimate.  In some
sense this makes it reminiscent of the simpler idea of `XSPH',
originally introduced by \citet{Monaghan1989} and later derived in
more accurate form from a Lagrangian \citep{Monaghan2002}.  In this
scheme the particles are moved with a SPH interpolated version of the
velocity, and not with the individual particle velocities.  This has
been used in some simulations to reduce problems of particle
interpenetration.

Finally, a higher order formulation of SPH has also been proposed
\citep{Oger2007}. It has been shown to eliminate the tensile
instability, but at the price of somewhat larger errors in the
gradient estimates.

\subsection{Improved Artificial Viscosity} 

The idea to use a time-variable coefficient for the parametrization of
the artificial viscosity is in general very interesting. It can help
to restrict the viscosity to regions where it is really needed
(i.e. in shocks), though it should be reduced to very small levels
everywhere else, ideally to the lowest level still consistent with
keeping the velocity noise of the scheme under control. There is
clearly still room for improvement of the switches used at present to
increase or decrease the viscosity, as discussed in
Section~\ref{SecVisco}.

A step further go attempts to outfit SPH with a Riemann solver
\citep[e.g.][]{Inutsuka2002,Cha2003}, which makes SPH more similar to
the Godunov approach used in Eulerian hydrodynamics. If successful,
this may in principle eliminate the need for an artificial viscosity
entirely, which would be a very welcome feature.  So far, these
versions of Godunov-based SPH have not found widespread application
though, presumably an indication that still more work is required to
make them equally robust as the standard SPH formulation in all
situations.

\subsection{Modelling Mixing}

As discussed earlier, the standard form of SPH has no mechanism to
equilibrate the specific entropies of neighboring particles, a
property that does not allow different phases to mix into a perfectly
homogeneous phase where all particle have the same thermodynamic
properties. As seen above, this even causes an artificial surface
tension effect. It has been suggested that the issue of mixing is
ultimately the primary cause for the differences seen in the central
entropy profiles of galaxy clusters simulated with SPH and mesh-based
Eulerian techniques \citep{Mitchell2009}.

Recently, there have been a few suggestions to treat mixing in SPH by
invoking artificial heat conduction.  \citet{Price2008_KH} has shown
that this improves the results for Kelvin-Helmholtz instability tests
when they are started with sharp discontinuities in the initial
conditions.  Also, in a study by \citet{Rosswog2007}, it was found
that this improves the accuracy of very strong shocks in Sedov-Taylor
blast waves. One problem is to parameterize the heat conductivity such
that it only affects the calculations where it is really needed, and
not anywhere else. The parameterization by \citet{Price2008_KH}, for
example, also operates for static contact discontinuities, where it is
not clear that any mixing should occur in such a case. A more physical
parametrization of the diffusive mixing has been suggested by
\citet{Wadsley2008} who use an estimate of the local velocity shear to
parametrize the mixing rate.

Yet another approach to mixing has been recently proposed by
\citet{Read2009}, who do not attempt to make the thermodynamic
properties of a mixed fluid equal at the particle level. Instead, the
number of smoothing neighbors is drastically enlarged by an order of
magnitude, such that resolution elements of the mixed phases are
sampled by a very large number of particles.

\subsection{Alternative Fluid Particle Models}

One possibility to substantially modify SPH lies in replacing the
kernel interpolation technique of SPH by a different technique to
estimate densities. This idea has been explored in the fluid particle
model of \citet{Hess2009}, where an auxiliary Voronoi tessellation was
used to define a density estimate based on the mass of a particle
divided by its associated Voronoi volume.  The latter is simply the
region of space that lies closer to the point than to any other
particle. The same Lagrangian as in Eqn.~(\ref{eqndisctLgr}) can then
be used to derive the equations of motion of the new particle based
scheme, where now no smoothing kernel is necessary. This highlights
that the kernel interpolation technique in principle only really
enters in the density estimate of SPH. If the density estimate is
replaced with another construction, different particle-based
hydrodynamical schemes can be constructed that are nevertheless
similar in spirit to SPH.  The Voronoi based scheme of
\citet{Hess2009} does not show the surface tension effect and has the
advantage of leading to a consistent partitioning of the volume. It
may also be more amenable to Riemann-solver based approaches to avoid
the need of artificial viscosity, because an unambiguous interaction
area between two particles (namely the area of the facet shared by the
two Voronoi cells) is given. However, it is not clear whether it can
significantly reduce the noise inherent in multidimensional SPH
calculations.

A considerable step further is the moving mesh technique devised by
\citet{Springel2009}. This also exploits a Voronoi mesh, but solves
the fluid equations with the finite-volume approach of the traditional
Eulerian Godunov approach, where a Riemann solver is used to work out
fluxes across boundaries. This is really a Eulerian scheme, but
adjusted to work with particle-based fluid cells.  In fact, as
\citet{Springel2009} discusses, this approach is identical the well
known Eulerian MUSCL scheme \citep{Leer1984} when the mesh-generating
points are kept fixed. If they are allowed to move with the flow, the
scheme behaves in a Lagrangian way similarly to SPH. Nevertheless, in
this mode the conceptual differences with respect to SPH also become
most apparent. The first difference is that the errors in the discrete
kernel sum are avoided through the use of an explicit second-order
accurate spatial reconstruction. The second difference is that the
cells/particles may exchange not only momentum, but also mass and
specific entropy, allowing already for the mixing that may happen in
multi-dimensional flow.

\section{CONCLUSIONS}  \label{SecConclusions}

Smoothed particle hydrodynamics is a remarkably versatile approach to
model gas dynamics in astrophysical simulations. The ease with which
it can provide a large dynamic range in spatial resolution and
density, as well as an automatically adaptive resolution, are
unmatched in Eulerian methods.  At the same time, SPH has excellent
conservation properties, not only for energy and linear momentum, but
also for angular momentum. The latter is not automatically guaranteed
in Eulerian codes, even though it is usually fulfilled at an
acceptable level for well-resolved flows. When coupled to
self-gravity, SPH conserves the total energy exactly, which is again
not manifestly true in mesh-based approaches to hydrodynamics.
Finally, SPH is Galilean-invariant and free of any errors from
advection alone, which is another significant advantage compared to
Eulerian approaches.

Thanks to its completely mesh-free nature, SPH can easily deal with
complicated geometric settings and large regions of space that are
completely devoid of particles. Implementations of SPH in a numerical
code tend to be comparatively simple and transparent. For example, it
is readily possible to include passively advected scalars in SPH (for
example chemical composition) in a straightforward and simple way.  At
the same time, the scheme is characterized by remarkable
robustness. For example, negative densities or negative temperatures,
sometimes a problem in mesh-based codes, can not occur in SPH by
construction. Although shock waves are broadened in SPH, the
properties of the post-shock flow are correct. Also, contact
discontinuities can even be narrower than in mesh-based codes.

All of these features make SPH a very interesting method for many
astrophysical problems. Indeed, a substantial fraction of the
simulation progress made in the past two decades on understanding
galaxy formation, star formation and planet formation has become
possible thanks to SPH.

The main disadvantage of SPH is clearly its limited accuracy in
multi-dimensional flows. One source of noise originates in the
approximation of local kernel interpolants through discrete sums over
a small set of nearest neighbours. While in 1D the consequences of
this noise tend to be quite benign, particle motion in multiple
dimensions has a much higher degree of freedom. Here the mutually
repulsive forces of pressurized neighbouring particle pairs do not
easily cancel in all dimensions simultaneously, especially not given
the errors of the discretized kernel interpolants. As a result, some
`jitter' in the particle motions readily develops, giving rise to
velocity noise up to a few percent of the local sound speed.  This
noise in SPH is likely also a primary cause for the mixed results
obtained thus far for MHD techniques implemented on top of SPH,
despite the considerable effort invested to make them work accurately.

In some sense it may seem surprising that SPH still models acceptable
fluid behaviour despite the presence of this comparatively large
noise. However, because SPH accurately respects the conservation laws
of fluid dynamics as described by the Lagrangian, the local
fluctuations tend to average out, enforcing the correct large-scale
motion of the fluid.  Fulfilling the conservation laws is hence
clearly more important than a high order of the underlying
scheme. However, as a consequence of this noise, individual SPH
particles cannot be readily interpreted as precise tracers of the
local state of the fluid at their representative location. Instead one
needs to carry out some sort of averaging or binning procedure
first. In this sense, SPH then indeed exhibits a kind of `Monte-Carlo'
character, as manifested also in its comparatively slow convergence
rate for multi-dimensional flow.

The noise inherent in SPH can be reduced by using a larger artificial
viscosity and/or a higher number of smoothing neighbors. Both
approaches do work at some level, but they are not without caveats and
hence need to be used with caution. The standard artificial viscosity
makes the simulated gas slightly viscous, which can affect the
calculated solutions. The results will then in fact not converge to
the solution expected for an inviscid gas, but to those of a slightly
viscous gas, which is really described by the appropriate
Navier-Stokes equations and not the Euler equations. However, improved
schemes for artificial viscosity, such as a time-dependent artificial
viscosity with judiciously chosen viscosity triggers, can improve on
this substantially.  Simply using a larger number of smoothing
neighbors is computationally more costly and invokes the danger of
suffering from artificial particle clumping. Here modified kernel
shapes and different viscosity prescriptions may provide a
satisfactory solution.

In the future, it will be of primary importance to make further
progress in understanding and improving the accuracy properties of SPH
in order to stay competitive with the recently matured
adaptive-mesh-refining and moving-mesh methodologies. If this can be
achieved, the SPH technique is bound to remain one of the primary
workhorses in computational astrophysics.

\section*{DISCLOSURE STATEMENT}

The author is not aware of any affiliations, memberships, funding, or
financial holdings that might be perceived as affecting the
objectivity of this review.

\section*{ACKNOWLEDGMENTS}

I apologize in advance to all researchers whose work could not be
cited due to space limitations.  I would like to thank Lars Hernquist,
Simon White, Stephan Rosswog, and Klaus Dolag for very helpful
comments on this review.
 
\bibliography{paper}

\begin{thebibliography}{}
\expandafter\ifx\csname natexlab\endcsname\relax\def\natexlab#1{#1}\fi

\bibitem[{{Agertz} et~al.(2007){Agertz}, {Moore}, {Stadel}, {Potter}, {Miniati}
  et~al.}]{Agertz2007}
{Agertz} O, {Moore} B, {Stadel} J, {Potter} D, {Miniati} F, et~al. 2007.
\newblock \textit{\mnras} 380:963--978

\bibitem[{{Altay}, {Croft} \& {Pelupessy}(2008)}]{Altay2008}
{Altay} G, {Croft} RAC, {Pelupessy} I. 2008.
\newblock \textit{\mnras} 386:1931--1946

\bibitem[{{Ayal} et~al.(2001){Ayal}, {Piran}, {Oechslin}, {Davies} \&
  {Rosswog}}]{Ayal2001}
{Ayal} S, {Piran} T, {Oechslin} R, {Davies} MB, {Rosswog} S. 2001.
\newblock \textit{\apj} 550:846--859

\bibitem[{{Bagla} \& {Khandai}(2009)}]{Bagla2009}
{Bagla} JS, {Khandai} N. 2009.
\newblock \textit{\mnras} 396:2211--2227

\bibitem[{{Balsara}(1995)}]{Balsara1995}
{Balsara} DS. 1995.
\newblock \textit{Journal of Computational Physics} 121:357--372

\bibitem[{{Barnes} \& {Hut}(1986)}]{Barnes1986}
{Barnes} J, {Hut} P. 1986.
\newblock \textit{\nat} 324:446--449

\bibitem[{{Barnes} \& {Hernquist}(1991)}]{Barnes1991}
{Barnes} JE, {Hernquist} LE. 1991.
\newblock \textit{\apjl} 370:L65--L68

\bibitem[{{Bate}(1998)}]{Bate1998}
{Bate} MR. 1998.
\newblock \textit{\apjl} 508:L95--L98

\bibitem[{{Bate} \& {Bonnell}(2005)}]{Bate2005}
{Bate} MR, {Bonnell} IA. 2005.
\newblock \textit{\mnras} 356:1201--1221

\bibitem[{{Bate} \& {Burkert}(1997)}]{Bate1997}
{Bate} MR, {Burkert} A. 1997.
\newblock \textit{\mnras} 288:1060--1072

\bibitem[{{Bate} et~al.(2003){Bate}, {Lubow}, {Ogilvie} \& {Miller}}]{Bate2003}
{Bate} MR, {Lubow} SH, {Ogilvie} GI, {Miller} KA. 2003.
\newblock \textit{\mnras} 341:213--229

\bibitem[{{Benz} \& {Asphaug}(1999)}]{Benz1999}
{Benz} W, {Asphaug} E. 1999.
\newblock \textit{Icarus} 142:5--20

\bibitem[{{Benz} et~al.(1990){Benz}, {Cameron}, {Press} \& {Bowers}}]{Benz1990}
{Benz} W, {Cameron} AGW, {Press} WH, {Bowers} RL. 1990.
\newblock \textit{\apj} 348:647--667

\bibitem[{{Benz} \& {Hills}(1987)}]{Benz1987}
{Benz} W, {Hills} JG. 1987.
\newblock \textit{\apj} 323:614--628

\bibitem[{{Benz}, {Slattery} \& {Cameron}(1986)}]{Benz1986}
{Benz} W, {Slattery} WL, {Cameron} AGW. 1986.
\newblock \textit{Icarus} 66:515--535

\bibitem[{{Borgani} et~al.(2004){Borgani}, {Murante}, {Springel}, {Diaferio},
  {Dolag} et~al.}]{Borgani2004}
{Borgani} S, {Murante} G, {Springel} V, {Diaferio} A, {Dolag} K, et~al. 2004.
\newblock \textit{\mnras} 348:1078--1096

\bibitem[{{B{\o}rve}, {Omang} \& {Trulsen}(2001)}]{Borve2001}
{B{\o}rve} S, {Omang} M, {Trulsen} J. 2001.
\newblock \textit{\apj} 561:82--93

\bibitem[{{Brandenburg}(2010)}]{Brandenburg2010}
{Brandenburg} A. 2010.
\newblock \textit{\mnras} 401:347--354

\bibitem[{{Bromm}, {Coppi} \& {Larson}(2002)}]{Bromm2002}
{Bromm} V, {Coppi} PS, {Larson} RB. 2002.
\newblock \textit{\apj} 564:23--51

\bibitem[{{Brookshaw}(1985)}]{Brookshaw1985}
{Brookshaw} L. 1985.
\newblock \textit{Proceedings of the Astronomical Society of Australia}
  6:207--210

\bibitem[{{Cabez{\'o}n}, {Garc{\'{\i}}a-Senz} \&
  {Rela{\~n}o}(2008)}]{Cabezon2008}
{Cabez{\'o}n} RM, {Garc{\'{\i}}a-Senz} D, {Rela{\~n}o} A. 2008.
\newblock \textit{Journal of Computational Physics} 227:8523--8540

\bibitem[{{Cha} \& {Whitworth}(2003)}]{Cha2003}
{Cha} S, {Whitworth} AP. 2003.
\newblock \textit{\mnras} 340:73--90

\bibitem[{{Clark}, {Glover} \& {Klessen}(2008)}]{Clark2008}
{Clark} PC, {Glover} SCO, {Klessen} RS. 2008.
\newblock \textit{\apj} 672:757--764

\bibitem[{{Cleary} \& {Monaghan}(1999)}]{Cleary1999}
{Cleary} PW, {Monaghan} JJ. 1999.
\newblock \textit{Journal of Computational Physics} 148:227--264

\bibitem[{{da Silva} et~al.(2000){da Silva}, {Barbosa}, {Liddle} \&
  {Thomas}}]{daSilva2000}
{da Silva} AC, {Barbosa} D, {Liddle} AR, {Thomas} PA. 2000.
\newblock \textit{\mnras} 317:37--44

\bibitem[{{Dav{\'e}} et~al.(2001){Dav{\'e}}, {Cen}, {Ostriker}, {Bryan},
  {Hernquist} et~al.}]{Dave2001}
{Dav{\'e}} R, {Cen} R, {Ostriker} JP, {Bryan} GL, {Hernquist} L, et~al. 2001.
\newblock \textit{\apj} 552:473--483

\bibitem[{{Dav{\'e}} et~al.(1999){Dav{\'e}}, {Hernquist}, {Katz} \&
  {Weinberg}}]{Dave1999}
{Dav{\'e}} R, {Hernquist} L, {Katz} N, {Weinberg} DH. 1999.
\newblock \textit{\apj} 511:521--545

\bibitem[{{Dedner} et~al.(2002){Dedner}, {Kemm}, {Kr{\"o}ner}, {Munz},
  {Schnitzer} \& {Wesenberg}}]{Dedner2002}
{Dedner} A, {Kemm} F, {Kr{\"o}ner} D, {Munz} C, {Schnitzer} T, {Wesenberg} M.
  2002.
\newblock \textit{Journal of Computational Physics} 175:645--673

\bibitem[{{Di Matteo}, {Springel} \& {Hernquist}(2005)}]{DiMatteo2005}
{Di Matteo} T, {Springel} V, {Hernquist} L. 2005.
\newblock \textit{\nat} 433:604--607

\bibitem[{{Dolag}, {Bartelmann} \& {Lesch}(1999)}]{Dolag1999}
{Dolag} K, {Bartelmann} M, {Lesch} H. 1999.
\newblock \textit{\aap} 348:351--363

\bibitem[{{Dolag}, {Bartelmann} \& {Lesch}(2002)}]{Dolag2002}
{Dolag} K, {Bartelmann} M, {Lesch} H. 2002.
\newblock \textit{\aap} 387:383--395

\bibitem[{{Dolag} et~al.(2008){Dolag}, {Borgani}, {Schindler}, {Diaferio} \&
  {Bykov}}]{Dolag2008Review}
{Dolag} K, {Borgani} S, {Schindler} S, {Diaferio} A, {Bykov} AM. 2008.
\newblock \textit{Space Science Reviews} 134:229--268

\bibitem[{{Dolag} et~al.(2005{\natexlab{a}}){Dolag}, {Grasso}, {Springel} \&
  {Tkachev}}]{Dolag2005}
{Dolag} K, {Grasso} D, {Springel} V, {Tkachev} I. 2005{\natexlab{a}}.
\newblock \textit{Journal of Cosmology and Astro-Particle Physics} 1:9

\bibitem[{{Dolag} et~al.(2004){Dolag}, {Jubelgas}, {Springel}, {Borgani} \&
  {Rasia}}]{Dolag2004}
{Dolag} K, {Jubelgas} M, {Springel} V, {Borgani} S, {Rasia} E. 2004.
\newblock \textit{\apjl} 606:L97--L100

\bibitem[{{Dolag} \& {Stasyszyn}(2009)}]{Dolag2009}
{Dolag} K, {Stasyszyn} F. 2009.
\newblock \textit{\mnras} 398:1678--1697

\bibitem[{{Dolag} et~al.(2005{\natexlab{b}}){Dolag}, {Vazza}, {Brunetti} \&
  {Tormen}}]{Dolag2005_turbulence}
{Dolag} K, {Vazza} F, {Brunetti} G, {Tormen} G. 2005{\natexlab{b}}.
\newblock \textit{\mnras} 364:753--772

\bibitem[{{Eckart}(1960)}]{Eckart1960}
{Eckart} C. 1960.
\newblock \textit{Physics of Fluids} 3:421--427

\bibitem[{{Evans} \& {Hawley}(1988)}]{Evans1988}
{Evans} CR, {Hawley} JF. 1988.
\newblock \textit{\apj} 332:659--677

\bibitem[{{Faber} \& {Rasio}(2000)}]{Faber2000}
{Faber} JA, {Rasio} FA. 2000.
\newblock \textit{\prd} 62:064012

\bibitem[{{Forgan} et~al.(2009){Forgan}, {Rice}, {Stamatellos} \&
  {Whitworth}}]{Forgan2009}
{Forgan} D, {Rice} K, {Stamatellos} D, {Whitworth} A. 2009.
\newblock \textit{\mnras} 394:882--891

\bibitem[{{Frenk} et~al.(1999){Frenk}, {White}, {Bode}, {Bond}, {Bryan}
  et~al.}]{Frenk1999}
{Frenk} CS, {White} SDM, {Bode} P, {Bond} JR, {Bryan} GL, et~al. 1999.
\newblock \textit{\apj} 525:554--582

\bibitem[{{Fryer}, {Rockefeller} \& {Warren}(2006)}]{Fryer2006}
{Fryer} CL, {Rockefeller} G, {Warren} MS. 2006.
\newblock \textit{\apj} 643:292--305

\bibitem[{{Fulk}(1996)}]{Fulk1996}
{Fulk} D. 1996.
\newblock \textit{Journal of Computational Physics} 126:165--180

\bibitem[{{Gingold} \& {Monaghan}(1977)}]{Gingold1977}
{Gingold} RA, {Monaghan} JJ. 1977.
\newblock \textit{\mnras} 181:375--389

\bibitem[{{Gingold} \& {Monaghan}(1982)}]{Gingold1982}
{Gingold} RA, {Monaghan} JJ. 1982.
\newblock \textit{Journal of Computational Physics} 46:429--453

\bibitem[{{Gnedin} \& {Abel}(2001)}]{Gnedin2001}
{Gnedin} NY, {Abel} T. 2001.
\newblock \textit{New Astronomy} 6:437--455

\bibitem[{{Governato} et~al.(2007){Governato}, {Willman}, {Mayer}, {Brooks},
  {Stinson} et~al.}]{Governato2007}
{Governato} F, {Willman} B, {Mayer} L, {Brooks} A, {Stinson} G, et~al. 2007.
\newblock \textit{\mnras} 374:1479--1494

\bibitem[{{Gresho} \& Chan(1990)}]{Gresho1990}
{Gresho} PM, Chan ST. 1990.
\newblock \textit{International Journal for Numerical Methods in Fluids}
  11:621--659

\bibitem[{Hairer, Lubich \& Wanner(2002)}]{Hairer2002}
Hairer E, Lubich C, Wanner G. 2002.
\newblock \textit{Geometric numerical integration}.
\newblock Springer Series in Computational Mathematics. Springer, Berlin

\bibitem[{{Hernquist}(1989)}]{Hernquist1989Natur}
{Hernquist} L. 1989.
\newblock \textit{\nat} 340:687--691

\bibitem[{{Hernquist}(1993)}]{Hernquist1993}
{Hernquist} L. 1993.
\newblock \textit{\apj} 404:717--722

\bibitem[{{Hernquist} \& {Katz}(1989)}]{Hernquist1989}
{Hernquist} L, {Katz} N. 1989.
\newblock \textit{\apjs} 70:419--446

\bibitem[{{Hernquist} et~al.(1996){Hernquist}, {Katz}, {Weinberg} \&
  {Miralda-Escud{\'e}}}]{Hernquist1996}
{Hernquist} L, {Katz} N, {Weinberg} DH, {Miralda-Escud{\'e}} J. 1996.
\newblock \textit{\apjl} 457:L51

\bibitem[{{He{\ss}} \& {Springel}(2010)}]{Hess2009}
{He{\ss}} S, {Springel} V. 2010.
\newblock \textit{\mnras} 406:2289--2311

\bibitem[{{Hockney} \& {Eastwood}(1981)}]{Hockney1981}
{Hockney} RW, {Eastwood} JW. 1981.
\newblock \textit{{Computer Simulation Using Particles}}

\bibitem[{{Hopkins} et~al.(2006){Hopkins}, {Hernquist}, {Cox}, {Di Matteo},
  {Robertson} \& {Springel}}]{Hopkins2006}
{Hopkins} PF, {Hernquist} L, {Cox} TJ, {Di Matteo} T, {Robertson} B, {Springel}
  V. 2006.
\newblock \textit{\apjs} 163:1--49

\bibitem[{{Hopkins} et~al.(2005){Hopkins}, {Hernquist}, {Martini}, {Cox},
  {Robertson} et~al.}]{Hopkins2005}
{Hopkins} PF, {Hernquist} L, {Martini} P, {Cox} TJ, {Robertson} B, et~al. 2005.
\newblock \textit{\apjl} 625:L71--L74

\bibitem[{{Imaeda} \& {Inutsuka}(2002)}]{Imaeda2002}
{Imaeda} Y, {Inutsuka} S. 2002.
\newblock \textit{\apj} 569:501--518

\bibitem[{{Inutsuka}(2002)}]{Inutsuka2002}
{Inutsuka} S. 2002.
\newblock \textit{Journal of Computational Physics} 179:238--267

\bibitem[{{Jubelgas}, {Springel} \& {Dolag}(2004)}]{Jubelgas2004}
{Jubelgas} M, {Springel} V, {Dolag} K. 2004.
\newblock \textit{\mnras} 351:423--435

\bibitem[{{Jubelgas} et~al.(2008){Jubelgas}, {Springel}, {En{\ss}lin} \&
  {Pfrommer}}]{Jubelgas2008}
{Jubelgas} M, {Springel} V, {En{\ss}lin} T, {Pfrommer} C. 2008.
\newblock \textit{\aap} 481:33--63

\bibitem[{{Katz}, {Weinberg} \& {Hernquist}(1996)}]{Katz1996}
{Katz} N, {Weinberg} DH, {Hernquist} L. 1996.
\newblock \textit{\apjs} 105:19

\bibitem[{{Kitsionas} \& {Whitworth}(2002)}]{Kitsionas2002}
{Kitsionas} S, {Whitworth} AP. 2002.
\newblock \textit{\mnras} 330:129--136

\bibitem[{{Kitsionas} \& {Whitworth}(2007)}]{Kitsionas2007}
{Kitsionas} S, {Whitworth} AP. 2007.
\newblock \textit{\mnras} 378:507--524

\bibitem[{{Klessen}, {Burkert} \& {Bate}(1998)}]{Klessen1998}
{Klessen} RS, {Burkert} A, {Bate} MR. 1998.
\newblock \textit{\apjl} 501:L205

\bibitem[{{Klessen}, {Heitsch} \& {Mac Low}(2000)}]{Klessen2000}
{Klessen} RS, {Heitsch} F, {Mac Low} M. 2000.
\newblock \textit{\apj} 535:887--906

\bibitem[{{Kotarba} et~al.(2009){Kotarba}, {Lesch}, {Dolag}, {Naab},
  {Johansson} \& {Stasyszyn}}]{Kotarba2009}
{Kotarba} H, {Lesch} H, {Dolag} K, {Naab} T, {Johansson} PH, {Stasyszyn} FA.
  2009.
\newblock \textit{\mnras} 397:733--747

\bibitem[{{Laguna}, {Miller} \& {Zurek}(1993)}]{Laguna1993}
{Laguna} P, {Miller} WA, {Zurek} WH. 1993.
\newblock \textit{\apj} 404:678--685

\bibitem[{{Landau} \& {Lifshitz}(1959)}]{Landau1959}
{Landau} LD, {Lifshitz} EM. 1959.
\newblock \textit{{Fluid mechanics}}

\bibitem[{{Lanzafame}, {Molteni} \& {Chakrabarti}(1998)}]{Lanzafame1998}
{Lanzafame} G, {Molteni} D, {Chakrabarti} SK. 1998.
\newblock \textit{\mnras} 299:799--804

\bibitem[{Liska \& Wendroff(2003)}]{Liska2003}
Liska R, Wendroff B. 2003.
\newblock \textit{SIAM J. Sci. Comput.} 25:995--1017

\bibitem[{{Lombardi} et~al.(1999){Lombardi}, {Sills}, {Rasio} \&
  {Shapiro}}]{Lombardi1999}
{Lombardi} JC, {Sills} A, {Rasio} FA, {Shapiro} SL. 1999.
\newblock \textit{Journal of Computational Physics} 152:687--735

\bibitem[{{Lucy}(1977)}]{Lucy1977}
{Lucy} LB. 1977.
\newblock \textit{\aj} 82:1013--1024

\bibitem[{{Lufkin} et~al.(2004){Lufkin}, {Quinn}, {Wadsley}, {Stadel} \&
  {Governato}}]{Lufkin2004}
{Lufkin} G, {Quinn} T, {Wadsley} J, {Stadel} J, {Governato} F. 2004.
\newblock \textit{\mnras} 347:421--429

\bibitem[{{Marri} \& {White}(2003)}]{Marri2003}
{Marri} S, {White} SDM. 2003.
\newblock \textit{\mnras} 345:561--574

\bibitem[{{Mayer} et~al.(2002){Mayer}, {Quinn}, {Wadsley} \&
  {Stadel}}]{Mayer2002}
{Mayer} L, {Quinn} T, {Wadsley} J, {Stadel} J. 2002.
\newblock \textit{Science} 298:1756--1759

\bibitem[{{Mayer} et~al.(2004){Mayer}, {Quinn}, {Wadsley} \&
  {Stadel}}]{Mayer2004}
{Mayer} L, {Quinn} T, {Wadsley} J, {Stadel} J. 2004.
\newblock \textit{\apj} 609:1045--1064

\bibitem[{{McCarthy} et~al.(2010){McCarthy}, {Schaye}, {Ponman}, {Bower},
  {Booth} et~al.}]{McCarthy2009}
{McCarthy} IG, {Schaye} J, {Ponman} TJ, {Bower} RG, {Booth} CM, et~al. 2010.
\newblock \textit{\mnras} 406:822--839

\bibitem[{{Mihos} \& {Hernquist}(1994)}]{Mihos1994}
{Mihos} JC, {Hernquist} L. 1994.
\newblock \textit{\apjl} 431:L9--L12

\bibitem[{{Mihos} \& {Hernquist}(1996)}]{Mihos1996}
{Mihos} JC, {Hernquist} L. 1996.
\newblock \textit{\apj} 464:641

\bibitem[{{Mitchell} et~al.(2009){Mitchell}, {McCarthy}, {Bower}, {Theuns} \&
  {Crain}}]{Mitchell2009}
{Mitchell} NL, {McCarthy} IG, {Bower} RG, {Theuns} T, {Crain} RA. 2009.
\newblock \textit{\mnras} 395:180--196

\bibitem[{{Monaghan}(1985)}]{Monaghan1985}
{Monaghan} J. 1985.
\newblock \textit{Computer Physics Reports} 3:71--124

\bibitem[{{Monaghan}(1989)}]{Monaghan1989}
{Monaghan} JJ. 1989.
\newblock \textit{Journal of Computational Physics} 82:1--15

\bibitem[{{Monaghan}(1992)}]{Monaghan1992}
{Monaghan} JJ. 1992.
\newblock \textit{\araa} 30:543--574

\bibitem[{{Monaghan}(1997)}]{Monaghan1997}
{Monaghan} JJ. 1997.
\newblock \textit{Journal of Computational Physics} 136:298--307

\bibitem[{{Monaghan}(2000)}]{Monaghan2000}
{Monaghan} JJ. 2000.
\newblock \textit{Journal of Computational Physics} 159:290--311

\bibitem[{{Monaghan}(2002)}]{Monaghan2002}
{Monaghan} JJ. 2002.
\newblock \textit{\mnras} 335:843--852

\bibitem[{{Monaghan}(2005)}]{Monaghan2005}
{Monaghan} JJ. 2005.
\newblock \textit{Reports on Progress in Physics} 68:1703--1759

\bibitem[{{Monaghan} \& {Gingold}(1983)}]{Monaghan1983}
{Monaghan} JJ, {Gingold} RA. 1983.
\newblock \textit{Journal of Computational Physics} 52:374

\bibitem[{{Monaghan} \& {Price}(2001)}]{Monaghan2001}
{Monaghan} JJ, {Price} DJ. 2001.
\newblock \textit{\mnras} 328:381--392

\bibitem[{{Morris}(1997)}]{Morris1997}
{Morris} J. 1997.
\newblock \textit{Journal of Computational Physics} 136:41--50

\bibitem[{{M{\"u}ller} \& {Steinmetz}(1995)}]{Mueller1995}
{M{\"u}ller} E, {Steinmetz} M. 1995.
\newblock \textit{Computer Physics Communications} 89:45--58

\bibitem[{{Nayakshin}, {Cha} \& {Hobbs}(2009)}]{Nayakshin2009}
{Nayakshin} S, {Cha} S, {Hobbs} A. 2009.
\newblock \textit{\mnras} 397:1314--1325

\bibitem[{{Nelson} \& {Papaloizou}(1994)}]{Nelson1994}
{Nelson} RP, {Papaloizou} JCB. 1994.
\newblock \textit{\mnras} 270:1

\bibitem[{{Oechslin}, {Janka} \& {Marek}(2007)}]{Oechslin2007}
{Oechslin} R, {Janka} H, {Marek} A. 2007.
\newblock \textit{\aap} 467:395--409

\bibitem[{{Oger} et~al.(2007){Oger}, {Doring}, {Alessandrini} \&
  {Ferrant}}]{Oger2007}
{Oger} G, {Doring} M, {Alessandrini} B, {Ferrant} P. 2007.
\newblock \textit{Journal of Computational Physics} 225:1472--1492

\bibitem[{{Oppenheimer} \& {Dav{\'e}}(2006)}]{Oppenheimer2006}
{Oppenheimer} BD, {Dav{\'e}} R. 2006.
\newblock \textit{\mnras} 373:1265--1292

\bibitem[{{O'Shea} et~al.(2005){O'Shea}, {Nagamine}, {Springel}, {Hernquist} \&
  {Norman}}]{Oshea2005}
{O'Shea} BW, {Nagamine} K, {Springel} V, {Hernquist} L, {Norman} ML. 2005.
\newblock \textit{\apjs} 160:1--27

\bibitem[{{Owen}(2004)}]{Owen2004}
{Owen} JM. 2004.
\newblock \textit{Journal of Computational Physics} 201:601--629

\bibitem[{{Owen} et~al.(1998){Owen}, {Villumsen}, {Shapiro} \&
  {Martel}}]{Owen1998}
{Owen} JM, {Villumsen} JV, {Shapiro} PR, {Martel} H. 1998.
\newblock \textit{\apjs} 116:155

\bibitem[{{Pakmor} et~al.(2010){Pakmor}, {Kromer}, {R{\"o}pke}, {Sim}, {Ruiter}
  \& {Hillebrandt}}]{Pakmor2009}
{Pakmor} R, {Kromer} M, {R{\"o}pke} FK, {Sim} SA, {Ruiter} AJ, {Hillebrandt} W.
  2010.
\newblock \textit{\nat} 463:61--64

\bibitem[{{Pawlik} \& {Schaye}(2008)}]{Pawlik2008}
{Pawlik} AH, {Schaye} J. 2008.
\newblock \textit{\mnras} 389:651--677

\bibitem[{{Pearce} \& {Couchman}(1997)}]{Pearce1997}
{Pearce} FR, {Couchman} HMP. 1997.
\newblock \textit{New Astronomy} 2:411--427

\bibitem[{{Pearce} et~al.(1999){Pearce}, {Jenkins}, {Frenk}, {Colberg}, {White}
  et~al.}]{Pearce1999}
{Pearce} FR, {Jenkins} A, {Frenk} CS, {Colberg} JM, {White} SDM, et~al. 1999.
\newblock \textit{\apjl} 521:L99--L102

\bibitem[{{Petkova} \& {Springel}(2009)}]{Petkova2009}
{Petkova} M, {Springel} V. 2009.
\newblock \textit{\mnras} 396:1383--1403

\bibitem[{{Pfrommer} et~al.(2007){Pfrommer}, {En{\ss}lin}, {Springel},
  {Jubelgas} \& {Dolag}}]{Pfrommer2007}
{Pfrommer} C, {En{\ss}lin} TA, {Springel} V, {Jubelgas} M, {Dolag} K. 2007.
\newblock \textit{\mnras} 378:385--408

\bibitem[{{Pfrommer} et~al.(2006){Pfrommer}, {Springel}, {En{\ss}lin} \&
  {Jubelgas}}]{Pfrommer2006}
{Pfrommer} C, {Springel} V, {En{\ss}lin} TA, {Jubelgas} M. 2006.
\newblock \textit{\mnras} 367:113--131

\bibitem[{{Phillips} \& {Monaghan}(1985)}]{Phillips1985}
{Phillips} GJ, {Monaghan} JJ. 1985.
\newblock \textit{\mnras} 216:883--895

\bibitem[{{Price}(2008)}]{Price2008_KH}
{Price} DJ. 2008.
\newblock \textit{Journal of Computational Physics} 227:10040--10057

\bibitem[{{Price}(2009)}]{Price2009_MHDvec}
{Price} DJ. 2009.
\newblock \textit{\mnras} :1651

\bibitem[{{Price} \& {Bate}(2007)}]{Price2007}
{Price} DJ, {Bate} MR. 2007.
\newblock \textit{\mnras} 377:77--90

\bibitem[{{Price} \& {Monaghan}(2004{\natexlab{a}})}]{Price2004_MHD1}
{Price} DJ, {Monaghan} JJ. 2004{\natexlab{a}}.
\newblock \textit{\mnras} 348:123--138

\bibitem[{{Price} \& {Monaghan}(2004{\natexlab{b}})}]{Price2004_MHD2}
{Price} DJ, {Monaghan} JJ. 2004{\natexlab{b}}.
\newblock \textit{\mnras} 348:139--152

\bibitem[{{Price} \& {Monaghan}(2005)}]{Price2005_MHD3}
{Price} DJ, {Monaghan} JJ. 2005.
\newblock \textit{\mnras} 364:384--406

\bibitem[{{Price} \& {Monaghan}(2007)}]{Price2007_adapt}
{Price} DJ, {Monaghan} JJ. 2007.
\newblock \textit{\mnras} 374:1347--1358

\bibitem[{{Puchwein}, {Sijacki} \& {Springel}(2008)}]{Puchwein2008}
{Puchwein} E, {Sijacki} D, {Springel} V. 2008.
\newblock \textit{\apjl} 687:L53--L56

\bibitem[{{Rasio}(2000)}]{Rasio2000}
{Rasio} FA. 2000.
\newblock \textit{Progress of Theoretical Physics Supplement} 138:609--621

\bibitem[{Rasio \& Shapiro(1991)}]{Rasio1991}
Rasio FA, Shapiro SL. 1991.
\newblock \textit{ApJ} 377:559

\bibitem[{{Read}, {Hayfield} \& {Agertz}(2010)}]{Read2009}
{Read} JI, {Hayfield} T, {Agertz} O. 2010.
\newblock \textit{\mnras} 405:1513--1530

\bibitem[{{Ritchie} \& {Thomas}(2001)}]{Ritchie2001}
{Ritchie} BW, {Thomas} PA. 2001.
\newblock \textit{\mnras} 323:743--756

\bibitem[{{Robertson} et~al.(2009){Robertson}, {Kravtsov}, {Gnedin}, {Abel} \&
  {Rudd}}]{Robertson2009}
{Robertson} BE, {Kravtsov} AV, {Gnedin} NY, {Abel} T, {Rudd} DH. 2009.
\newblock \textit{\mnras} :1774

\bibitem[{{Rosswog}(2005)}]{Rosswog2005}
{Rosswog} S. 2005.
\newblock \textit{\apj} 634:1202--1213

\bibitem[{{Rosswog}(2009)}]{Rosswog2009}
{Rosswog} S. 2009.
\newblock \textit{New Astronomy Review} 53:78--104

\bibitem[{{Rosswog} \& {Price}(2007)}]{Rosswog2007}
{Rosswog} S, {Price} D. 2007.
\newblock \textit{\mnras} 379:915--931

\bibitem[{{Saitoh} \& {Makino}(2009)}]{Saitoh2009}
{Saitoh} TR, {Makino} J. 2009.
\newblock \textit{\apjl} 697:L99--L102

\bibitem[{{Scannapieco} et~al.(2008){Scannapieco}, {Tissera}, {White} \&
  {Springel}}]{Scannapieco2008}
{Scannapieco} C, {Tissera} PB, {White} SDM, {Springel} V. 2008.
\newblock \textit{\mnras} 389:1137--1149

\bibitem[{{Schoenberg}(1969)}]{Schoenberg1969}
{Schoenberg} I. 1969.
\newblock \textit{Journal of Approximation Theory} 2:167--206

\bibitem[{{Shapiro} et~al.(1996){Shapiro}, {Martel}, {Villumsen} \&
  {Owen}}]{Shapiro1996}
{Shapiro} PR, {Martel} H, {Villumsen} JV, {Owen} JM. 1996.
\newblock \textit{\apjs} 103:269

\bibitem[{{Siegler} \& {Riffert}(2000)}]{Siegler2000}
{Siegler} S, {Riffert} H. 2000.
\newblock \textit{\apj} 531:1053--1066

\bibitem[{{Sijacki} et~al.(2008){Sijacki}, {Pfrommer}, {Springel} \&
  {En{\ss}lin}}]{Sijacki2008}
{Sijacki} D, {Pfrommer} C, {Springel} V, {En{\ss}lin} TA. 2008.
\newblock \textit{\mnras} 387:1403--1415

\bibitem[{{Sijacki} \& {Springel}(2006)}]{Sijacki2006}
{Sijacki} D, {Springel} V. 2006.
\newblock \textit{\mnras} 371:1025--1046

\bibitem[{{Simpson}(1995)}]{Simpson1995}
{Simpson} JC. 1995.
\newblock \textit{\apj} 448:822

\bibitem[{{Smith}, {Clark} \& {Bonnell}(2009)}]{Smith2009}
{Smith} RJ, {Clark} PC, {Bonnell} IA. 2009.
\newblock \textit{\mnras} 396:830--841

\bibitem[{{Springel}(2005)}]{Springel2005}
{Springel} V. 2005.
\newblock \textit{\mnras} 364:1105--1134

\bibitem[{{Springel}(2009)}]{Springel2009}
{Springel} V. 2009.
\newblock \textit{\mnras} :1655

\bibitem[{{Springel}, {Di Matteo} \& {Hernquist}(2005)}]{Springel2005_BH}
{Springel} V, {Di Matteo} T, {Hernquist} L. 2005.
\newblock \textit{\mnras} 361:776--794

\bibitem[{{Springel} \& {Hernquist}(2002)}]{Springel2002}
{Springel} V, {Hernquist} L. 2002.
\newblock \textit{\mnras} 333:649--664

\bibitem[{{Springel} \& {Hernquist}(2003)}]{Springel2003_multi}
{Springel} V, {Hernquist} L. 2003.
\newblock \textit{\mnras} 339:289--311

\bibitem[{{Springel}, {Yoshida} \&
  {White}(2001{\natexlab{a}})}]{Springel2001gadget}
{Springel} V, {Yoshida} N, {White} SDM. 2001{\natexlab{a}}.
\newblock \textit{New Astronomy} 6:79--117

\bibitem[{{Springel}, {Yoshida} \& {White}(2001{\natexlab{b}})}]{Springel2001}
{Springel} V, {Yoshida} N, {White} SDM. 2001{\natexlab{b}}.
\newblock \textit{New Astronomy} 6:79--117

\bibitem[{{Steinmetz}(1996)}]{Steinmetz1996}
{Steinmetz} M. 1996.
\newblock \textit{\mnras} 278:1005--1017

\bibitem[{{Steinmetz} \& {Mueller}(1993)}]{Steinmetz1993}
{Steinmetz} M, {Mueller} E. 1993.
\newblock \textit{\aap} 268:391--410

\bibitem[{{Stone} et~al.(2008){Stone}, {Gardiner}, {Teuben}, {Hawley} \&
  {Simon}}]{Stone2008}
{Stone} JM, {Gardiner} TA, {Teuben} P, {Hawley} JF, {Simon} JB. 2008.
\newblock \textit{\apjs} 178:137--177

\bibitem[{{Swegle}(1995)}]{Swegle1995}
{Swegle} J. 1995.
\newblock \textit{Journal of Computational Physics} 116:123--134

\bibitem[{{Tasker} et~al.(2008){Tasker}, {Brunino}, {Mitchell}, {Michielsen},
  {Hopton} et~al.}]{Tasker2008}
{Tasker} EJ, {Brunino} R, {Mitchell} NL, {Michielsen} D, {Hopton} S, et~al.
  2008.
\newblock \textit{\mnras} 390:1267--1281

\bibitem[{{Thacker} et~al.(2000){Thacker}, {Tittley}, {Pearce}, {Couchman} \&
  {Thomas}}]{Thacker2000}
{Thacker} RJ, {Tittley} ER, {Pearce} FR, {Couchman} HMP, {Thomas} PA. 2000.
\newblock \textit{\mnras} 319:619--648

\bibitem[{{Toomre} \& {Toomre}(1972)}]{Toomre1972}
{Toomre} A, {Toomre} J. 1972.
\newblock \textit{\apj} 178:623--666

\bibitem[{{Toro}(1997)}]{Toro1997}
{Toro} E. 1997.
\newblock \textit{Riemann solvers and numerical methods for fluid dynamics}.
\newblock Springer

\bibitem[{{van Leer}(1984)}]{Leer1984}
{van Leer} B. 1984.
\newblock \textit{SIAM J. Sci. Stat. Comput.} 5:1

\bibitem[{{Viau}, {Bastien} \& {Cha}(2006)}]{Viau2006}
{Viau} S, {Bastien} P, {Cha} S. 2006.
\newblock \textit{\apj} 639:559--570

\bibitem[{{Wadsley}, {Stadel} \& {Quinn}(2004)}]{Wadsley2004}
{Wadsley} JW, {Stadel} J, {Quinn} T. 2004.
\newblock \textit{New Astronomy} 9:137--158

\bibitem[{{Wadsley}, {Veeravalli} \& {Couchman}(2008)}]{Wadsley2008}
{Wadsley} JW, {Veeravalli} G, {Couchman} HMP. 2008.
\newblock \textit{\mnras} 387:427--438

\bibitem[{{Wetzstein} et~al.(2009){Wetzstein}, {Nelson}, {Naab} \&
  {Burkert}}]{Wetzstein2009}
{Wetzstein} M, {Nelson} AF, {Naab} T, {Burkert} A. 2009.
\newblock \textit{\apjs} 184:298--325

\bibitem[{{Whitehouse} \& {Bate}(2004)}]{Whitehouse2004}
{Whitehouse} SC, {Bate} MR. 2004.
\newblock \textit{\mnras} 353:1078--1094

\bibitem[{{Whitehouse} \& {Bate}(2006)}]{Whitehouse2006}
{Whitehouse} SC, {Bate} MR. 2006.
\newblock \textit{\mnras} 367:32--38

\bibitem[{{Whitehouse}, {Bate} \& {Monaghan}(2005)}]{Whitehouse2005}
{Whitehouse} SC, {Bate} MR, {Monaghan} JJ. 2005.
\newblock \textit{\mnras} 364:1367--1377

\bibitem[{{Whitworth}(1998)}]{Whitworth1998}
{Whitworth} AP. 1998.
\newblock \textit{\mnras} 296:442--444

\bibitem[{{Williams}, {Churches} \& {Nelson}(2004)}]{Williams2004}
{Williams} PR, {Churches} DK, {Nelson} AH. 2004.
\newblock \textit{\apj} 607:1--19

\bibitem[{{Wood} et~al.(2005){Wood}, {Robertson}, {Simpson}, {Kawaler},
  {O'Brien} et~al.}]{Wood2005}
{Wood} MA, {Robertson} JR, {Simpson} JC, {Kawaler} SD, {O'Brien} MS, et~al.
  2005.
\newblock \textit{\apj} 634:570--584

\bibitem[{{Yoshida} et~al.(2006){Yoshida}, {Omukai}, {Hernquist} \&
  {Abel}}]{Yoshida2006}
{Yoshida} N, {Omukai} K, {Hernquist} L, {Abel} T. 2006.
\newblock \textit{\apj} 652:6--25

\bibitem[{{Yukawa}, {Boffin} \& {Matsuda}(1997)}]{Yukawa1997}
{Yukawa} H, {Boffin} HMJ, {Matsuda} T. 1997.
\newblock \textit{\mnras} 292:321

\end{thebibliography}

\end{document}